\documentclass[amsmath,amssymb,aps,floats,amsfonts,notitlepage,superscriptaddress,eqsecnum,nofootinbib,prd,a4paper,longbibliography]{revtex4-1}
\allowdisplaybreaks

\usepackage[utf8]{inputenc}
\usepackage[]{graphicx}
\usepackage{hyperref}
\usepackage{url}
\usepackage{color}
\usepackage[usenames,dvipsnames,svgnames,table]{xcolor}
\usepackage{multirow}
\usepackage{mathrsfs}
\usepackage{orcidlink}
\usepackage{subcaption}
\usepackage[normalem]{ulem}
\usepackage{enumitem}

\newcommand{\Disc}{\delta}

\newcommand{\Gret}{G_{\rm ret}}

\newcommand{\scri}{{\mathscr I}}
\newcommand{\hor}{{\mathscr H}}

\newcommand{\indmode}{\ell m}%{\ell,,m\omega} %mode index for the radial functions and coefficients

\newcommand{\Slm}{ S_{\ell m\omega}}

\newcommand{\Ruphat}[1]{{}_{#1}\hat{R}^{\text{up}}_{\ell m}}
\newcommand{\Ruphatmm}[1]{{}_{#1}\hat{R}^{\text{up}}_{\ell ,-m}}

\newcommand{\Rin}[1]{{}_{#1}R^{\text{in}}_{\ell m}}
\newcommand{\Rinmm}[1]{{}_{#1}R^{\text{in}}_{\ell ,-m}}
\newcommand{\Rup}[1]{{}_{#1}R^{\text{up}}_{\ell m}}

\newcommand{\RinST}[1]{{}_{#1}R^{\text{in,ST}}_{\ell m}}

\newcommand{\Rupphat}{\hat{R}^{\text{up},+}_{\ell m}}
\newcommand{\Rupmhat}{\hat{R}^{\text{up},-}_{\ell m}}

\newcommand{\Rinup}[1]{{}_{#1}R^{\text{in/up}}_{\ell m}}
\newcommand{\Ctra}{C^{\text{tra}}}
\newcommand{\Cinc}{C^{\text{inc}}}
\newcommand{\Cref}{C^{\text{ref}}}
\newcommand{\Btra}{B^{\text{tra}}}
\newcommand{\Binc}{B^{\text{inc}}}
\newcommand{\Bref}{B^{\text{ref}}}

\newcommand{\Bincp}{B^{\text{inc},+}}
\newcommand{\Brefp}{B^{\text{ref},+}}

\newcommand{\Bincreftra}{B^{\text{inc/ref/tra}}}
\newcommand{\Cincreftra}{C^{\text{inc/ref/tra}}}

\newcommand{\nn}{\nonumber}
\newcommand{\an}[1]{a_{#1}^{\nu}}

\newcommand{\Wbp}{W^+}
\newcommand{\Wbm}{W^-}

\newcommand{\reft}{r}
\newcommand{\tra}{t}
\newcommand{\logt}{l}
\newcommand{\logtt}{ll}
\newcommand{\Q}{Q}

\newcommand{\SLett}{S}
\newcommand{\St}[1]{\SLett_{#1}}

\newcommand{\Bi}{\Binc}
\newcommand{\Br}{\Bref}
\newcommand{\Bt}{\Btra}

\newcommand{\BiW}[1]{\Bi_{#1}}

\newcommand{\BrW}[1]{\Br_{#1}}

\newcommand{\BtW}[1]{\Bt_{#1}}

\newcommand{\qW}[1]{\Q_{#1}}
\newcommand{\qWLog}[1]{\Q_{#1}^{\rm \logt}}
\newcommand{\qWref}[1]{\Q_{#1}^{\rm \reft}}
\newcommand{\qWreflog}[1]{\Q_{#1}^{\rm \reft,\logt}}
\newcommand{\qWrefloglog}[1]{\Q_{#1}^{\rm \reft,\logtt}}
\newcommand{\qWtra}[1]{\Q_{#1}^{\rm \tra}}
\newcommand{\qWtralog}[1]{\Q_{#1}^{\rm \tra,\logt}}

\newcommand{\g}[1]{g_{#1}}
\newcommand{\gLog}[1]{g_{#1}^{\rm \logt}}
\newcommand{\gtra}[1]{g_{#1}^{\rm \tra}}
\newcommand{\gtralog}[1]{g_{#1}^{\rm \tra,\logt}}
\newcommand{\gref}[1]{g_{#1}^{\rm \reft}}
\newcommand{\greflog}[1]{g_{#1}^{\rm \reft,\logt}}
\newcommand{\grefloglog}[1]{g_{#1}^{\rm \reft,\logtt}}

\newcommand{\G}[1]{G_{#1}}
\newcommand{\GLog}[1]{G_{#1}^{\rm \logt}}
\newcommand{\Grp}[1]{G_{#1}^{(+)}}
\newcommand{\GrpLog}[1]{G_{#1}^{(+,{\rm \logt})}}
\newcommand{\Ginf}[1]{G_{#1}^{(\infty)}}
\newcommand{\GinfLog}[1]{G_{#1}^{(\infty,{\rm \logt})}}
\newcommand{\GinfLogLog}[1]{G_{#1}^{(\infty,{\rm \logtt})}}

\newcommand{\rST}{\mathfrak{r}}
\newcommand{\ExpQ}{\alpha}
\newcommand{\ExpQRef}{\beta}
\newcommand{\ExpQTra}{\delta}
\newcommand{\Expr}{\zeta}

\newcommand{\Glm}{G_{\indmode}}

\newcommand{\tort}{r_*}
\newcommand{\fNIA}{\sigma}

\newcommand{\pdiff}[2]  {\frac{\partial #1}{\partial #2}}
\newcommand{\spdiff}[2] {\frac{\partial^2 #1}{\partial #2^2}}

\begin{document}

\title{High-order gravitational late-time tails in Kerr spacetime}

\author{Marc Casals\, \orcidlink{0000-0002-8914-4072}}
\email{marc.casals@uni-leipzig.de }
\affiliation{
%\address{
Institut f\"ur Theoretische Physik, Universit\"at Leipzig,\\ Br\"uderstra{\ss}e 16, 04103 Leipzig, Germany}
\affiliation{School of Mathematics and Statistics, University College Dublin, Belfield D04 N2E5, Dublin 4, Ireland}
\affiliation{Centro Brasileiro de Pesquisas F\'isicas (CBPF), Rio de Janeiro, CEP 22290-180, Brazil}

\author{Chris Kavanagh\,\orcidlink{0000-0002-2874-9780}}
\email{chris.kavanagh1@ucd.ie}
\affiliation{School of Mathematics and Statistics, University College Dublin, Belfield D04 N2E5, Dublin 4, Ireland}
\author{Jakob Neef\,\orcidlink{0000-0002-3215-5694}}
\email{jakob.neef@ucdconnect.ie}
\affiliation{School of Mathematics and Statistics, University College Dublin, Belfield D04 N2E5, Dublin 4, Ireland}
\author{Adrian Ottewill\,\orcidlink{0000-0003-3293-8450}}
\email{adrian.ottewill@ucd.ie}
\affiliation{School of Mathematics and Statistics, University College Dublin, Belfield D04 N2E5, Dublin 4, Ireland}

\begin{abstract}
We calculate high-order late-time tails of the retarded Green function of the  Teukolsky equation
for  linear field perturbations of  (subextremal) Kerr spacetime.
We calculate these tails at  a fixed  spheroidal harmonic $\ell$ and azimuthal number $m$
up to the first three orders  for the field point:
at finite radius (away from the event horizon) for large Boyer-Lindquist time $t$; along the future event horizon $\hor^+$  for large ingoing Eddington-Finkelstein coordinate $v$; and along future null infinity $\scri^+$ for large outgoing Eddington-Finkelstein coordinate $u$. 
We obtain the tail powers for generic integer field spin $s$ and the tail coefficients specifically for gravitational ($s=-2$) perturbations.
Our asymptotics include the known leading power-law (generic) tails, respectively,  $t^{-2\ell-3}$, $e^{im\Omega_H v}v^{-2\ell-3-b}$ (where
$b=1$ for $s>0, m=0$ and
$b=0$ otherwise, and where $\Omega_H$ is the angular velocity of the event horizon) and $u^{-\ell+s-2}$, as well as their
higher-order
logarithmic corrections:
  $t^{-2\ell-5}\ln t$,
  $e^{im\Omega_H v}v^{-2\ell-5-b}\ln v$
  and $u^{-\ell+s-3}\ln u$
  (as well as $u^{-\ell+s-4}\ln^2 u$).
Since we obtain the high-order expansions for modes for generic $\ell$ and $m$, we can readily infer the explicit expansions of the {\it full} retarded Green function for $s=-2$ (and its decay powers for generic integer $s$).
We obtain the late-time asymptotics from small-frequency expansions of the Fourier modes of the retarded Green function in the frequency domain.  
Accordingly, we also provide small-frequency expansions of various quantities of interest in the scattering theory. We also attach two notebooks  which provide expansions for specific values of $s$: one notebook provides them to the first three leading orders for generic $\ell$ and the other one to arbitrary order for specific values of $\ell$.
\end{abstract}

\maketitle

%---------------------------------------------------------------------------------------------------------
%---------------------------------------------------------------------------------------------------------

\section{Introduction}

It is well-known that linear field perturbations of asymptotically-flat black hole spacetimes decay at {\it leading} order for late times as a {\it power} law (see, e.g.,~\cite{Price:1971fb,Price:1972pw} in Schwarzschild and   \cite{1999PhRvD..61b4033H,hod2000mode,PhysRevLett.84.10,PhysRevD.61.024026,barack1999late} in Kerr).
This late-time decay may refer to any of the following: (a) large Boyer-Lindquist time $t$ at finite radius away from the horizon; (b) large 
ingoing Eddington-Finkelstein coordinate
$v$ along the future horizon $\hor^+$;
(c) large 
outgoing Eddington-Finkelstein coordinate
$u$ along future null infinity $\scri^+$.

There are
%, however, 
many reasons  why
going beyond leading order and
knowing higher-order terms in the late-time decay may be useful 
and we next list a few.
Firstly, the application to the self-force programme for modelling the inspiral of a small compact body into a supermassive black hole (extreme or intermediate mass-ratio inspirals). It has been demonstrated that the self-force can be calculated (see, e.g., the review \cite{Poisson:2011nh}) via a worldline-integration of the retarded Green function (GF) of the equation satisfied by the field perturbation. 
In the case of the scalar field, it has been found that the higher-order tail is required in order to obtain the self-field/force to high enough precision both in Schwarzschild in~\cite{CDOW13} and in Kerr~\cite{GFKerr} black hole background spacetimes.
In the case of gravitational and electromagnetic perturbations, the equations for the metric perturbation and electromagnetic potential do not decouple and so, instead, we shall henceforth consider the decoupled Teukolsky~\cite{Teukolsky:1973ha} (which is the focus of this paper) or, in the case of Schwarzschild, also Regge-Wheeler~\cite{Regge:1957td} equations, both of which also separate by variables.

Secondly, within the context of modelling black hole mergers, high order tails are relevant in obtaining high precision ringdown waveforms for inspirals at any mass ratio.
With upcoming gravitational-wave interferometers, it has recently become of relevance  to model the ringdown to high precision and match it to a separate calculation of the waveform in the preceding merger phase. While the ringdown phase is well modelled by the so-called quasinormal modes (QNMs), with respect to which  tails are typically subdominant (see, e.g.,~\cite{Leaver:1986}), 
there is the possibility that, in certain settings, the high order tail may be useful for achieving high precision
(e.g., Ref.~\cite{rosato2026singularstructurescausalityschwarzschild} claims that the subleading tail 
dominates over 
QNMs 
with overtones $n>1$
%) 
for waveforms in Schwarzschild, at least for sources outside the photon orbit). Recent work has also demonstrated that in certain cases, predominantly head on collisions and highly eccentric mergers, the late time tails following a merger can be  amplified, making their contribution more significant \cite{Albanesi:2023bgi,Islam:2024vro,DeAmicis:2024not,DeAmicis:2024eoy,Becker:2025zzw}.

A third reason for going beyond leading order in the tail is the investigation of the regularity properties of the Cauchy horizon, which are of relevance for the Strong Cosmic Censorship hypothesis~\cite{1979grec.conf..581P,Christodoulou:2008nj} and are known to  intrinsically depend on the 
asymptotics of the 
%\MC{linear?} 
field along the future horizon (see, e.g.,~\cite{DafermosLuk2017,gurriaran2026nonlinearinstabilitykerrcauchy,luk2026formationweaknullsingularity}).
In the case of subextremal black holes, it is only the leading order in the tail along $\mathcal{H}^+$ which dictates the degree of regularity on the Cauchy horizon. 
In the case of extremal black holes, on the other hand,
the dependence between the regularity on the Cauchy horizon and the tail along $\mathcal{H}^+$ seems to be particularly delicate: it is shown in~\cite{Gajic:2015csa,angelopoulos2018late} that, for scalar field perturbations of extremal Reissner-Nordstr\"om, 
the next-to-leading order term in the asymptotics along the horizon is relevant for the regularity properties of the Cauchy horizon.
In the case of the Cauchy horizon in extremal Kerr, so far only the $C^0$-extendibility has been shown ~\cite{2015arXiv151208953G} for axisymmetric scalar field perturbations using their leading order asymptotics along $\mathcal{H}^+$; whether higher orders along $\mathcal{H}^+$ will be relevant for the regularity of the extremal Cauchy horizon in the non-axisymmetric case remains to be seen.

As said, for the regularity properties of the Cauchy horizon, it is the tail (b) along $\hor^+$ that matters.
On the other hand, for
the self-force, it is the tail (a) at finite radius away from the horizon that matters, whereas for the ringdown waveform, it is the tail (c) along $\scri^+$.

Interestingly, it is known that the late-time expansion of the field is not given by a pure power law in the inverse of time but that non-analytic terms like a logarithm in time  appears at some order. Specifically, in the context of post-Newtonian theory the presence of logarithms signal the effect of long range back scattering of the field and so-called tail and iterated tail effects, see e.g. \cite{Blanchet:2019rjs}. In the context of self-force and black hole perturbation theory, when working in the frequency-$\omega$ domain, one can also see from a weak field expansion of the GF that the presence of $\ln(\omega)$ type terms also leads directly to non-localities in the solutions for the field \cite{Bini:2024icd}. 

Indeed, one way of obtaining the late-time tails is by going to the frequency domain. The Fourier frequency modes of the GF are known (e.g.,~\cite{Leaver:1986,Leaver:1986a,CKO}) to possess a branch cut (BC) down the negative imaginary axis (NIA). The late-time asymptotics of the GF may then be obtained from the small-frequency asymptotics along the BC of its Fourier modes.
We obtain the small-frequency expansions using the so-called MST method (after its authors: Mano, Suzuki and Takasugi; see \cite{Sasaki:2003xr} for a review), which consists of infinite series representations for the radial solutions and their scattering coefficients. 
The expansions for small-frequency of the various scattering quantities in black hole perturbation theory are useful, not only for the late-time tail, but for other problems as well. For example, 
in classical GR they are relevant for calculating the tidal deformability and Love numbers of Kerr black holes (e.g.,~\cite{2021PhRvL.126m1102L,PhysRevD.103.084021,2021PhRvD.104b4013C,2023arXiv231003660P,2024PhRvD.109f4058S}) as well as for analytic weak-field self-force calculations describing small mass-ratio binaries (e.g.~\cite{Fujita:2012cm,Kavanagh:2015lva, Kavanagh:2016idg, Munna:2023wce,Castillo:2024isq}).
The small-frequency expansions of scattering coefficients are also relevant in  Quantum Field Theories: from obtaining the %asymptotics 
behaviour
of renormalized expectation values near the Cauchy horizon 
of 
black holes (e.g.,~\cite{Lanir:2018vgb} in Reissner-Nordstr\"om and~\cite{AZCO:2026} in Kerr) to
calculating the Hawking flux 
for near-extremal Reissner-Nordstr\"om~\cite{2021PhRvD.104b4066Z}
to checking that various integrals are well-defined (see, e.g.,~\cite{hollands2020quantum} in Reissner-Nordstr\"om-de Sitter and~\cite{2022PhRvD.106l5011Z} in Kerr).

Let us now briefly review some of the results for high-order black hole tails which have so far been obtained.
Ref.~\cite{angelopoulos2018late} rigorously obtained (thus confirming and extending  previous numerical and asymptotic results in~\cite{lucietti2013horizon}) the leading and next-to-leading  (where a logarithm first appears) order terms in the late-time asymptotics of a scalar field in extremal Reissner-Nordstr\"om
at fixed radius --on and outside the event horizon-- as well as along $\scri^+$, both for compact and non-compact initial data.
For non-extremal black holes, Ref.~\cite{Angelopoulos:2017iop} rigorously obtained the leading and next-to-leading order terms (where a logarithm first appears) in the late $u$-time decay along $\scri^+$ of a scalar field in  Schwarzschild and subextremal Reissner-Nordstr\"om both for non-compact and, in the case of a spherically-symmetric field,
compact  initial data.

In the specific case of Schwarzschild, two of the authors obtained~\cite{Casals:Ottewill:2015,PhysRevLett.109.111101} the late-time asymptotics of the spherical $\ell$-modes of the GF 
for both the Teukolsky and Regge-Wheeler equations for generic field spin $s$ up to third leading order (where a logarithm first appears)   in $t$  for finite radius outside the horizon.
It readily follows from equations therein\footnote{\label{ftn:Schw,scri}Specifically, from Eqs.~(2.33), (2.34), (5.4) and (6.11)-(6.13) in~\cite{Casals:Ottewill:2015}.} that a $\ln(u)$ term appears first  at next-to-leading order for large $u$ along $\scri^+$  (and for generic finite radius of the source point) for the Teukolsky equation for generic spin, and similarly for the Regge-Wheeler equation using 
the corresponding equations\footnote{Namely, Eqs.~(2.15) and (2.13) together with the Chandrasekhar transformation Eqs.~(B4) and (2.40) combined with the mentioned Teukolsky results of (5.4) and (6.11)--(6.13) (which are just Taylor series, and so they extend to arbitrary order  in the frequency as a power series), all equations in~\cite{Casals:Ottewill:2015}.} in the same paper 
(this latter result has been recently corroborated by \cite{rosato2026singularstructurescausalityschwarzschild}). 
For the asymptotics  along $\hor^+$, one may merely take the trivial horizon-asymptotics for the ingoing solution \footnote{\label{ftn:Schw,H}Specifically, Eqs.~(2.11) and (2.32) in~\cite{Casals:Ottewill:2015} with ${}_sR^{\text{in,tra}}_{\ell}=1$ for, respectively, the Teukolsky and Regge-Wheeler equations.} in the result for generic finite radius,  thus yielding a $\ln(v)$ appearing at third order.

We finally move on to Kerr. Three of the authors obtained~\cite{CKO} the late-time asymptotics of the GF for fixed, arbitrary spheroidal harmonic $\ell$ and azimuthal number $m$ for the scalar wave equation up to third leading order (where a logarithm first appears) for finite radius outside the horizon.
It readily follows from equations therein \footnote{Specifically, from Eqs.~(2.6), (5.8), (6.7), (6.8) and (6.10) in~\cite{CKO} together with a generic power series expansion in the frequency for the ingoing  radial solution, such as that from Eqs.~(7.7)--(7.15) in~\cite{CKO}.} that a $\ln(u)$ term appears first  at next-to-leading order for large $u$ along $\scri^+$  (and for generic finite radius of the source point).
Again, for the asymptotics of the scalar field modes along $\hor^+$, one may merely take the trivial horizon-asymptotics for the ingoing solution \footnote{Namely, Eq.~(2.9) in~\cite{CKO}.} in the result for generic finite radius, thus yielding a $\ln(v)$ appearing at third order.

Even though~\cite{CKO} carried out the explicit expansions in the scalar case only, it also laid out the whole high-order tail formalism, based on small-frequency expansions of BC quantities, for the Teukolsky equation for generic integer field spin $s$.
In this paper, we use this formalism in order to derive late-time asymptotics up to third leading order of the gravitational GF for Teukolsky spin $s=-2$ for fixed, arbitrary $\ell$ and $m$ in all three cases of: (a) at finite radius away from the horizon, (b) along $\hor^+$ and (c) along $\scri^+$.
Similarly to the cases of $s=0$ in Kerr and generic-integer-$s$ for both the Teukolsky and Regge-Wheeler equations in Schwarzschild mentioned earlier, 
a logarithm in time appears first at third leading order at finite radius but only at second order along $\scri^+$.
Along $\scri^+$, a quadratic logarithm appears at third leading order.
We note that, within our terminology, we do not consider that the presence of a logarithmic factor yields a different ``order" in an asymptotic expansion.
We check our late-time expansion  at finite radius away from the horizon (for the mode $\ell=2$ and $m=0,2$) against numerical calculations.

Although the explicit expansions for small-frequency or, correspondingly, for late-time in the time domain, that we give are in the specific case of spin $s=-2$ for most of the quantities, we give for generic $s=0,\pm 1,\pm 2$  the  powers in the leading order of all the relevant quantities (as well as the explicit small-frequency expansion of the angular eigenfunctions and eigenvalues).

For $s=-2$, since we obtain the high-order expansions for generic spheroidal $\ell$ and azimuthal $m$, we can readily infer the high-order expansions of only the $m$ modes of the GF as well of the {\it full} $3+1$-dimensional GF. For generic integer $s$, we do not calculate the coefficients in the expansion, but we can readily infer the decay powers (at least to leading order) for the $m$ modes of the GF as well of the $3+1$-dimensional GF.

In obtaining the late-time asymptotics, we also provide explicit small-frequency expansions of the scattering coefficients and the BC GF modes for Teukolsky spin $s=-2$, which may be useful for  other purposes as mentioned earlier.

The rest of this paper is organized as follows. In Sec.~\ref{sec:GF}, we introduce the GF of the Teukolsky equation and give expressions for the contribution to the GF from the BC down the NIA.
This contribution consists of angular  factors (spin-weighted spheroidal harmonics) and radial factors (corresponding to  the scattering theory).
In Sec.~\ref{eq:MST} we provide MST expressions for 
quantities relevant for the  radial factors.
We then
 provide small-frequency expansions of the angular and radial factors in Secs.~\ref{sec:SWSH} and ~\ref{sec:Small-w factors}, respectively.
In Sec.~\ref{sec:Small-w BC modes} we put together the expansions of the various radial quantities in order  to obtain high-order small-frequency expansions of the BC modes.
In Sec.~\ref{sec:tails}, we then integrate  these small-frequency expansions for the angular and radial factors  in order to obtain the high-order late-time tails for the GF. We  show our numerical validation in Sec.~\ref{sec:tails}.
We finish in Sec.~\ref{sec:conclusions} with some conclusions.
We also have three appendices.
In the first App.~\ref{sec:more_plots} we provide extra plots comparing the numerical GF modes and analytical tail, highlighting various physical and technical aspects.
In  App.~\ref{sec:small-w InCoeffs} we provide the small-frequency expansions of the scattering coefficients of the ingoing radial solution (which we use to calculate the expansions of the GF). Additionally, we attach two Mathematica notebooks containing 
small-frequency and late-time expansions for many quantities used in this paper.

We choose units $c=G=1$. Also, from Sec.~\ref{sec:Small-w factors} onwards we set Kerr black hole mass $M=1$.
 Throughout, we use the ‘Little-o’ $o$ and `Big-o' $\mathcal{O}$ asymptotic notations (see, e.g.,~\cite{Olver:1974}).
Finally, all the ``$\log$" symbols are understood to be Naperian logarithms ($\ln$).

%---------------------------------------------------------------------------------------------------------
%---------------------------------------------------------------------------------------------------------

\section{Retarded Green function and branch cut contribution}\label{sec:GF}

%---------------------------------------------------------------------------------------------------------

%---------------------------------------------------------------------------------------------------------

\subsection{Green function}\label{sec:GF}

We consider Kerr spacetime with mass $M$, angular momentum per unit mass  $a$ and Boyer-Lindquist coordinates $(t,r,\theta,\phi)\in \mathbb{R}\times (r_+,\infty)\times \mathbb{S}^2$, where $r_+=M+\sqrt{M^2-a^2}$ is the radius of the event horizon.
The radius of the Cauchy horizon is  $r_-= M- \sqrt{M^2-a^2}$ and we shall consider the subextremal black hole case ($a<M$). 

The Teukolsky equation~\cite{Teukolsky:1973ha} describes massless, linear spin-$s$ field perturbations ($s=0$, $\pm 1/2$, $\pm 1$ and $\pm 2$ for, respectively, scalar,  fermion, electromagnetic and gravitational fields) of Kerr spacetime.
In particular, and of most relevance in this paper, $s=+2$ and $s=-2$ correspond to linear perturbations of the Weyl scalars $\psi_0$ and $\psi_4$, respectively.
Given two spacetime points  $x$ and $x'$, the retarded Green function (GF) $\Gret(x,x')$ of the Teukolsky equation satisfies 
\begin{align}\label{eq:GF eq}
\mathcal{O}\,\Gret(x,x')=4\pi\Sigma\cdot \delta_{4}(x,x'),
\end{align}	
where  
\begin{align}
	\mathcal{O}\equiv -\left[\frac{(r^2+a^2)^2}{\Delta}-a^2\sin^2\theta\right]\spdiff{}{t}-\frac{4 M a r}{\Delta}&\frac{\partial^2 }{\partial t \partial \phi}-
	\left[\frac{a^2}{\Delta}-\frac{1}{\sin^2\theta}\right]\spdiff{}{\phi} 
	+\Delta^{-s}\pdiff{}{r}\left(\Delta^{s+1}\pdiff{}{r}\right) +\frac{1}{\sin \theta}\pdiff{}{\theta}\left(\sin\theta \pdiff{}{\theta}\right)  \nonumber\\
	+2s\left[\frac{a(r-M)}{\Delta}+\frac{i \cos\theta}{\sin^2\theta}\right]&\pdiff{}{\phi}
	+2 s \left[\frac{M(r^2-a^2)}{\Delta}-r -i a \cos\theta\right]\pdiff{}{t}-(s^2\cot^2\theta-s)
	\label{Eq:TeukMaster}
\end{align}
is the Teukolsky operator,
 $\Delta(r)\equiv (r-r_+)(r-r_-)$, 
$\delta_{4}(x,x')\equiv \delta_{4}(x-x')/\sqrt{|g|}$ is 
an invariant 4-dimensional Dirac delta distribution,  $g=-\Sigma^2\sin^2\theta$ is the determinant of the metric and $\Sigma\equiv r^2+a^2 \cos^2\theta$.
The GF satisfies causal boundary conditions:  $\Gret(x,x')$ is zero if the field point $x$ is not in the causal future of the base point $x'$.
Henceforth we 
take $t'=\phi'=0$, which is without loss of generality because of the stationarity and axisymmetry   of Kerr spacetime.

The Teukolsky equation separates by variables and so the GF admits the following decomposition\footnote{We note that the corresponding Eq.~(2.3) in Ref.~\cite{CKO} is valid with two caveats: (i) under the causal boundary condition that $G(x,x')$ is zero if $x$ is not in the causal future of  $x'$, not the reverse as erroneously stated above Eq.~(2.3) in~\cite{CKO}; (ii) with the normalization of the angular functions as $\int_0^{\pi} d\theta\sin\theta \left({}_s\Slm\right)^2=\frac{1}{2\pi}$, not as $\int_0^{\pi} d\theta\sin\theta \left|{}_s\Slm\right|^2=\frac{1}{2\pi}$ as erroneously stated below Eq.~(2.3) in~\cite{CKO}. Also, see Sec.~\ref{sec:BC modes} here for a comment on the complex-conjugation of an angular factor in Eq.~(2.3) in~\cite{CKO}.}:
\begin{equation}\label{eq:Gret=lmode}
\Gret(x,x')=
\sum_{\ell=|s|}^{\infty}G_{\ell}(x,x'),
\end{equation}
 with
\begin{equation}\label{eq:Gl}
G_{\ell}\equiv -2\Delta^s(r')\sum_{m=-\ell}^{\ell}e^{ i m \phi} 
\mathcal{G}_{\ell m}(r,r';\theta,\theta';t),
\end{equation}
and 
\begin{equation}
\mathcal{G}_{\ell m}\equiv  \int_{\mathbb{R}}d\omega\, e^{-i\omega t}\Glm(r,r';\omega){}_s\Slm(\theta){}_s\Slm(\theta') .
\label{eq:Glm_real_axis}
\end{equation}
The spin-weighted spheroidal harmonics (SWSHs) ${}_s\Slm(\theta)$~\cite{Berti:2005gp,berti2006erratum} satisfy, for $\omega\in\mathbb{R}$, a Sturm-Liouville problem with eigenvalue ${}_{s}\lambda_{\ell m\omega}$. 
We normalize these angular functions as
$\int_0^{\pi} d\theta\sin\theta \left({}_s\Slm\right)^2=\frac{1}{2\pi}$.
In their turn, the  GF Fourier modes $\Glm$ obey the following radial Green function equation:
\begin{equation} \label{eq:radial teuk. eq.}
% eq.2 Detw'76
\left[\Delta^{-s }\frac{d}{dr}\left(\Delta^{s+1}\frac{d}{dr}\right)+\frac{K^2-2is (r-M)K}{\Delta}+4is \omega r-{}_{s}\lambda_{\ell m\omega}\right]
G_{\ell m}(r,r';\omega)=-\delta(r-r'),
\end{equation}
where $K\equiv (r^2+a^2)\omega-am$.
They can thus be obtained as
\begin{align}
\Glm(r,r';\omega)=-\frac{\Rin{s}(r_<,\omega) \Rup{s}(r_>,\omega)}{W(\omega)
}, \label{eq:rGF}
\end{align}
 where $r_<\equiv\min(r,r'), r_>\equiv\max(r,r')$, $\Rinup{s}$ are  linearly independent, homogeneous solutions of Eq.~\eqref{eq:radial teuk. eq.} and
\begin{align}\label{eq:Wronsk}
W(\omega)
\equiv
\Delta^{s+1}\bar W\left(\Rin{s},\Rup{s}\right),
\quad
\bar W\left(\Rin{s},\Rup{s}\right)=
\Rin{s}\frac{d \Rup{s}}{dr}-\Rup{s}\frac{d\Rin{s}}{dr}.
\end{align}
Here, $\bar W\left(\Rin{s},\Rup{s}\right)$
is  the Wronskian of the In and Up  solutions  and we shall refer to $W$, which is constant, as the invariant Wronskian.
The causality condition on the GF requires that the solutions
$\Rinup{s}$ satisfy the following, specific boundary conditions:
\begin{align}\label{eq:bc Rin}
\Rin{s}(r,\omega)&\sim \left\{\begin{array}{l l}
\Btra\Delta^{-s}e^{-i \tilde{\omega}r_*}, & r\rightarrow r_+, \\
r^{-2s-1} \Bref e^{i \omega r_*} +r^{-1} \Binc e^{-i \omega r_*}, & r \rightarrow \infty,
\end{array}
\right. \\
\label{eq:bc Rup}
\Rup{s}(r,\omega)&\sim\left\{\begin{array}{l l}
\Cinc e^{i \tilde{\omega}r_*}+\Cref\Delta^{-s}e^{-i \tilde{\omega}r_*}, & r\rightarrow r_+, \\
r^{-2s-1} \Ctra e^{i \omega r_*}, & r \rightarrow \infty.
\end{array}
\right.
\end{align}
Here $\tilde{\omega}\equiv \omega-m\Omega_H$, $\Omega_H\equiv a/(r_+^2+a^2)$ is the angular velocity of the black hole,
 $B^{\text{inc}/\text{ref}/\text{tra}}$ and $C^{\text{inc}/\text{ref}/\text{tra}}$  are the
incidence/reflection/transmission
 complex-valued scattering coefficients of, respectively, the In and Up solutions, and we have defined the tortoise coordinate $\tort$ via $\dfrac{d\tort}{dr}=\dfrac{(r^2+a^2)}{\Delta}$ as
\begin{equation}\label{eq:tortoise}
%p.10WOF
\tort=
%Eq.21ST
r+\frac{2M}{r_+-r_-}\left(r_+\ln\left|\frac{r-r_+}{2M}\right|-r_-\ln\left|\frac{r-r_-}{2M}\right|\right).
\end{equation}
The  invariant Wronskian is given in terms of the scattering coefficients by 
\begin{align}\label{eq:Wronsk}
W
=2
i\omega
\Binc \Ctra.
\end{align}
The GF Fourier modes $\Glm$ in Eq.~\eqref{eq:rGF} are of course independent of the normalizations of the In and Up solutions. We find it useful to henceforth choose $\Ctra=1$ and  we denote by $\Ruphat{s}$ the Up solution  $\Rup{s}$ normalized in this manner.

For later purposes, it is also useful to define another  homogeneous solution $\hat{R}_+^\nu(r,\omega)$ of Eq.~\eqref{eq:radial teuk. eq.} by the following boundary condition:
\begin{equation}
\hat{R}_+^\nu(r,\omega)\sim \frac{ e^{-i \omega \tort}}{r},\quad r\to \infty.
\end{equation}

%---------------------------------------------------------------------------------------------------------

\subsection{Summation order more suitable for an initial value problem}

One may swap the order of the $\ell$- and $m$-sums in Eqs.~\eqref{eq:Gret=lmode} and \eqref{eq:Gl} in order to better suit an initial value problem.
Swapping the sums, yields
\begin{equation}\label{eq:Gret=mmode}
\Gret(x,x')=
\sum_{m=-\infty}^{\infty} e^{im \phi} G_{m}(r,r';\theta,\theta';t),
\end{equation}
with
\begin{equation}\label{eq:Gm}
G_m\equiv -2\Delta^s(r')\sum_{\ell=\ell_0}^{\infty}
\mathcal{G}_{\ell m}(r,r';\theta,\theta';t),
\end{equation}
where $\ell_0\equiv \max(|m|,|s|)$.
Because of the axisymmetry of Kerr spacetime, the $m$-modes $G_{m}$ clearly satisfy a $2+1$-dimensional Green function equation resulting from  replacing  $\partial_\phi \to im$ in Eq.~\eqref{Eq:TeukMaster} and $\delta(\phi-\phi')\to \frac{1}{2\pi}$ in $\delta_{4}(x,x')$ in the $3+1$-dimensional Teukolsky Eq.~\eqref{eq:GF eq}.
Thus, late-time asymptotics for $G_m$ correspond to the late-time evolution of some characteristic initial data.

%---------------------------------------------------------------------------------------------------------

\subsection{Complex frequency plane}

In this subsection we will consider the analytical properties on the complex-frequency plane of the various quantities that make up the GF Fourier modes $\Glm$ in Eq.~\eqref{eq:Glm_real_axis} and we will introduce a deformation of the real-frequency Fourier integral on the complex-frequency plane.

%---------------------------------------------------------------------------------------------------------

\subsubsection{Angular quantities}

The SWSHs  ${}_s\Slm$ 
appearing in Eq.~\eqref{eq:Glm_real_axis}, as well as the eigenvalues ${}_{s}\lambda_{\ell m\omega}$
can be analytically continued to $\omega\in\mathbb{C}$ except for a finite number of 
  `angular BCs'~\cite{oguchi1970eigenvalues, BONGK:2004}.
Since the angular eigenvalue ${}_{s}\lambda_{\ell m\omega}$ also appears in the radial equation \eqref{eq:radial teuk. eq.}, these angular BCs might naturally carry over to the radial solutions as well as to their scattering coefficients.
However, these angular BCs  are spurious in the sense that they cancel out when summing the integrand for $\mathcal{G}_{\ell m}$ with that for $\mathcal{G}_{\ell+2,m}$ (see App.~A in~\cite{,CKO}). This implies, in particular, that there is no contribution from the angular BCs to the $G_m$ in Eq.~\eqref{eq:Gm} nor to the full GF.
Furthermore, these angular BCs
stem from branch points which do not lie on the real axis (since the SWSHs are complete for $\omega\in\mathbb{R}$ -- see~\cite{stewart1975stability}).
This means that, not only the angular BCs do not contribute at all to the $m$-modes $G_m$ nor to the GF, but also that, as we shall see in the following subsections,  %the angular BCs 
they
should play no role in the late-time tail (at any order) of the $\mathcal{G}_{\ell m}$ or the $G_{\ell}$.
We will thus  no longer consider these angular BCs and we will ignore them in our expressions.

As will be useful for later, we note that, from the angular equation and boundary conditions satisfied by the SWSHs, it readily follows that  their eigenvalues are symmetric under  complex-conjugation together with the transformation $(m,\omega)\to (-m,-\omega^*)$:
\begin{equation}\label{eq:ang-symm}
{}_{s}\lambda_{\ell m\omega}=\left({}_{s}\lambda_{\ell,-m,-\omega^*}\right)^*.
\end{equation}

%---------------------------------------------------------------------------------------------------------

\subsubsection{Radial quantities}

It is known~\cite{Leaver:1986a,Leaver:1986,CKO} that the Up solution $\Ruphat{s}$, viewed as a function of complex $\omega$, possesses a branch point at $\omega=0$.
Henceforth, whenever we plainly refer to a BC (without any extra qualification, such as `angular') in a quantity, we will mean the associated BC, namely, a BC on the complex-$\omega$ plane stemming from $\omega=0$.
Whenever such a BC exists we will  take it to run down the NIA of the complex-$\omega$ plane.

As for the In solution as a function of complex $\omega$, $\Rin{s}$ for $r<\infty$ possesses no BC 
as long as a choice of $\Btra$ is made such that this coefficient itself has no BC.  
For now, we make such a choice, so that $\Btra$ and $\Rin{s}$ for $r<\infty$ have no BC.
Despite this, the In coefficients $\Bref$ and $\Binc$ may generally possess a BC.

The BC of $\Ruphat{s}$ is inherited by the invariant Wronskian $W(\omega)$ in \eqref{eq:Wronsk} and by the GF Fourier modes $\Glm(r,r';\omega)$ in \eqref{eq:rGF}.
We define 
\begin{align*}
    \sigma\equiv i\omega,
\end{align*} so that $\sigma>0$ when $\omega\in NIA$, and henceforth we shall denote by a ``-/+" superscript on a quantity possessing a BC when it is being
evaluated on the left/right side of the BC down the NIA (namely, on the 3rd/4th quadrant of the complex-$\omega$ plane). 
For example,
$\Glm^{\pm}(r,r';\omega)=\lim_{\epsilon\to 0^+}\Glm(r,r';-i\sigma\pm\epsilon)$
with
$\fNIA>0$.

It is also known that, as functions of complex $\omega$, the Fourier modes $\Glm$ possess poles at zeros of the invariant Wronskian $W(\omega)$.
These poles lie below the
real axis (\cite{whiting1989mode} and~\cite{Andersson:2016epf} showed that there are none above  nor on the real axis, respectively) and they correspond to the QNM frequencies.

Note that, similarly to the symmetry in Eq.~\eqref{eq:ang-symm} for the angular eigenvalues and making use of it, the radial Eq.~\eqref{eq:radial teuk. eq.} is symmetric under complex-conjugation together with the transformation $(m,\omega)\to (-m,-\omega^*)$.
 Our normalization for the Up solution $\Ruphat{s}$ trivially satisfies the same symmetry (since $\Ctra=1$) and, therefore, so does the Up solution itself and all its scattering coefficients:
 \begin{equation}\label{eq:symm-Up}
 \Ruphat{s}(r,\omega)=\left(\Ruphatmm{s}(r,-\omega^*)\right)^*,\quad \Cincreftra=\left(\left.\Cincreftra\right|_{m\to -m, \omega\to -\omega^*}\right)^*.
 \end{equation}
We will also choose in Sec.~\ref{sec:Radial} a normalization for the In solution which preserves this symmetry, so that our In solution and its scattering coefficients satisfy:
 \begin{equation}\label{eq:symm-In}
\Rin{s}(r,\omega)=\left(\Rinmm{s}(r,-\omega^*)\right)^*,\quad 
\Bincreftra=\left(\left.\Bincreftra\right|_{m\to -m, \omega\to -\omega^*}\right)^*.
 \end{equation}

%---------------------------------------------------------------------------------------------------------

\subsubsection{Deformation of the integration contour on the complex-$\omega$ plane}\label{sec:deform}

The real-frequency integral in Eq.~\eqref{eq:Glm_real_axis} for $\mathcal{G}_{\ell m}$ may be deformed on the lower half of the complex-$\omega$ plane~\cite{Leaver:1986}.
Using the residue theorem, 
and taking into account the above analytical properties of %the various quantities 
the Fourier modes $\Glm$
as functions of complex-$\omega$,
it  follows that, in subextremal Kerr, $\mathcal{G}_{\ell m}$ may be obtained (at least at `sufficiently' late times~\footnote{At `early' times, the various contributions may separately diverge -- see, e.g., Ref.~\cite{Casals:2011aa} in Schwarzschild spacetime.}) from the following contributions: (i) an integral around the BC; (ii) a sum over the residues at the QNM frequencies;
and (iii) an integral along a high-frequency arc (see Fig.1 in~\cite{CKO} for a schematic illustration).

Since the poles of $\Glm$ 
lie below the
branch point $\omega=0$ %(see~\cite{whiting1989mode,Andersson:2016epf}),
(and there are no angular branch points on the real axis),
it follows using a generalized Watson’s lemma (see the later Eq.~\eqref{eq:Watson}) for the BC integral and Jordan's lemma for the high-frequency arc integral~\footnote{To be more precise and rigorous, Proposition 8.1 in Ref.~\cite{Stucker-PhD-2026} shows that the contribution to the GF (after a Fourier decomposition but without decomposing into $\ell$ and $m$ modes) from a square %(instead or circular) 
contour on the lower-$\omega$ plane, such that the sides of the square go to infinity and the lower side has a specific finite value, decays exponentially in time. This is shown for the  scalar field in Kerr but it is believed that it should similarly hold for a general-spin Teukolsky field (furthermore, one should be able to remove the restriction on the value for the lower side of the square).}, that the late-time asymptotics of 
$\mathcal{G}_{\ell m}$ are given by the small-frequency asymptotics of the integrand in the BC integral.

In the following sections
we shall calculate small-frequency expansions of the BC integrand and the corresponding late-time tails of $\mathcal{G}_{\ell m}$ as the field point approaches future timelike infinity $i^+$. 
The time in the late-time tails
will be one of the following: (a) Boyer-Lindquist $t$ at finite radius $r>r_+$; (b) the  ingoing Eddington-Finkelstein coordinate $v\equiv t+r_*$ along the future event horizon $\hor^+$; (c)
the outgoing Eddington-Finkelstein coordinate $u \equiv t-r_*$ along  future future null infinity $\scri^+$.
While $\scri^+$ is parameterized by $(u,\theta,\phi)$, $\hor^+$ is parameterized by $(v,\theta,\phi_+)$, where
$\phi_+\equiv \phi-\Omega_H t$.

We provide the coefficients in the expansions for $s=-2$ and the powers for generic integer $s$. We note that, in principle, one should be able to obtain the coefficients in the expansion for $s=+2$ from those provided here for $s=-2$ by applying the so-called
Teukolsky-Starobinsky identities~\cite{Starobinskil:1974nkd,teukolsky1974perturbations}.

%---------------------------------------------------------------------------------------------------------

\subsection{Contribution to the GF from the branch cut}
\label{sec:BC modes}

The contribution from the BC along the NIA to the 
 modes $\mathcal{G}_{\ell m}$ of the GF in
Eq.~\eqref{eq:Glm_real_axis},
is given by
%\footnote{We note that there is the wrong sign in Eq.~(5.7) in~\cite{CKO}  and in Eq.~(2.19) in~\cite{Casals:Ottewill:2015}.}
\begin{equation}\label{eq:Disc-m}
\Disc \mathcal{G}_{\ell m}(r,r';\theta,\theta';t)\equiv -i \int_{0}^{\infty}d\fNIA\, e^{-\fNIA t}\Disc \Glm(r,r';\fNIA)\left.{}_s\Slm(\theta){}_s\Slm(\theta')\right|_{\omega=-i\sigma}.
\end{equation}
Here, the BC mode $\Disc \Glm\equiv \Glm^+-\Glm^-$ is the discontinuity across the BC of the  radial Green function modes $\Glm(r,r';\omega)$.

%\CK{\textbf{Derivation sketch:} With the standard contour picture in mind, Cauchy's theorem will tell us
%\begin{align}
%\int_{\mathbb{R}}d\omega+\int_{{\rm arc}}d\omega+\int_{-i\infty}^0 d\omega\, G^{+}+\int_0^{-i\infty}d\omega \,G^{-}=\sum {\rm QNM}
%\end{align}
%At late times this becomes
%\begin{align*}
%   \int_{\mathbb{R}}d\omega +\int_{-i\infty}^0 d\omega\, G^{+}+\int_0^{-i\infty}d\omega \,G^{-}&\approx0\\
%\end{align*}
%giving
%\begin{align}
%\mathcal{G}_{\ell m}&\approx-\int_{-i\infty}^0 d\omega\, G^{+}-\int_0^{-i\infty}d\omega \,G^{-} \\
%&=i \int_{\infty}^0 d\sigma\, G^{+}+i \int_{0}^\infty d\sigma\, G^{-} \\
%&=-i\int_{0}^\infty d\sigma(G^{+}-G^{-})
%\end{align}}
Before providing the expressions for the BC modes $\Disc \Glm$, we make one remark on Eq.~\eqref{eq:Disc-m}.
In 
some expressions in the literature (e.g., Eq.~(2.3) in~\cite{CKO}, Eq.~(2.3) in~\cite{Yang:2013shb},
Eq.~(2) in~\cite{gralla2016transient},
Eq.~(14) in~\cite{casals2016horizon} and Eq.~(4.4) in~\cite{2025arXiv250523895B}, together with their corresponding versions for complex frequencies) for the Kerr GF involving an integral in the complex frequency domain, the SWSH factor ${}_s\Slm(\theta')$ is complex conjugated.
Of course, since ${}_s\Slm(\theta')$ is real-valued for $\omega\in\mathbb{R}$, such complex conjugation would make no difference in the real-frequency integrand in
Eq.~\eqref{eq:Glm_real_axis}. The issue arises when trying to analytically continue the real-frequency integrand into the complex domain as in \eqref{eq:Disc-m}.
If ${}_s\Slm$ is analytic in $\omega\in\mathbb{C}$ (by which we mean at least away from  the angular BCs, which go away when summing over $\ell$ anyway), then ${}_s\Slm^{*}$ (which involves $\omega^*$) is not analytic in $\omega$, which would mean that if a SWSH were complex-conjugated in \eqref{eq:Glm_real_axis}, then its integrand would (generally) not be analytic anywhere in $\omega\in\mathbb{C}$ and so we would not be able to analytically continue it into the complex-frequency plane (and apply the residue theorem, which we need it in order to deform the contour and obtain \eqref{eq:Disc-m}).
On the other hand, without the complex-conjugation as  \eqref{eq:Glm_real_axis} is, its integrand is analytic in $\omega\in\mathbb{C}$ and so it can be analytically-continued into the complex-frequency plane. This is an analytical reason for writing Eqs.~\eqref{eq:Glm_real_axis} and \eqref{eq:Disc-m} as they are without any complex-conjugated SWSH.
In the later Fig.~\ref{Fig:residuals m=2,with/without cc}, we provide further numerical support for it:
the plots show that the late-time residuals by subtracting our analytical small-frequency expansions from  a real-frequency integration are better when not including the complex-conjugation in a SWSH than when including it.
We note that in the case of $s=0$ on the NIA, a complex-conjugation  on the SWSH makes no difference since, from symmetries of the angular equation:: $\left.{}_0\Slm(\theta)\right|_{\omega=-i\sigma}=\left(\left.{}_0\Slm(\theta)\right|_{\omega=-i\sigma}\right)^*\in\mathbb{R}$. This means that the results in Ref.~\cite{CKO} for $s=0$ on the NIA remain intact.

Orthogonality with $\ell$  in the complex-$\omega$ plane  without the complex conjugation in the SWSH now follows straightforwardly. The function 
\begin{align}
    I_{\ell\ell'}(\omega)\equiv \int_0^{\pi} d\theta \sin\theta {}_s S_{\ell m\omega}(\theta){}_s S_{\ell'm\omega}(\theta)
\end{align}
is identically $0$  whenever $\ell'\neq \ell$  for $\omega\in\mathbb{R} $ (from standard Sturm-Liouville theory). Since $I_{\ell\ell'}(\omega)$ is analytic in $\omega$, it therefore must be $0$ anywhere we can analytically continue to when $\ell'\neq \ell$ (in agreement with Ref.~\cite{2023PhRvD.107d4056L}), which is the complex plane modulo any angular branch points.  Thus orthogonality will hold for our contour choice (see Fig.~1 of \cite{CKO}).

We now move on to provide expressions for the BC modes separately for the field point 
(i) with
finite radius $r>r_+$, (ii) for $r\to r_+$ and 
(iii) for $r\to \infty$.
In all cases, the source point has finite radius $r'>r_+$.

\subsubsection{BC modes at finite radius $r\geq r_+$}\label{sec:BC modes,finite r}

The following expression for the BC modes at finite radius $r\geq r_+$ is derived from Eq.~\eqref{eq:rGF} in~\cite{CKO}:
\begin{align}
% p.15 WOB
\Disc \Glm(r,r';\fNIA)
&=-2 i \sigma 
\qW{\indmode}(\sigma)
\Rin{s}(r,-i\fNIA) \Rin{s}(r',-i\fNIA), 
\quad \fNIA>0.
 \label{Eq:GBC}
\end{align}
The new radius-independent factor in Eq.~\eqref{Eq:GBC} is 
\begin{equation}\label{eq:Q}
\qW{\indmode}(\sigma) \equiv \frac{q(\sigma)}{\Wbp\Wbm},
 \end{equation}
where the so-called BC strength $q(\sigma)$ 
defined via
\begin{align}
\Disc\Ruphat{s}(r,\fNIA)=i\, q(\sigma) \hat{R}_+^\nu(r,-i\fNIA), \label{Eq:R BC}
\end{align}
where $\Disc\Ruphat{s}\equiv \Rupphat-\Rupmhat$ is the discontinuity across the BC of the Up radial solution $\Ruphat{s}$.

In its turn, the Wronskian factor in Eq.~\eqref{eq:Q} can be expressed using Eq.~(5.10) in~\cite{CKO}\footnote{There is a typographical error in Eq.~(5.10) in~\cite{CKO} for $W^+W^-$: the sign of the second term should be `$+$' instead of `$-$' (see Eq.~\eqref{Eq:WpWm}  in this paper for the corrected version). Correspondingly, the sign of the second term in the last expression in Eq.~(5.9) in~\cite{CKO} should be `$+$' instead of `$-$'. These typographical errors in Eqs.~(5.9) and (5.10) in~\cite{CKO} did not have any consequences on the rest of the expressions and results in~\cite{CKO} (which were obtained using the correct expression for $W^+W^-$).} multiplied across by $\left(\Btra\right)^2$ so as to accommodate for our normalization in Eq.~\eqref{eq:bc Rin} with a generic $\Btra$ which does not have a BC:
\begin{align}
\Wbp \Wbm=
\left(2 \sigma \Bincp\right)^2+4 i \sigma^2 q(\sigma)\Bincp \Brefp. \label{Eq:WpWm}
\end{align}

In the specific case $r\to r_+$,
we insert into Eq.~\eqref{Eq:GBC} the boundary condition in Eq.~\eqref{eq:bc Rin} for $\Rin{s}(r)$ as $r\to r_+$, yielding:
\begin{align}
% p.15 WOB
\Disc \Glm(r,r';\fNIA)
& \sim-2 i \sigma 
\qWtra{\indmode}(\sigma)
\Rin{s}(r',-i\fNIA) 
\left(\Delta(r)\right)^{-s} 
e^{im\Omega_H r_*}e^{-{\sigma}r_*},
\quad r\to r_+, 
\quad \fNIA>0,
 \label{Eq:GBC,rp}
\end{align}
where $\qWtra{\indmode}(\sigma)\equiv \qW{\indmode}(\sigma)\left.\Btra\right|_{\omega=-i\sigma}$.

%---------------------------------------------------------------------------------------------------------
\subsubsection{BC modes as $r\to \infty$}
\label{sec:BC at scri}

Setting the field point $x$ on $\scri^+$ implies, from Eq.~\eqref{eq:rGF}, that $\Rup{s}$ can be replaced by its boundary condition in Eq.~\eqref{eq:bc Rup} as $r\to \infty$. Since we have chosen $\Ctra=1$ ($\Rup{s}=\Ruphat{s}$), this means that the BC modes are then given from Eq.~\eqref{eq:rGF} by
\begin{equation}
\Disc \Glm(r,r';\fNIA)\sim -\left(\frac{1}{W^+}-\frac{1}{W^-}\right)r^{-2s-1}e^{\sigma r_*}\Rin{s}(r',-i\fNIA), \quad r\to\infty, \quad \fNIA>0.
\end{equation}
Now using Eq.~(5.9) in~\cite{CKO} (taking cognizance of the fact that the Wronskian there was for an In solution normalized with $\Btra=1$, whereas so far here we have an arbitrary  $\Btra$, except for it having no BC), we obtain:
\begin{equation}\label{eq:dGlm,r->infty}
\Disc \Glm(r,r';\fNIA)\sim 
2i\sigma
\qWref{\indmode}(\sigma)
r^{-2s-1}e^{\sigma r_*}\Rin{s}(r',-i\fNIA), \quad r\to\infty, \quad \fNIA>0,
\end{equation}
where $\qWref{\indmode}(\sigma)\equiv \qW{\indmode}(\sigma)\Brefp $.

%---------------------------------------------------------------------------------------------------------
%---------------------------------------------------------------------------------------------------------

\section{MST Method}\label{eq:MST}

The MST method (see the original \cite{Mano:Suzuki:Takasugi:1996} and the review \cite{Sasaki:2003xr})
provides infinite series representations for  $\Rin{s}$, $\Rup{s}$ and their scattering coefficients which, even if valid for arbitrary frequencies, naturally lend themselves to obtaining small-frequency expansions.
In this section we briefly introduce the MST method and provide the main
expressions.
Note that, in principle, the MST expressions are only valid for $\Re(\omega)\geq 0$, and one may use the symmetry  under $(m,\omega)\to (-m,-\omega^*)$ together with complex conjugation (see Eqs.~\eqref{eq:symm-Up}, \eqref{eq:symm-In} and \eqref{eq:ang-symm}) for obtaining expressions valid for $\Re(\omega)< 0$. However, if a certain MST expression or expansion contains only functions which are manifestly holomorphic in $\omega$ (for $\Re(\omega)\geq 0$), then they should also be valid for all $\omega\in\mathbb{C}$.

The MST method uses the following dimensionless quantities\footnote{The symbol ``$q$" is used to denote both $a/M$ as well as the BC strength via Eq.~\eqref{Eq:R BC}, as is common in the literature. However, it is only  within Sec.~\ref{sec:BC modes,finite r} that $q$ is used to denote the BC strength and, furthermore, in that case it  always comes with the argument $\sigma$, so that it serves to disambiguate between the two meanings.}: $\epsilon\equiv2 M \omega$, $\kappa\equiv\sqrt{1-q^2}$, $q\equiv a/M$, $\tau\equiv(\epsilon-m q)/\kappa$ and $\epsilon_+\equiv (\epsilon+\tau)/2$.

%-------------------------------------------------------------------------------------------------------------------------------------------------------------------------------------------

\subsection{In radial solution}\label{sec:Radial}

We compute the In radial solution as 
\begin{align}
    &\Rin{s}(r,\omega)= \epsilon^{-\nu}(R_\mathrm{C}^\nu + K R_\mathrm{C}^{-\nu-1}),
    \label{eq:RIn}
\end{align}
where $K$ is the so-called tidal response function given by
%\begin{subequations}
%\begin{align}
\begin{equation}\label{eq:K}
    K\equiv\frac{K_{-\nu-1}}{K_\nu},
    \end{equation}
    with
    \begin{align}
        K_\nu &\equiv 
        (\epsilon  \kappa )^{s-\nu -\rST}\frac{2^{-\nu-\rST } i^{\rST} e^{i \epsilon  \kappa }  \Gamma (1-s-2 i \epsilon_+)\Gamma (\rST+2 \nu +2)}{\Gamma (\rST+\nu +1-s+i \epsilon) \Gamma (\rST+\nu+1 +i \tau ) \Gamma (\rST+\nu+1 +s+i \epsilon )} 
    \frac{}{} \notag\\
    &\times
    \frac{        \displaystyle \sum _{n=\rST}^{\infty} a_n^\nu
    \frac{(-1)^n \Gamma(n+\rST+2\nu+1) \Gamma(n+\nu +1+s+i\epsilon) \Gamma(n+\nu +1+i\tau)}{(n-\rST)! \Gamma(n+\nu+1-s -i \epsilon)  \Gamma(n+\nu+1-i\tau)}}{        \displaystyle \sum _{n=-\infty}^\rST a_n^\nu\frac{(-1)^n (\nu +s-i \epsilon +1)_n}{(\rST-n)! (2 \nu +\rST+2)_n (\nu -s+i \epsilon +1)_n}}.
    \label{eq:Kν}
\end{align}
%\label{eq:K}
%\end{subequations}
Here, $\rST$ is an arbitrary integer value (the value of $K_\nu$ does not depend on the value of $\rST$). 
The new radial functions $R_\mathrm{C}^\nu$ in Eq.~\eqref{eq:RIn}   
can be calculated using
Eq.~(162) in~\cite{Sasaki:2003xr}:
\begin{subequations}
\begin{align}
   &R_\mathrm{C}^\nu \equiv e^{-i \hat{z}} 2^\nu (\epsilon \kappa)^{-s-i \epsilon_+} \hat{z}^{\nu+i \epsilon_+} \left( \frac{\hat{z}}{\epsilon \kappa}-1 \right)^{-s-i\epsilon_+} \sum_{n=-\infty}^{\infty} \sum_{j=0}^{\infty} D_{n,j} \hat{z}^{n+j},
   \label{eq:RC}
   \\
   &D_{n,j} \equiv \frac{(-1)^n (2 i)^{j+n} a^\nu_n \Gamma (\nu +n-s+i \epsilon +1) (\nu +s-i \epsilon +1)_n (\nu +n-s+i \epsilon +1)_j}{j!\, \Gamma (2 \nu +2 n+2) (2 \nu +2 n+2)_j (\nu -s+i \epsilon +1)_n},
\end{align}
\end{subequations}
where 
 $\hat{z}\equiv \omega \, (r-r_-)$ and $(x)_n$ is the Pochhammer symbol. 
In practice we have found our low frequency expansion code (which we later use to obtain the late time tails) to run most efficiently for $\rST=0$.
In their turn, the MST series coefficients $a^\nu_n$ are calculated using a three-term recurrence relation given by:
\begin{equation}
\alpha_n^\nu \an{n+1}+\beta_n^\nu \an{n}+\gamma_n^\nu \an{n-1}=0,
\label{Eq:anrecursion}
\end{equation}
where
\begin{align}
\alpha_n^\nu&\equiv\frac{i\epsilon\kappa(n+\nu+1+s+i\epsilon)(n+\nu+1+s-i\epsilon)(n+\nu+1+i\tau)}{(n+\nu+1)(2 n+2 \nu+3)} ,\label{Eq:alpha}\\
\beta_n^\nu&\equiv-{}_{s}\lambda_{\ell m\omega}-s(s+1)+(n+\nu)(n+\nu+1)+\epsilon^2+\epsilon(\epsilon-m q)+\frac{\epsilon(\epsilon-m q)(s^2+\epsilon^2)}{(n+\nu)(n+\nu+1)},%\label{Eq:beta} 
\nonumber
\\
\gamma_n^\nu&\equiv-\frac{i\epsilon\kappa(n+\nu-s+i\epsilon)(n+\nu-s-i\epsilon)(n+\nu-i\tau)}{(n+\nu)(2 n+2 \nu-1)}. 
%\label{Eq:gamma}
\nonumber
\end{align} 
We choose the normalization $a^\nu_{0}=1$. 
It is clear from Eqs.~\eqref{Eq:anrecursion}--\eqref{Eq:alpha} that the series coefficients satisfy $a^{-\nu-1}_n = a^\nu_{-n}$. 
It can be shown (see Eqs.~(129) and (130) in~\cite{Sasaki:2003xr}) that the recurrence relation \eqref{Eq:anrecursion} possesses  a (unique) minimal solution  as $n\to +\infty$ and a (unique) minimal solution as $n\to-\infty$.
The value of the so-called renormalized angular momentum parameter $\nu$ is chosen so that the minimal solutions as $n\to +\infty$ and as $n\to-\infty$ coincide. The MST coefficients $\an{n}$  are then chosen to be 
this  minimal solution,
which in its turn guarantees that the double-ended infinite series in \eqref{eq:RC} converges for all $r\in (r_+,\infty)$.
In practise, this may be achieved by requiring 
$\nu$  to satisfy  the following implicit equation in terms of infinite continued fractions  (see Eq.~(133) in Ref.~\cite{Sasaki:2003xr}):
\begin{equation}
R_nL_{n-1}=1,
\end{equation}
for an arbitrary choice of $n\in\mathbb{Z}$,
where
\begin{equation}\label{eq:Rn}
R_n=-\frac{\gamma_n^\nu}{        \displaystyle \beta_n^\nu-\frac{\alpha_n^\nu\gamma_{n+1}^\nu}{\displaystyle \beta_{n+1}^\nu-\frac{\alpha_{n+1}^\nu\gamma_{n+2}^\nu}{\beta_{n+2}^{\nu}-\dots}}},\quad
L_n=-\frac{\alpha_n^\nu}{\displaystyle \beta_n^\nu-\frac{\alpha_{n-1}^\nu\gamma_{n}^\nu}{\displaystyle \beta_{n-1}^\nu-\frac{\alpha_{n-2}^\nu\gamma_{n-1}^\nu}{\beta_{n-2}^{\nu}-\dots}}}.
\end{equation}
The minimal solution $a^{\nu}_n$ for all $n\in\mathbb{Z}$ can then be obtained from $a^{\nu}_0=1$ via $a^{\nu}_{n}=R_na^{\nu}_{n-1}$ for $n>0$ and via $a^{\nu}_{n}=L_na^{\nu}_{n+1}$ for $n<0$.
The MST coefficients $a^{\nu}_n$ and the renormalized angular momentum $\nu$ as functions of $\omega\in\mathbb{C}$ do not have a BC (see Sec.~V in~\cite{CKO}). 

Eq.~\eqref{eq:RIn} for the In radial  solution corresponds to Eq.~(166) in Ref.~\cite{Sasaki:2003xr} but with a different normalization, which we next specify.
Let us denote by $\RinST{s}$ the In radial solution $\Rin{s}$ normalized with a $\Btra$ given by Eq.~(167) in~\cite{Sasaki:2003xr}. Then, our $\Rin{s}$ in Eq.~\eqref{eq:RIn} is equal to $\epsilon^{-\nu}\RinST{s}/K_\nu$. The reason for dividing by $K_\nu$ is that it significantly simplifies the small-frequency expressions which we give later for $\qW{\indmode}$.
%in terms of $a$.
The reason for further multiplying by $\epsilon^{-\nu}$ is that, otherwise, $\Rin{s}$ would have a  BC (in the form of $\ln \omega$ factors) along the NIA, which would make calculations more complicated (e.g., Eq.~\eqref{Eq:WpWm} was derived for $\Rin{s}$ without a BC).
With our current normalization, $\Btra$ has no BC and so neither does $\Rin{s}$ for $r<\infty$, as we next justify.
That $\RinST{s}$ has no BC follows readily from Eqs.~(116) and (120) in~\cite{Sasaki:2003xr} or, equivalently, from its transmission coefficient in Eq.~(167) in~\cite{Sasaki:2003xr}.
On the other hand, $K_\nu$ has a BC which  comes only from the overall factor $\epsilon^{-\nu}$ in \eqref{eq:Kν}.
This already implies that our $\Rin{s}$, which  is equal to $\epsilon^{-\nu}\RinST{s}/K_\nu$, has no BC. It is also instructive to see this property alternatively from Eq.~\eqref{eq:RIn}.
Since $K_{\nu}$ has a BC coming only from an overall factor $\epsilon^{-\nu}$,
 $K$ in \eqref{eq:K} has a BC which  comes only from a factor $\epsilon^{\nu+1}/\epsilon^{-\nu}=\epsilon^{2\nu+1}$ (with the `1' playing no part in the BC). In its turn,
  $R_\mathrm{C}^\nu$ has a BC which  comes only  from the factor $\epsilon^{\nu}$ (coming from the $\hat{z}^\nu$) in \eqref{eq:RC};
  thus, $K R_\mathrm{C}^{-\nu-1}$ has a BC which  comes only  from $\epsilon^{2\nu+1}\epsilon^{-\nu-1}=\epsilon^{\nu}$.
  Together, this means that $R_\mathrm{C}^{\nu}+K R_\mathrm{C}^{-\nu-1}$ has a BC which  comes only from an overall factor $\epsilon^{\nu}$, which is cancelled out by the prefactor $\epsilon^{-\nu}$ in \eqref{eq:RIn}, so that $\Rin{s}$, indeed, has no BC. 
  
It is interesting to note that all the mentioned BCs in intermediate quantities are  logarithmic BCs: a small-$\epsilon$ expansion of $\epsilon^\nu$ with $\nu=\ell+\sum_{n=1}^{\infty}\nu_n\epsilon^n$ (see Eq.~(173) in~\cite{Sasaki:2003xr} for generic $s$ or Eq.~\eqref{eq:nu-exp} below to higher order for $s=-2$ ), for  some coefficients $\nu_n$, contains terms like $\epsilon^{\nu_0+p}\ln^q\epsilon$, where $p,q\in\mathbb{Z}_{>0}$ with $q_0\leq q\leq p$ for some $q_0>0$.

Furthermore, our normalization for the In solution is symmetric
 under complex-conjugation together with $(m,\omega)\to (-m,-\omega)$, so that the symmetries  in \eqref{eq:symm-In} of $\Rin{s}(r,\omega)$ and its scattering coefficients $\Bincreftra$ are indeed satisfied. 
 Using the corresponding symmetry for the eigenvalue in \eqref{eq:ang-symm}, it follows that this  symmetry also applies to the  renormalized angular momentum and to the MST coefficients (whose normalization $a^\nu_{0}=1$ trivially satisfies the symmetry):
 \begin{equation}
 \nu=\left(\left.\nu\right|_{m\to -m, \omega\to -\omega^*}\right)^*,\quad
 a^\nu_{n}=\left(\left.a^\nu_{n}\right|_{m\to -m, \omega\to -\omega^*}\right)^*. 
\end{equation}

\subsection{``In" scattering coefficients}

The scattering coefficients in Eq.~\eqref{eq:bc Rin} of the In solution,  normalized as indicated  in Sec.~\ref{sec:Radial}, admit the following MST expressions:
\begin{align}\label{eq:Binc/ref/tra}
\Btra=&\left(\frac{\epsilon \kappa}{\omega}\right)^{2 s} \frac{\epsilon^{-\nu}}{K_\nu} e^{i \kappa \epsilon_+(1+\frac{2\ln\kappa}{1+\kappa})}\sum_{n=-\infty}^{\infty} \an{n}\nn,\\
B^{\rm inc}
=&2M \epsilon^{-\nu-1} \left[1-
ie^{-i\pi\nu} \frac{\sin \pi(\nu-s+i\epsilon)}
{\sin \pi(\nu+s-i\epsilon)}
{K}\right]A_{+}^{\nu} e^{-i(\epsilon\ln\epsilon -\frac{1-\kappa}{2}\epsilon)},
\nn\\
B^{\rm ref}
=& (2M)^{1+2s}\epsilon^{-\nu-1-2s}\left[1
+ie^{i\pi\nu} K\right]A_{-}^{\nu}
e^{i(\epsilon\ln\epsilon -\frac{1-\kappa}{2}\epsilon)}, \nn\\
\end{align}
where
\begin{align}\label{eq:A+/-}
&A_{+}^\nu\equiv e^{-{\pi\over 2}\epsilon}e^{{\pi\over 2}i(\nu+1-s)}
2^{-1+s-i\epsilon}{\Gamma(\nu+1-s+i\epsilon)\over 
\Gamma(\nu+1+s-i\epsilon)}\sum_{n=-\infty}^{+\infty}\an{n},\\
&A_{-}^\nu\equiv 2^{-1-s+i\epsilon}e^{-{\pi\over 2}i(\nu+1+s)}e^{-{\pi\over 2}\epsilon}
\sum_{n=-\infty}^{+\infty}(-1)^n{(\nu+1+s-i\epsilon)_n\over 
(\nu+1-s+i\epsilon)_n}\an{n}. 
\nonumber
\end{align}
Following the same argument as at the end of  Sec.~\ref{sec:Radial}, it is clear that $A_{\pm}^\nu$ given by \eqref{eq:A+/-} and $\Btra$ by \eqref{eq:Binc/ref/tra} have no BC.
On the other hand,  $\Binc$ and $\Bref$ in \eqref{eq:Binc/ref/tra}  have logarithmic BCs coming from the BCs of $K_\nu$, $K_{-\nu-1}$ and $e^{\mp i\epsilon\ln\epsilon}$.
These BCs in $\Binc$ and $\Bref$, which do not factor out, lead to $\Bref/\Binc$ also having a logarithmic BC.

\subsection{BC strength}

An expression for the BC strength is
given in 
 Eq.~(5.19) in~\cite{CKO}\footnote{It is understood that, on the right hand side of Eq.~(5.19) in~\cite{CKO}, one should replace $\omega=-i\fNIA$.},
\begin{align}
%p.8WOF
q(\fNIA)=(-1)^si\frac{A_+^\nu}{A_-^\nu}\fNIA^{2 s}\epsilon^{-2 i \epsilon}e^{i \epsilon(1- \kappa)}\left(1-e^{2\pi(\epsilon-i\nu)}\right). \label{Eq:q eqtn}
\end{align}

We next proceed to obtain small-frequency expansions of the various  factors appearing in the integrand in Eq.~\eqref{eq:Disc-m} for
$\Disc \mathcal{G}_{\ell m}$, namely expansions of the SWSHs ${}_s\Slm$ and its eigenvalues as well as expansions of the BC modes $\Disc \Glm$. We start with the former, since the eigenvalues ${}_s\lambda_{\ell m \omega}$ are required for the latter.

%---------------------------------------------------------------------------------------------------------

\section{SWSHs and angular eigenvalues}\label{sec:SWSH}

Apart from the BC modes $\Disc \Glm$, the integrand in Eq.~\eqref{eq:Disc-m}  contains the SWSHs ${}_s\Slm$; furthermore, the radial Eq.~\eqref{eq:radial teuk. eq.}, and so also the radial quantities, depend on the angular  eigenvalues ${}_s\lambda_{\ell m \omega}$. The SWSHs and the eigenvalues are analytic at $\omega=0$  and so they admit a Taylor series expansion about $\omega=0$ (on the other hand, as mentioned, they possess branch points away from the origin~\cite{oguchi1970eigenvalues, BONGK:2004}).

Secs.~III.B.4 and App.~B in~\cite{Kavanagh:2016idg} 
%\MC{Adrian and Chris - Eqs.3.23,3.24~\cite{Kavanagh:2016idg} seem to yield a (strictly) wrong ${}_s\Slm=\sum_{k=|s|}^{\infty}{}_sY_{km}$ for $\omega=0$. Do you wish to write anything about typos there?} 
derive expansions of the SWSHs for small $|a\omega|$ and in terms of spin-weighted {\it spherical} harmonics and of the angular eigenvalues ${}_{s}\lambda_{\ell m\omega}$ for small-$a\omega$.\footnote{The description of this expansion in Eqs.~(3.23) and (3.24) of \cite{Kavanagh:2016idg} is unfortunately incorrect although the correct expansion was implemented in the calculation through code now available as part of  \texttt{SpinWeightedSpheroidalHarmonics} package of the \texttt{BlackHolePerturbationToolkit} (BHPT)~\cite{BHPToolkit}.  For a correct description of the expansion algorithm see Appendix A of \cite{Hughes:2000} and erratum \cite{Hughes:2000E2}.}
We will use\footnote{These small $|a\omega|$ expansions are also the ones that were used to obtain the late-time asymptotics for the scalar field in~\cite{CKO}.} in  the integrand in Eq.~\eqref{eq:Disc-m} these expansions  of the SWSHs and angular eigenvalues for small absolute value  $|a\omega|$ of the spheroidicity. Since $a\omega=(a/M)(M\omega)=-i(a/M)(M\sigma)$, small $|a\omega|$ along the NIA can be thought of being small $M\sigma$ for fixed $a/M$.
Thus, in the language of the later Sec.~\ref{sec:Small_freq}, an expansion for small $|a\omega|$  up to $n$ terms beyond leading order is an $n$PS expansion.

The small $|a\omega|$ expansions of the SWSHs and angular eigenvalues can easily be obtained from the \\ \texttt{SpinWeightedSpheroidalHarmonics} package of the \texttt{BlackHolePerturbationToolkit} (BHPT)~\cite{BHPToolkit}, which essentially uses the method in Secs.~III.B.4 and App.~B in~\cite{Kavanagh:2016idg}. We give first the expansions of the SWSH up to 2PS:
\begin{subequations}
\begin{align}
\label{eq:SWSH,small-c}
    {}_s\Slm(\theta)=&\St{0}(\theta)+\St{1}(\theta)\,  \omega +\St{2}(\theta)\, \omega^2 +O( a^3\omega^3 )
    \,, \\
    \St{0}(\theta) \equiv& {}_sY_{\indmode}(\theta) 
    \,,\\
    \St{1}(\theta)\equiv &a\sum_{i=-1}^1 c^{(1)}_i \, {}_{s}Y_{(\ell+i) m} (\theta)
    \,,\\
    \St{2}(\theta)\equiv &a^2\sum_{i=-2}^2 c^{(2)}_i \, {}_{s}Y_{(\ell+i) m} (\theta)
    \,,\\
    c_{-1}^{(1)} =& \frac{s \sqrt{\ell ^2-m^2} \sqrt{\ell ^2-s^2} }{\ell ^2 \sqrt{2 \ell -1} \sqrt{2 \ell +1}}
    \,,\\
    c_{0}^{(1)} =& 0
    \,,\\
    c_1^{(1)} =&   -\frac{s \sqrt{(\ell +1)^2-m^2} \sqrt{(\ell +1)^2-s^2} }{(\ell +1)^2 \sqrt{2 \ell +1} \sqrt{2 \ell +3}}
    \,,\\
    c_{-2}^{(2)} =& -\frac{\sqrt{(\ell -1)^2-m^2} \sqrt{\ell ^2-m^2} \sqrt{(\ell -1)^2-s^2} \left(\ell -2 s^2\right) \sqrt{\ell ^2-s^2}}{4 (1-2 \ell )^2 (\ell -1) \ell ^2 \sqrt{2 \ell -3} \sqrt{2 \ell +1}}
    \,, \\
    c_{-1}^{(2)} =& \frac{m \sqrt{\ell ^2-m^2} \sqrt{\ell ^2-s^2} \left(s \ell ^2-2 s^3\right)}{2 \ell ^4 \sqrt{2 \ell -1} \sqrt{2 \ell +1} \left(\ell ^2-1\right)}
    \,,\\
    c_{0}^{(2)} = & \left( \frac{s^4 \left(\ell ^2 (\ell +1)^2 (2 \ell  (\ell +1)+3)-m^2 \left(2 \ell  (\ell +1) \left(\ell ^2+\ell +4\right)+3\right)\right)}{4 \ell ^4 (\ell +1)^4 (4 \ell  (\ell +1)-3)} \right.
    \notag\\
    &+ \left.\frac{s^2 \left(m^2 (2 \ell  (\ell +1)+3)-2 \ell ^2 (\ell +1)^2\right)}{4 \ell ^2 (\ell +1)^2 (4 \ell  (\ell +1)-3)} \right)
    \,,\\
    c_{1}^{(2)} =&-\frac{m s \sqrt{(\ell +1)^2-m^2} \left((\ell +1)^2-2 s^2\right) \sqrt{(\ell +1)^2-s^2} }{2 \ell  (\ell +1)^4 (\ell +2) \sqrt{2 \ell +1} \sqrt{2 \ell +3}}
    \,,\\
    c_{2}^{(2)} =&\frac{\sqrt{(\ell +1)^2-m^2} \sqrt{(\ell +2)^2-m^2} \left(2 s^2+\ell +1\right) \sqrt{(\ell +1)^2-s^2} \sqrt{(\ell +2)^2-s^2} }{4 (\ell +1)^2 (\ell +2) \sqrt{2 \ell +1} (2 \ell +3)^2 \sqrt{2 \ell +5}}
    \,.
\end{align}
\end{subequations}
Note that ${}_s\Slm^{*}(\theta)=
{}_sS_{\ell m \omega^*}(\theta)$, so that $c^{(n)}_i\in\mathbb{R}$ for all $n$ and $i$. The spin-weighted spheroidal eigenvalue ${}_s \lambda_{\ell m }$ admits a similar PS expansion. 
We give it up to 2PS explicitly,
\begin{subequations}
\label{eq:SWSHEV_PS}
\begin{align}
    {}_s\lambda_{\ell m \omega} =& {}_s\lambda_{\ell m}^{(0)} + a \omega {}_s\lambda_{\ell m}^{(1)} + (a \omega)^2 {}_s\lambda_{\ell m}^{(2)} + \mathcal{O}(a^3 \omega^3)
    \,,\\
    {}_s\lambda_{\ell m}^{(0)} =& \ell  (\ell +1)-s (s+1)
    \,,\\
    {}_s\lambda_{\ell m}^{(1)} =&  -\frac{2 m \left(s^2+\ell ^2+\ell \right)}{\ell  (\ell +1)}
    \,,\\
    {}_s\lambda_{\ell m}^{(2)} =& \frac{s^4 \left(2 m^2 \left(5 \ell ^2+5 \ell +3\right)-6 \ell ^2 (\ell +1)^2\right)}{\ell ^3 (\ell +1)^3 \left(4 \ell ^2+4 \ell -3\right)}+\frac{4 s^2 \left(-3 m^2+\ell ^2+\ell \right)}{\ell  (\ell +1) \left(4 \ell ^2+4 \ell -3\right)}+\frac{2 \left(m^2+\ell ^2+\ell -1\right)}{4 \ell ^2+4 \ell -3} 
    \,.
\end{align}
\end{subequations}
As mentioned, the expansions of both the SWSH and the angular eigenvalues are really expansions in the spheroidicity parameter $a\omega$. However, we prefer to write Eq.~\eqref{eq:SWSH,small-c} for the SWSH with coefficients of $\omega$ rather than of $a\omega$, since that will make the final tail expressions (which arise from expansions for small $M\sigma$  in the frequency domain) in Sec.~\ref{sec:tails} a little simpler.

\section{Small-frequency expansions of factors in the BC modes}\label{sec:Small-w factors}

In this section we use the MST expressions provided in Sec.~\ref{eq:MST} in order to obtain small $|\sigma|$ expansions of the
various factors that make up the BC modes $\Disc \Glm$ in Eqs.~\eqref{Eq:GBC}, \eqref{Eq:GBC,rp} and \eqref{eq:dGlm,r->infty}. Namely, we provide expansions of
the In radial solution $\Rin{s}$,
the BC-strength-related factor $\qW{\indmode}$ as well as $\qW{\indmode}$ times the In transmission  coefficient, $\qWtra{\indmode}$, and $\qW{\indmode}$  times the In reflection  scattering coefficient, $\qWref{\indmode}$.
The MST expressions for all the quantities depend on the MST series coefficients $\an{n}$ and  renormalized angular momentum $\nu$, whose small-frequency expansions we provide first.
We provide in App.~\ref{sec:small-w InCoeffs} the small-frequency expansions of the scattering coefficients $\Binc$, $\Bref$ and $\Btra$,
which we use to obtain the expansions for $\qW{\indmode}$, $\qWtra{\indmode}$ and $\qWref{\indmode}$ given in this section.
While we provide the coefficients in the expansions explicitly for spin $s=-2$, we obtain the powers of $\sigma$ for generic $s=0,\pm 1,\pm 2$. While we don't need $B_{\rm inc}$ explicitly we do note that in our normalization it scales as $\sigma^{-\ell-1}$ for all possible $s$, $\ell$ and $m$.

The expansions which we provide for the various quantities for {\it generic} $\ell$ and $m$ were obtained using a new version of the BHPT's \texttt{Teukolsky} package \cite{BHPToolkit}. This version expands the  low-frequency MST capabilities of the current online version of the package to generic $\ell$ and $m$ and gives easier access to all amplitudes. It will be merged into the main branch in the near future and accompanied by a publication. In the meantime it is available through the \texttt{PN} branch of the github page, though we warn that this branch is in active development.

Henceforth, we take units with $M=1$.

%---------------------------------------------------------------------------------------------------------

\subsection{Small-frequency expansion of series coefficients and renormalized angular momentum} \label{sec:small-w an-nu}

Here we provide the small $|\omega|=|\sigma|$ expansions of the MST series coefficients $\an{n}$ as well as of the renormalized angular momentum $\nu$ introduced in Sec.~\ref{sec:Radial} and which form the bedrock of  the MST series representations for all the various quantities. 
These expressions are valid for all $\ell \geq 2$ and $|m|\leq \ell$, however higher orders (like those given electronically)  will not be valid for lower $\ell$. In practice we find that the series  up to including $\omega^n$ is valid for $\ell \geq n$, though it is often the case that it is only the $a_{n}^\nu$ for negative $n$ that cause problems   while $\nu$ might be valid even below that threshold. 

For $s=-2$ and generic $\ell$ and $m$, we find the following expansions for $\nu$  and $\an{n}$, $n=-2\to 2$:
\begin{subequations}
\begin{align}\label{eq:nu-exp}
    \nu =&  
        \ell -\frac{2 \omega ^2 \left(15 \ell ^4+30 \ell ^3+28 \ell ^2+13 \ell +24\right)}{\ell  (\ell +1) (2 \ell -1) (2 \ell +1) (2 \ell +3)}+O\left(\omega ^3\right),
    \\
    a_0^\nu =& 1,
   \\
    a_1^\nu =& 
        \frac{i \omega  (\ell +3)^2 \left(\sqrt{1-a^2} (\ell +1)+i a m\right)}{(\ell +1)^2 (2 \ell +1)}+\frac{2 \omega ^2 (\ell +3)^2 \left(4 a^2 m^2-4 i a \sqrt{1-a^2} m (\ell +1)+\ell  (\ell +1)^2 (\ell +2)\right)}{\ell  (\ell +1)^4 (\ell +2) (2 \ell +1)}+O\left(\omega ^3\right),
    \\
    a_{-1}^\nu=&
        \frac{\omega  (\ell -2)^2 \left(a m+i \sqrt{1-a^2} \ell \right)}{\ell ^2 (2 \ell +1)}-\frac{2 \omega ^2 \left((\ell -2)^2 \left(4 a^2 m^2+4 i a \sqrt{1-a^2} m \ell +\ell ^4-\ell ^2\right)\right)}{(\ell -1) \ell ^4 (\ell +1) (2 \ell +1)}+O\left(\omega ^3\right),
    \\
    a_{2}^\nu=&
        \frac{\omega ^2 (\ell +3)^2 (\ell +4)^2 \left(a^2 \left(m^2+\ell ^2+3 \ell +2\right)-i \sqrt{1-a^2} a m (2 \ell +3)-\ell ^2-3 \ell -2\right)}{(\ell +1)^2 (\ell +2) (2 \ell +1) (2 \ell +3)^2}+O\left(\omega ^3\right),
    \\
    a_{-2}^\nu=&
        \frac{\omega ^2 (\ell -3)^2 (\ell -2)^2 \left(a^2 \left(m^2+(\ell -1) \ell \right)+i \sqrt{1-a^2} a m (2 \ell -1)-(\ell -1) \ell \right)}{(1-2 \ell )^2 (\ell -1) \ell ^2 (2 \ell +1)}+O\left(\omega ^3\right).
\end{align}
\end{subequations}
The expansion of $\nu$ up to $O(\omega^2)$ for generic spin was already given in Eq.~(3.7) in~\cite{CKO} and it agrees, for $s=-2$, with Eq.~\eqref{eq:nu-exp} here.
To the best of our knowledge, the explicit expansion of the MST coefficients $a_{n}^\nu$ for $s=-2$ is only given here for the first time (in~\cite{CKO} it was only given for $s=0$). These expressions used the expansion of the spin-weighted spheroidal eigenvalue in Eq.~\eqref{eq:SWSHEV_PS}.

\subsection{Small-frequency expansion of the In radial solution}
\label{sec:Small_freq}

Let us now turn to an expansion of the In radial solution $\Rin{s}(r, -i\fNIA)$.
We write the following ansatz for the asymptotics for small $|\sigma|$ of our In radial solution at fixed radius (remember that we choose a $\Btra$ without a BC):
\begin{equation}\label{eq:Rin-generic}
\Rin{s}(r, -i\fNIA)=
%\sigma^{\Expr}\sum_{n=0}^{\infty}R_n(r)\,\sigma^n,
\sigma^{\Expr}\left(\sum_{n=0}^{2}R_n(r)\,\sigma^n+o(\sigma^2)\right),
\end{equation}
for some coefficients $R_n(r)$ and  power $\Expr = -s$ for all $s=0,\pm 1,\pm 2$ and generic $\ell\geq |s|$ and $|m|\leq \ell$. 
Note that even if we write Eq.~\eqref{eq:Rin-generic} in terms of $\sigma$ instead of $\omega$, this equation --and all equations in this subsection, hold for all $\sigma\in\mathbb{C}$.
We refer to an expansion for small $|\sigma|$ without scaling of other variables as \textit{post-static} (PS), where $n$PS refers to an expansion up to $n$ terms beyond the leading order. 
The task of obtaining the coefficients $R_n(r)$ can be approached in a number of ways:
\begin{enumerate}[label=\textbf{\alph*)}, ref=\alph*)]
    \item 
    \label{it:Barnes}
        \textbf{Mellin-Barnes approach:} The coefficients $R_n(r)$ in Eq.~\eqref{eq:Rin-generic} can be obtained from the MST series representation for $\Rin{s}$ provided by Eqs.~(116) and (120) in \cite{Sasaki:2003xr}. We shall repeat it here (in our normalization) for convenience,
        \begin{subequations}
        \begin{align}
            &{}_sR^{\rm in}_{\ell m}(r,\omega) = \frac{\epsilon^{-\nu}}{K_\nu} e^{i \epsilon \kappa x}(-x)^{-s-i(\epsilon+\tau)/2} (1-x)^{i(\epsilon-\tau)/2}p_\mathrm{in}(x)
            \,,\\
            & p_{\rm in}(x) = \sum_{n=-\infty}^{\infty} a_n^\nu\, {}_2 F_1 (n+\nu+1-i \tau,-n-\nu-i\tau;1-s-i\epsilon-i\tau;x)
            \,,\\
            &x= \frac{1-r+\kappa}{2\kappa}
            \,.
        \end{align}
        \label{eq:p_in}
        \end{subequations}
        This expansion converges for all $r\in [r_+,\infty)$. It can be expanded in low frequency by way of the Mellin-Barnes representation of the hypergeometric ${}_2F_1$ functions. This is indeed what was done in~\cite{CKO} for the scalar perturbations of Kerr and in~\cite{Casals:Ottewill:2015} for the general-spin perturbations of Schwarzschild. However, this yielded coefficients $R_n(r)$, most of which could not be obtained in closed form but only as integral representations which need to be evaluated numerically.
    \item 
        \label{it:Der_2F1}
        \textbf{Derivatives of $\mathbf{{}_2F_1}$:} If one naively expands  Eq.~\eqref{eq:p_in} in low frequency one encounters  derivatives of hypergeometric functions with respect to their first arguments, e.g., $\partial_a\,{}_{2}F_1 (a,b,c;x)$. The \texttt{HypExp} package \cite{HypExp} allows their computation (to high order) for $\{a,b,c\} \in \mathbb{Z}$, which is the case for a Schwarzschild background. However, in a Kerr background one needs to compute these derivatives for arbitrary complex entries, which is not currently supported. It seems, however, likely that one could resum the integer-valued-argument results and analytically continue to obtain an expression valid for all $\{a,b,c\}\in  \mathbb{C}$. We leave this topic to future work. In the meantime one can evaluate the derivatives numerically  (after plugging in values for the intrinsic angular momentum $a$ and radius $r$). This is possible with \texttt{Mathematica} albeit slow. Furthermore it will result in a GF that is virtually impossible to integrate over $r$ analytically whenever convoluted with a physical source. 
    \item
        \label{it:PN}
        \textbf{PN expansion:} We may construct large-radius asymptotics of the coefficients $R_n(r)$ in Eq.~\eqref{eq:Rin-generic} via a post-Newtonian (PN) expansion of Eq.~\eqref{eq:RC}. To do so we will, as is standard in perturbation theory, introduce an order counting parameter $\eta$ which we can treat as being small and set equal to 1 at the end of the problem (see e.g. Chapter 7 of \cite{Bender:Orszag}).  We treat $r/M$ to be large in the same sense as if we were describing the near-zone in  PN theory for a two-body problem, and $\omega$ were tied to an orbital frequency via Kepler's law. Thus, it is $r/M\sim (M\omega)^{2/3}$, motivating the rescaling $r\rightarrow r  \,\eta^{-2},\omega\rightarrow \omega\, \eta^3$. Setting $\eta$ as formally small simultaneously enforces small-frequency and large radius (keeping $r\,\omega^{-2/3}$ finite) assumptions. Expansions using this method are also available within the BHPT's \texttt{Teukolsky} package \cite{BHPToolkit}.
\end{enumerate}

The first two approaches \ref{it:Barnes} and \ref{it:Der_2F1} 
are expansions for small frequency only, and so they are pure PS expansions. As such, they should yield exactly the same result (for expansions to the same order in $\sigma$) obtained in two somewhat different ways. On the other hand, the PN approach \ref{it:PN} should yield an (analytical) approximation for large radius to the small-frequency result from \ref{it:Barnes} and \ref{it:Der_2F1} (for an expansion to the same order in $\sigma$).

\begin{figure}
    \centering
    \includegraphics[width=.9 \linewidth]{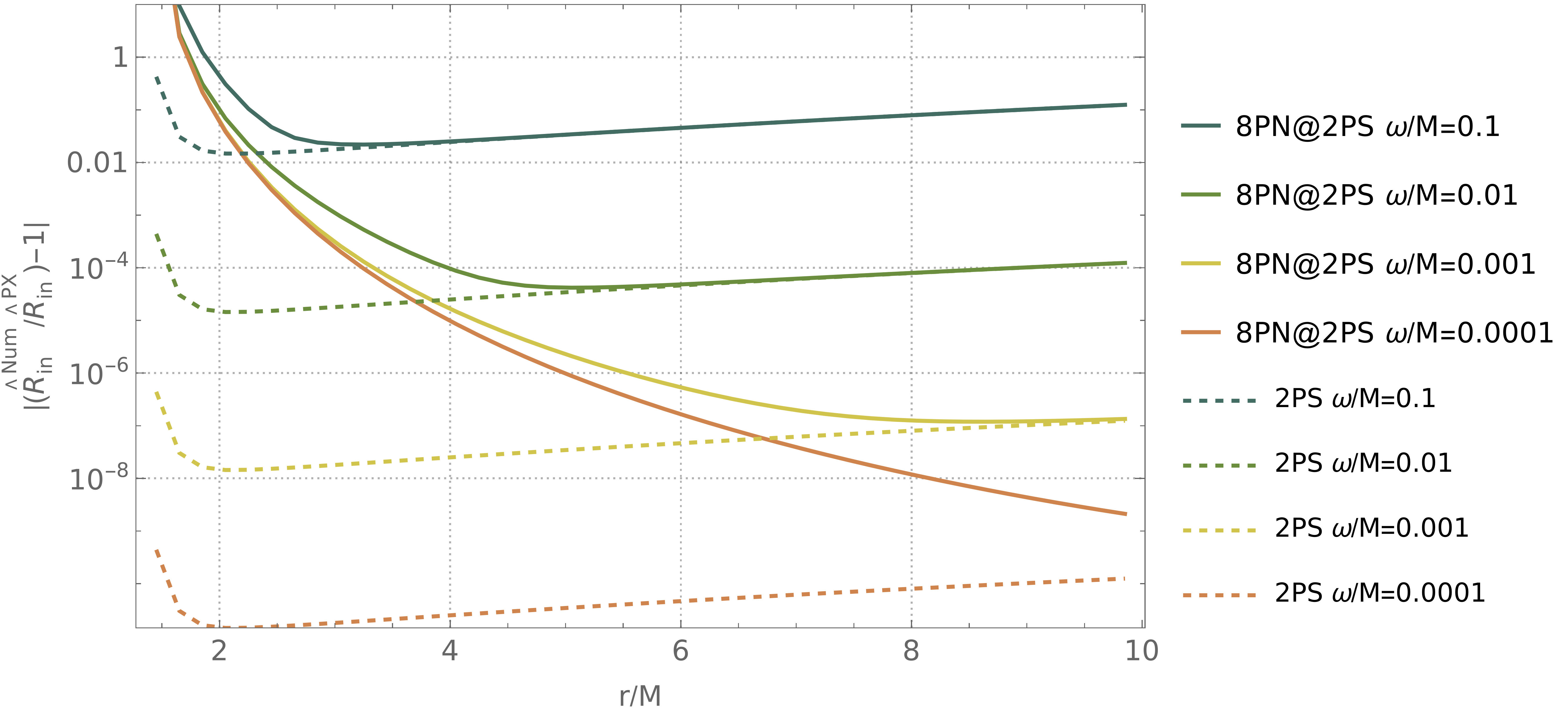}
    \caption{\textbf{PN vs. PS $\mathbf{{}_{-2}R_{\mathbf{\ell} m}^{\mathrm{\bf in}}}$:} Relative error of the approximations to the radial function ${}_{-2}\hat{R}_{\ell m}^{\rm in}\equiv {}_{-2}R_{\ell m}^{\rm in}/\Btra$  with $\{\ell,m\}=\{2,2\}$, for $a=0.9$ over radius $r$ (from the horizon $r_+ \approx 1.44$ to $r=10$) for different values of $\omega$. The solid lines are the 8PN expansion truncated at $\omega^2$ beyond leading, i.e. next-to-next-to leading order (NNLO), (8PN@2PS) according to \ref{it:PN} in Sec.~\ref{sec:Small_freq} (c.f. Eq.~\eqref{eq:Rin-PN,s=-2,exp}). The dashed lines are the pure low frequency (PS) expansion up to NNLO, where the derivatives of the ${}_2F_1$ are evaluated numerically according to \ref{it:Der_2F1} in the same section. This figure illustrates that even though the PN expansion is a large $r$ expansion, it converges  reasonably well as $\omega$ decreases for a fixed radius near the horizon.}
    \label{fig:RelErr}
\end{figure}

Here we choose to follow the PN approach. Using the representation for $\Rin{s}$ provided by Eqs.~\eqref{eq:RIn} and \eqref{eq:RC} with the PN rescalings leads directly to large-radius expansions of the coefficients $R_n(r)$. The advantage of this is that homogeneous Teukolsky solutions in this expansion are purely polynomial and logarithmic in $r$ and $\omega$  (although  $\Rin{s}$ in our normalization does not contain logarithms in $\omega$), and thus our expressions are in closed form. Moreover, in practice we find that the PN expansion converges well even in the strong field (c.f. Figs.~\ref{fig:RelErr},~\ref{fig:PN_tail_convergence_r3} and~\ref{fig:residuals_r3}).

Ultimately, our expressions for the high-order tails will be formally given  in Sec.~\ref{sec:tails} in terms of generic coefficients $R_n(r)$, regardless of how they are obtained -- whether as exact integral representations following the Mellin-Barnes approach as in~\cite{CKO,Casals:Ottewill:2015} (which one could extend to the $s=-2$ case here) or as the large-radius PN expansion which we next provide. In our explicit results of Sec. VII A, we will maintain for computational efficiency a consistent PN expansion through to the Green function.

Henceforth, we shall refer to an expansion being to $n$PN order, for a certain value of $n$, as meaning that the expansion includes up to terms of order $\eta^{2n}$ beyond the leading order. 
So, for example, 2PN corresponds to a series up to $\eta^4$ if the leading order is $\eta^0$. 
We note that, within our terminology, we do not consider $\log(\eta)$ factors to constitute a new ``order", so that a term of order $\eta^j$ could in fact also include  terms with $\eta^j\log\eta^k$ for arbitrary $k\in\mathbb{Z}_{\geq 0}$.
At a given PN order,  both infinite sums in \eqref{eq:RC}  truncate to finite number of terms and one is left with polynomials of $\sigma$, $r$, $\log r$  and $\log \sigma$ (the $\log \sigma$'s in $R_C^{\nu}$ in \eqref{eq:RC}, however, do not not survive in our $\Rin{s}$,
as explained below \eqref{eq:Rn}). 
%\MCt{Henceforth, whenever within the context of a PN expansion we refer to or write a term at a certain power of $\eta$ or $r$ or $\sigma$, it will be assumed that its coefficient may also contain powers of $\log\eta$, $\log r$ or $\log\sigma$.}

We give here explicitly the In radial solution for $s=-2$ and all $\ell\geq 2$ and $|m|\leq \ell$ up to 2PN: 
\begin{subequations}
\label{eq:Rin_PN}
\begin{align}
    {}_{-2}R_{\ell m}^{\rm in} =& {}_{-2}R_{\ell m}^{ \mathrm{in}(0)} \left( 1 + \eta \, {}_{-2}R_{\ell m}^{\mathrm{in} (1)} +\eta^2 {}_{-2}R_{\ell m}^{\mathrm{in} (2)} + \eta^3 {}_{-2}R_{\ell m}^{\mathrm{in} (3)} + \eta^4 {}_{-2}R_{\ell m}^{\mathrm{in} (4)}+ o(\eta^4) \right) 
    \,,\\
    {}_{-2}R_{\ell m}^{\mathrm{in} (0)} =&
    -\frac{\sigma ^2 r^{\ell +2} \eta ^{2-2 \ell } \Gamma (\ell +3)}{\Gamma (2 (\ell +1))}
    \,,\\
    {}_{-2}R_{\ell m}^{\mathrm{in} (1)} =&
    \frac{2 r \sigma }{\ell +1}
    \,,\\
    {}_{-2}R_{\ell m}^{\mathrm{in} (2)} =&
    \frac{r^2 \sigma ^2 (\ell +9)}{2 (\ell +1) (2 \ell +3)}-\frac{\frac{2 i a m}{\ell }+\ell +2}{r}
    \,,\\
    {}_{-2}R_{\ell m}^{\mathrm{in} (3)} =&
    \frac{r^3 \sigma ^3 (\ell +4)}{(\ell +1) (\ell +2) (2 \ell +3)}+\sigma  \left(\frac{4 i \sqrt{a^2-1}}{\ell ^2+\ell }-\frac{2 i a m (\ell +2)}{\ell ^2 (\ell +1)}-\frac{2 (\ell +3)}{\ell +1}+2 \psi ^{(0)}(\ell +3)\right)
    \,,\\
    {}_{-2}R_{\ell m}^{\mathrm{in} (4)} =&
    \sigma ^2 \left(\frac{8 i \sqrt{a^2-1} r}{\ell  (\ell +1)^2}-\frac{i a m r \left(\ell ^4+5 \ell ^3+19 \ell ^2+31 \ell +24\right)}{\ell ^2 (\ell +1)^3 (2 \ell +3)}-\frac{r \left(\ell ^2+5 \ell -8 (2 \ell +3) \psi ^{(0)}(\ell +3)+24\right)}{2 (\ell +1) (2 \ell +3)}\right)
   \notag\\
    &+ \frac{a^2 m^2 (\ell -8)}{2 r^2 \ell  (2 \ell -1)}  +\frac{\left(\ell ^2+3 \ell +2\right) \left(a^2+2 \ell -2\right)}{2 r^2 (2 \ell -1)}+\frac{2 i a m (\ell +1)}{r^2 \ell }+\frac{r^4 \sigma ^4 \left(\ell ^2+19 \ell +50\right)}{8 (\ell +1) (\ell +2) (2 \ell +3) (2 \ell +5)}
    \,.
\end{align}
\end{subequations}
where $\psi^{(n)}(z)=\tfrac{d^{n+1}}{dz^{n+1}}\ln\Gamma(z)$ is the polygamma function. 
To this PN order, no $\log\eta$'s have yet appeared but we do see them appearing at higher orders.
Let us write the expansion of the In radial solution to $n$PN order as $\Rin{-2}(r,-i\fNIA)=\eta^{2(1-\ell)}\left(\sum_{j=0}^{2n}\sum_{k=0}^{k_{\text{max}}}c_{jk}(r,\sigma)\,\eta^j\log^k\eta+o(\eta^{2n})\right)$, for some coefficients $c_{jk}(r,\sigma)$ and upper indices $k_{\text{max}}=k_{\text{max}}(j)$. Then we find that $c_j(r,\sigma)$ is 
a polynomial in $\sigma$ of degree $j+2$ and whose smallest power is  equal to $2$ if $j$ is odd and equal to $3$ if $j$ is even.
Thus, the expansion of $\Rin{-2}(r,-i\fNIA)$ to $n$PN order is $\sigma^2$ times a polynomial in $\sigma$ of degree $2n$.

For the purposes of computing the late time tail, we can now set $\eta=1$ and re-order Eq.~\eqref{eq:Rin_PN} as a double expansion in $\sigma$ and $r$: 
\begin{align}
\label{eq:Rin-PN,s=-2,exp}
    \Rin{-2}(r,-i\fNIA)=& 
    -\sigma ^2 \frac{\Gamma (\ell +3)}{\Gamma (2 \ell +2)}\left[ r^{\ell+2} -r^{\ell+1} \frac{ (\ell  (\ell +2)+2 i a m)}{\ell }  \right.
    \notag\\
    &+ \left. r^{\ell} \left( \frac{a^2 m^2 (\ell -8)}{2 \ell  (2 \ell -1)}+\frac{\left(\ell ^2+3 \ell +2\right) \left(a^2+2 \ell -2\right)}{4 \ell -2} +  \frac{2 i a m (\ell +1)}{\ell }\right) + o\left(\frac{1}{r^{-\ell}}\right) \right]
    \notag\\
    &-\sigma ^3 \frac{4 \Gamma (\ell +3)}{\Gamma (2 \ell +3)} \left[ r^{\ell+3} + r^{\ell+2} \left(\frac{2 i \sqrt{a^2-1}}{\ell }-\frac{i a m (\ell +2)}{\ell ^2}-\ell +(\ell +1) \psi ^{(0)}(\ell +3)-3\right) +o\left( \frac{1}{r^{-\ell -2}} \right)\right]
    \notag\\
    & -\sigma ^4 \frac{(\ell +9) \Gamma (\ell +3)}{\Gamma (2 \ell +4)} \left[ r^{\ell+4} + r^{\ell+3} \left( \frac{16 i \sqrt{a^2-1} (2 \ell +3)}{\ell  (\ell +1) (\ell +9)}-\frac{2 i a m \left(\ell ^4+5 \ell ^3+19 \ell ^2+31 \ell +24\right)}{\ell ^2 (\ell +1)^2 (\ell +9)} \right. \right.
    \notag\\
    &- \left. \left. \frac{\ell ^2+5 \ell +24}{\ell +9}+\frac{8 (2 \ell +3) \psi ^{(0)}(\ell +3)}{\ell +9}\right) + o\left(\frac{1}{r^{-\ell-3}} \right) \right]
    \notag\\
    &- \sigma ^5\frac{4 (\ell +4)\Gamma (\ell +3)}{\Gamma (2 \ell +5)} \left[ r^{\ell +5} + o\left(\frac{1}{r^{-\ell-5}} \right) \right]
    \notag\\
    &-\sigma ^6 \frac{(\ell  (\ell +19)+50)\Gamma (\ell +3)}{2 \Gamma (2 \ell +6)} \left[  r^{\ell +6} + o\left( \frac{1}{r^{-\ell-6}} \right) \right] + o(\sigma^7)
\end{align}
As mentioned, a PS series involves a truncation in only $\sigma$, whereas a PN series involves a truncation in both $\sigma$ and $1/r$.
E.g., the 2PN  series in \eqref{eq:Rin-PN,s=-2,exp} contains an expansion for small $\sigma$ up to 4PS  where each coefficient is expanded up to a power of $1/r$ such that it reaches up to 4 powers of $\eta$ beyond the overall leading order. In the cases  where we use PN as an approximation of PS we throw away higher order terms in $\sigma$, while keeping the PN order for the $r$ expansion. For these cases we introduce the notation $n$PN@$m$PS, meaning the $n$PN series where all terms in $\sigma$ above $m$PS are truncated. E.g., if we were to expand \eqref{eq:Rin-PN,s=-2,exp} for 2PN@0PS it would only consist of the $\sigma^2$ term, i.e. the first two lines.
The coefficients $R_n(r)$, for $s=-2$ and expanded for large $r$, in Eq.~\eqref{eq:Rin-generic} are readily read off from Eq.~\eqref{eq:Rin-PN,s=-2,exp}.
Logarithms of $r$ do appear at an order for large radius higher than that explicitly shown in Eq.~\eqref{eq:Rin-PN,s=-2,exp}; specifically, for $-s=\ell=2$ they appear at 3PN.

In Fig.~\ref{fig:RelErr} we provide a plot of the relative difference between the PS or PN@PS expansion \eqref{eq:Rin-PN,s=-2,exp} and the `exact' numerical value for $\Rin{-2}$ obtained using the BHPT. One can see that with an expansion to 8PN order we achieve for moderate dimensionless radii, e.g. $r\sim6$, a relative error of $\sim10^{-6}$.

\subsection{Small-frequency expansion of the BC strength factor}

In order to obtain a small-frequency expansion of $\qW{\indmode}$ in Eq.~\eqref{eq:Q}, we use \eqref{Eq:WpWm} for $\Wbp \Wbm$ and \eqref{Eq:q eqtn} for the BC strength $q(\fNIA)$, together with
\eqref{eq:Binc/ref/tra} for $\Bincp$ and $\Brefp$ (see App.~\ref{sec:small-w InCoeffs} for the expansions of $\Binc$ and $\Bref$, as well as $\Btra$, for $s=-2$) and, finally,  \eqref{eq:A+/-} for $A_{\pm}^\nu$.
The latter contains an infinite series  which we expand for small $\epsilon$ 
using the expansions for the MST coefficients $\an{n}$ and renormalized angular momentum $\nu$ that we provided in Sec.~\ref{sec:small-w an-nu}

We find that, for all $s=0, \pm 1, \pm 2$ and generic $\ell\geq 2$ and $|m|\leq \ell$, the expansion of $\qW{\indmode}$ has the following form:
\begin{equation}\label{eq:Q-generic}
\qW{\indmode}=\sigma^{\ExpQ} \qW{\text{pre}} \left(1 +\qW{1}\sigma+\left(\qW{2}+\qW{2}^{\logt}\,\log(2 \sigma)\right)\sigma^2+o(\sigma^2)\right), 
\end{equation}
where $\ExpQ=2(s+\ell)+1$  in our chosen normalization (see Sec.~\ref{sec:Radial}), 
for some coefficients $\qW{0\to 2}$ and $\qW{2}^{\logt}$.
 For the specific case $s=-2$, these coefficients are given by 
\begin{subequations}\label{:eq:Q-exp}
\begin{align}
    \qW{\text{pre}} =& 
    \frac{(-1 )^{\ell }2^{2 \ell +2 } \pi}{(\ell -1) \ell  (\ell +1) (\ell +2)}, %\sigma^{2\ell-3}
    \\
    \qW{\text{1}} =&
  -\frac{4 \left(5 \ell ^3+6
   \ell ^2-9 \ell -5\right)}{(\ell -1) \ell  (\ell +1) (\ell +2)}  + \frac{143 \ell ^2+143 \ell -51}{(2 \ell -1) (2 \ell +1) (2 \ell +3)}-\frac{8 \kappa }{\ell  (\ell +1)}-4 \psi ^{(0)}(\ell )-\frac{4 i a m (\ell -1) (\ell +2)}{\ell ^2 (\ell +1)^2},
 \label{eq:Q1-exp}  \\
   \qW{\text{2}} =&\frac{1}{2} -\frac{2 \pi ^2}{3}+\frac{4 \left(47 \ell ^7+169 \ell ^6+41 \ell ^5-296 \ell
   ^4-94 \ell ^3+79 \ell ^2-42 \ell +204\right)}{3 (\ell -1)^2 \ell ^2 (\ell +1)^2 (\ell
   +2)^2}
    \nonumber\\&\ 
   -
   \frac{12032 \ell ^4+25864 \ell ^3+3164 \ell ^2-6618 \ell +3933}{6 (2 \ell
   -1)^2 (2 \ell +1) (2 \ell +3)^2}
    \nonumber \\&\  
   +4\psi ^{(0)}(\ell ) \left(
  \frac{4 \left(5 \ell ^3+6 \ell ^2-9 \ell -5\right)}{(\ell -1) \ell  (\ell +1) (\ell +2)}-
  \frac{\left(143 \ell ^2+143 \ell -51\right)}{(2 \ell -1) (2 \ell +1) (2 \ell +3)}
   \right)
    +8 \psi ^{(0)}(\ell )^2
    \nonumber\\&\ 
   +8 \kappa  \left(\frac{272 \ell ^2+272 \ell -279}{(2 \ell -1) (2 \ell +1) (2 \ell +3)}-\frac{34
   \ell ^5+85 \ell ^4-20 \ell ^3-109 \ell ^2+2 \ell +20}{(\ell -1) \ell ^2 (\ell +1)^2 (\ell
   +2)}+\frac{4\psi ^{(0)}(\ell ) }{\ell (\ell+1) }\right)
   \nonumber\\&\ 
   -a^2 \left(\frac{1}{2}-\frac{48 \left(\ell ^2+\ell
   -1\right)}{\ell ^2 (\ell +1)^2}+\frac{15 \left(92 \ell ^2+92 \ell -129\right)}{2 (2 \ell
   -1)^2 (2 \ell +3)^2}\right)
   \nonumber\\&\ 
  -4 i a m \left(\frac{476 \ell ^7+1667 \ell ^6+1065
   \ell ^5-1503 \ell ^4-1625 \ell ^3-116 \ell ^2+52 \ell -40}{(\ell -1) \ell ^3 (\ell +1)^3
   (\ell +2)}-\frac{2 \left(1904 \ell ^2+1904 \ell -1453\right)}{(2 \ell -1) (2 \ell +1) (2
   \ell +3)}\right)
   \nonumber\\&\ 
  +\frac{16 i a m (\ell -1) (\ell +2)}{\ell ^2 (\ell
   +1)^2} \psi ^{(0)}(\ell )-\frac{2 i a m \kappa \left(\ell ^6+3 \ell ^5-85
   \ell ^4-175 \ell ^3+168 \ell ^2+256 \ell -48\right)}{\ell ^3 (\ell +1)^3 (2 \ell -1) (2
   \ell +3)}
    \nonumber\\&\ 
   +2 a^2 m^2 \left(\frac{15 \left(92 \ell ^2+92 \ell -129\right)}{(2 \ell -1)^2 (2 \ell
   +3)^2}-\frac{2 \left(43 \ell ^6+129 \ell ^5+137 \ell ^4+59 \ell ^3-28 \ell ^2-36 \ell
   -4\right)}{\ell ^4 (\ell +1)^4}\right),\label{eq:Q2-exp}
   \\
   \qWLog{\text{2}} =&-\frac{32 (2
   \ell +1)}{\ell  (\ell +1)}+\frac{4 \left(143 \ell ^2+143 \ell -51\right)}{(2 \ell -1) (2 \ell +1) (2 \ell +3)}.\label{eq:Q2l-exp}
 \end{align}
\end{subequations}
In the specific mode $\ell=2$, the $s=-2$ coefficients become:
\begin{subequations}
\begin{align}
    \qW{\text{pre}}\Big|_{\ell=2} =& \frac{8 \pi }{3},
    \\
    \qW{1}\Big|_{\ell=2} =& 
    -\frac{4}{9} i a m-\frac{4 \kappa }{3}+4 \gamma -\frac{661}{210},
    \\
    \qW{2}\Big|_{\ell=2} =& 
    -\frac{a^2 \left(401 m^2+4077\right)}{3969}-\frac{1}{945} i a m (-610 \kappa +1680 \gamma -1007) +\frac{1322  }{315}\kappa-\frac{2}{105} \gamma  (280 \kappa +661)-\frac{2 \pi ^2}{3}
    \notag\\
    &+8 \gamma ^2+\frac{8735}{1764},
    \notag\\
    \\
    \qW{2}^{\logt}\Big|_{\ell=2} =& \frac{428}{105},
\end{align}
\end{subequations}
where $\gamma$ is Euler's constant.
We note in Eq.~\eqref{eq:Q-generic} the appearance of the first $\log(\sigma)$  at next-to-next-leading order.
That this is the case for $s=-2$ follows readily from Eq.~\eqref{:eq:Q-exp}; for the other integer values of $s$, we have verified it empirically by giving many values to $\ell$ and $m$.

\subsection{Small-frequency expansion of the BC strength factor times the transmission coefficient}

 In Eq.~\eqref{Eq:GBC,rp}  for the BC modes $\Disc \Glm$ as $r\to r_+$,
 we need the BC strength factor $\qW{\indmode}$ times the In transmission coefficient $\Btra$, i.e., $\qWtra{\indmode}=\qW{\indmode}\Btra$.
Since $\Btra$ possesses no BC, 
 the first $\log(\sigma)$ in $\qW{\indmode}\Btra$ appears, like for $\qW{\indmode}$ in Eq.~\eqref{eq:Q-generic}, at next-to-next-to-leading order. 
 We find that, for all $s=0, \pm 1, \pm 2$ and generic $\ell\geq 2$ and $|m|\leq \ell$, the expansion of $\qWtra{\indmode}$ has the following form:
\begin{equation}\label{eq:QBtra-generic}
\qWtra{\indmode}(\sigma)
=\sigma^{\ExpQTra}\qWtra{\text{pre}}\left(1 +\qWtra{1}\sigma+\left(\qWtra{2}+\qWtralog{2}\,\log(2 \sigma)\right)\sigma^2+o(\sigma^2)\right), 
\end{equation}
where $\ExpQTra=2\ell+s +1+b$ with $b=0$ except for $s>0$ and $m=0$, in which case it is $b=1$ (since $\qW{\indmode}$ is of order $\sigma^{2s+2\ell+1}$, and $\Btra$ is of $\sigma^{-s+b}$ for some coefficients $\qWtra{0\to 2}$ and $\qWtralog{2}$. 

 For the specific case $s=-2$,  the expansion for $\qW{\indmode}$ is given in \eqref{eq:Q-generic}--\eqref{:eq:Q-exp} and for $\Btra$ in \eqref{eq:Btra-exp,s=-2}--\eqref{eq:Qref coeffs}. 
 Multiplying them together, we obtain, for $s=-2$:
 \begin{subequations}
\begin{align}
    \qWtra{\text{pre}} &=
    \frac{(-1)^{\ell +1} e^{-\frac{1}{2} i a m
   \left(\frac{2 \log (\kappa )}{\kappa
   +1}+1\right)}2^{3 \ell } \pi \kappa
   ^{\ell -2}  \Gamma (\ell
   -1) \Gamma (\ell +3)  \Gamma \left(\ell +1+{i a
   m}/{\kappa }\right)}{\Gamma (2 \ell
   +1) \Gamma (2 \ell +2) \Gamma \left(3+{i a
   m}/{\kappa }\right)},\\
   \qWtra{1}&=\qW{1}+\BtW{1},\\
   \qWtra{2}&=\qW{2}+\qW{1}\BtW{1} + \BtW{2},\\
   \qWtralog{2}&=\qW{2}^{\logt},
\end{align}
 \end{subequations}
where the expansions of $\qW{1}$, $\qW{2}$,  $\qW{2}^{\logt}$, $\BtW{1}$ and $\BtW{2}$ are given in Eqs.~\eqref{eq:Q1-exp},~\eqref{eq:Q2-exp},~\eqref{eq:Q2l-exp},~\eqref{eq:B1tra-exp},~\eqref{eq:B2tra-exp}, respectively (the explicit expansions for $\qWtra{1}$ and $\qWtra{2}$ are very long and we provide them electronically).
For the specific mode of $-s=\ell =2$, it is $\delta=3$
\begin{subequations}
\begin{align}
    \qWtra{\text{pre} }\Big|_{\ell=2} =& -\frac{8}{15} \pi  e^{-\frac{i a m (\kappa +2\log (\kappa )+1)}{2 (\kappa +1)}},
    \\
    \qWtra{1}\Big|_{\ell=2} =& 
    2 \psi ^{(0)}\left(\frac{i a m}{\kappa }+3\right)-\frac{4}{9} i a m+\kappa +\log (\kappa )+4 \gamma
   -\frac{451}{210},
   \\
    \qWtra{2}\Big|_{\ell=2} =& 
    -\frac{2 \left(a^2-2 \kappa -1\right) \psi ^{(1)}\left(\frac{i a m}{\kappa }+3\right)}{-\kappa^2}
    +\frac{1}{315} (-280 i a m+630 \kappa +2520 \gamma -711) \psi ^{(0)}\left(\frac{i a m}{\kappa }+3\right) 
    \notag\\
    &+ 2 \log \left(\kappa\right) \left(\frac{1}{630} (-280 i a m+630 \kappa +2520 \gamma -711)+2 \psi ^{(0)}\left(\frac{i a m}{\kappa }+3\right)\right)+2 \psi ^{(0)}\left(\frac{i a m}{\kappa }+3\right)^2
    \notag\\
    &+\frac{a^2 \left(-802 m^2-5067\right)}{7938}+\frac{1}{2} \log ^2\left(1-a^2\right)+a \left(\frac{1}{945} i (1217-1680 \gamma ) m-\frac{4 i \kappa  m}{9}\right)
    \notag\\
    & +\left(4 \gamma -\frac{451}{210}\right) \kappa +8 \gamma ^2-\frac{158 \gamma }{35}-\frac{144278}{11025}+\frac{214 \log (2)}{105},
    \\
    \qWtralog{2}\Big|_{\ell=2} =&
    \frac{428}{105}.
\end{align}
\end{subequations}

\subsection{Small-frequency expansion of the BC strength factor times the reflection coefficient}

In Eq.~\eqref{eq:dGlm,r->infty} for the BC modes $\Disc \Glm$ as $r\to \infty$,
we need the BC strength factor $\qW{\indmode}$ times the In reflection coefficient $\Brefp$, i.e., $\qWref{\indmode}=\qW{\indmode}\Brefp$. We find that, for all $s=0, \pm 1, \pm 2$ and generic $\ell\geq 2$ and $|m|\leq \ell$, the expansion of $\qWref{\indmode}$ has the following form:

\begin{equation}\label{eq:QBref-generic}
\qWref{\indmode}(\sigma)
=
\sigma^{\ExpQRef}\qWref{\text{pre}}\left(1+\left(\qWref{1}+\qWreflog{1}\,\log(4 \sigma)\right)\sigma+\left(\qWref{2}+\qWreflog{2}\,\log(\sigma) + \qWrefloglog{2}\,\log(4 \sigma)^2\right)\sigma^2+o(\sigma^2)\right), 
\end{equation}
where $\ExpQRef=\ell$ (since $\Brefp$ and $\qW{\indmode}$ are of orders, respectively, $\sigma^{-2s -\ell -1}$ and $\sigma^{2(s+\ell)+1}$),
for some coefficients $\qWref{0\to 2}$ and $\qWreflog{2}$.

 For the specific case $s=-2$,  the expansion for $\qW{\indmode}$ is given in \eqref{eq:Q-generic}--\eqref{:eq:Q-exp} and for $\Bref$ in \eqref{eq:Bref}--\eqref{:eq:Bref-exp}. 
 The resulting expansion for $\qWref{\text{pre}}$ for $s=-2$ is
\begin{subequations}\label{eq:Qref coeffs}
\begin{align}
    %\beta=&\ell
   % \\
    \qWref{\text{pre}} =& 
    \frac{\pi  (-1)^{\ell+1} 2^{\ell +3}}{(\ell -1) \ell  (\ell +1) (\ell +2)},
    \\
    \qWref{1} =& 
  -\frac{4 i a m \left(\ell ^2+\ell -1\right)}{\ell ^2
   (\ell +1)^2}-\frac{4 \kappa }{\ell  (\ell
   +1)}+\frac{2}{\ell -1}-\frac{10}{\ell
   }-\frac{10}{\ell +1}-\frac{2}{\ell
   +2}+\frac{225}{32 (2 \ell -1)}+\frac{347}{16 (2
   \ell +1)}
    \notag\\
    &
    +\frac{225}{32 (2 \ell +3)}-4 \psi^{(0)}(\ell )-1,
    \\
    \qWreflog{1} =& 2,
    \\
    \qWref{2} =&
   a^2 m^2 \left(\frac{15 \left(124 \ell ^2+124 \ell
   -153\right)}{(2 \ell -1)^2 (2 \ell +3)^2}-\frac{4
   \left(29 \ell ^6+87 \ell ^5+98 \ell ^4+51 \ell
   ^3-5 \ell ^2-16 \ell -4\right)}{\ell ^4 (\ell
   +1)^4}\right)
   \notag\\
   &+a^2 \left(\frac{16 \left(2 \ell ^2+2
   \ell -1\right)}{\ell ^2 (\ell +1)^2}-\frac{15
   \left(124 \ell ^2+124 \ell -153\right)}{4 (2 \ell
   -1)^2 (2 \ell +3)^2}-\frac{1}{4}\right)+i a \kappa
    m \left(\frac{8 \left(7 \ell ^2+7 \ell
   +8\right)}{\ell ^2 (\ell +1)^2}-\frac{225}{(2 \ell
   -1) (2 \ell +3)}\right)
   \notag\\
   &
   +\psi ^{(0)}(\ell )
   \left(\frac{16 i a m \left(\ell ^2+\ell
   -1\right)}{\ell ^2 (\ell +1)^2}-\frac{4 \left(143
   \ell ^2+143 \ell -51\right)}{(2 \ell -1) (2 \ell
   +1) (2 \ell +3)}+\frac{16 \left(5 \ell ^3+6 \ell
   ^2-9 \ell -5\right)}{(\ell -1) \ell  (\ell +1)
   (\ell +2)}+\frac{16 \kappa }{\ell  (\ell
   +1)}+4\right)
   \notag\\
   &+i a m \left(\frac{40 \left(232 \ell
   ^2+232 \ell -189\right)}{(2 \ell -1) (2 \ell +1)
   (2 \ell +3)}-\frac{8 \left(145 \ell ^7+507 \ell
   ^6+314 \ell ^5-477 \ell ^4-493 \ell ^3-18 \ell
   ^2+26 \ell -10\right)}{(\ell -1) \ell ^3 (\ell
   +1)^3 (\ell +2)}\right)
   \notag\\
   &+\left(\frac{16 (2 \ell
   +1)}{\ell  (\ell +1)}-\frac{2 \left(143 \ell
   ^2+143 \ell -51\right)}{(2 \ell -1) (2 \ell +1) (2
   \ell +3)}\right) \log (2)
   \notag\\
  &
   -\frac{30928 \ell
   ^4+70856 \ell ^3+13696 \ell ^2-27582 \ell +4977}{4
   (2 \ell -1)^2 (2 \ell +1) (2 \ell
   +3)^2}
  % \notag\\
  %&
  +\frac{30928 \ell ^4+70856 \ell ^3+13696
   \ell ^2-27582 \ell +4977}{6 (2 \ell -1)^2 (2 \ell
   +1) (2 \ell +3)^2}
   \notag\\
  &
   +\kappa  \left(\frac{4 \left(272
   \ell ^2+272 \ell -279\right)}{(2 \ell -1) (2 \ell
   +1) (2 \ell +3)}-\frac{8 \left(17 \ell ^5+42 \ell
   ^4-11 \ell ^3-54 \ell ^2+2 \ell +10\right)}{(\ell
   -1) \ell ^2 (\ell +1)^2 (\ell +2)}\right)
   \notag\\
  &
  +\frac{4
   \left(60 \ell ^7+232 \ell ^6+111 \ell ^5-405 \ell
   ^4-327 \ell ^3+117 \ell ^2+144 \ell
   +176\right)}{(\ell -1)^2 \ell ^2 (\ell +1)^2 (\ell
   +2)^2}
   \notag\\
  &-\frac{4 \left(118 \ell ^7+467 \ell ^6+262
   \ell ^5-772 \ell ^4-728 \ell ^3+149 \ell ^2+324
   \ell +396\right)}{3 (\ell -1)^2 \ell ^2 (\ell
   +1)^2 (\ell +2)^2}+8 \psi ^{(0)}(\ell )^2-\frac{2
   \pi ^2}{3}+\frac{3}{4},
    \\
   \qWreflog{2}=&
   -\frac{8 i a m \left(\ell ^2+\ell -1\right)}{\ell ^2
   (\ell +1)^2}+\frac{4 \left(143 \ell ^2+143 \ell
   -51\right)}{(2 \ell -1) (2 \ell +1) (2 \ell
   +3)}-\frac{24 \left(3 \ell ^3+4 \ell ^2-5 \ell
   -3\right)}{(\ell -1) \ell  (\ell +1) (\ell
   +2)}-\frac{8 \kappa }{\ell  (\ell +1)}-8 \psi
   ^{(0)}(\ell )-2,\\
   \qWrefloglog{2}=& 2.
\end{align}
\end{subequations}
or explicitly for $\ell=2$
\begin{subequations}
\begin{align}
    \qWref{\text{pre}}\Big|_{\ell=2} =& -\frac{4 \pi }{3}
    \,,\\
    \qWref{1}\Big|_{\ell=2} =& 
    -\frac{2 \kappa}{3}-\frac{5}{9} i a m+4 \gamma -\frac{871}{210}
    \,, \\
    \qWreflog{1}\Big|_{\ell=2} =& 2
    \,,\\
    \qWref{2}\Big|_{\ell=2} =& 
    \frac{a^2 \left(1150 m^2+3069\right)}{7938}+a \left(\frac{1}{378} i m (-997+840 \gamma +840 \log (2))-\frac{25 i \kappa  m}{63}\right)
    \notag\\
    &+ \kappa  \left(-\frac{871}{315}+\frac{8 \gamma }{3}+\frac{8 \log (2)}{3}\right)+\frac{2 \pi ^2}{3}-8 \gamma ^2-\frac{72157}{8820}+\frac{1528 \log (2)}{105}+\gamma  \left(\frac{1742}{105}-16 \log (2)\right) 
    \,,\\
    \qWreflog{2}\Big|_{\ell=2} =& \frac{1}{315} (350 i a m+420 \kappa -2520 \gamma +1971)
    \,,\\
   \qWrefloglog{2}\Big|_{\ell=2}=& 2
   \,.
\end{align}
\end{subequations}
We note in Eq.~\eqref{eq:Qref coeffs} the appearance of the first $\log(\sigma)$  at next-to-leading order. 
That this is the case for $s=-2$ follows readily from Eq.~\eqref{:eq:Q-exp} (which, in its turn, follows from Eq.~\eqref{eq:Bref}); for the other integer values of $s$, we have verified it empirically by giving many values to $\ell$ and $m$.

%---------------------------------------------------------------------------------------------------------
%---------------------------------------------------------------------------------------------------------

%---------------------------------------------------------------------------------------------------------
%---------------------------------------------------------------------------------------------------------

\section{Small-frequency expansions of BC modes}\label{sec:Small-w BC modes}

In this section we put together the small-frequency expansions of the various quantities in the previous section in order to obtain small-frequency expansions of  the BC modes $\Disc \Glm$ in Eqs.~\eqref{Eq:GBC}, \eqref{Eq:GBC,rp} and \eqref{eq:dGlm,r->infty}.

\subsection{Small-frequency expansion of BC modes  for  $r\in (r_+,\infty)$}

Putting together the expansions in Eq.~\eqref{eq:Q-generic} for $\qW{\indmode}$ and in Eq.~\eqref{eq:Rin-generic} for both $\Rin{s}(r)$ and $\Rin{s}(r')$ into the expression Eq.~\eqref{Eq:GBC} for $\Disc \Glm$, we  have that, for $r,r'\in (r_+,\infty)$:
\begin{equation}\label{eq:BC-GFmodes-generic}
\Disc \Glm(r,r';\fNIA)=
-2 i \qW{\text{pre}}\,
\sigma^{\ExpQ+2\Expr+1}\left(\g{0}(r,r')+\g{1}(r,r')\sigma+\left(\g{2}(r,r')+\gLog{2}(r,r')\log(\sigma)\right)\sigma^2+o(\sigma^2)\right),
\end{equation}
where
\begin{align}\label{eq:gn}
 \g{0}(r,r')= & R_0(r) R_0(r'),\\
 \g{1}(r,r')= & R_1(r) R_0(r')+ R_0(r) R_1(r')+\qW{1} R_0(r) R_0(r'),\nonumber\\
 \g{2}(r,r')= &R_2(r) R_0(r')+ R_0(r) R_2(r')+R_1(r) R_1(r')+\qW{1} \left(R_1(r)
   R_0(r')+R_0(r) R_1(r')\right)+
  %\nonumber \\&
   \left(\qW{2}+\qWLog{2} \log 2\right)R_0(r) R_0(r'),\nonumber\\
 \gLog{2}(r,r')  =  & \qWLog{2} R_0(r) R_0(r').\nonumber
\end{align}
We note that the leading power in Eq.~\eqref{eq:BC-GFmodes-generic} is $\ExpQ+2\Expr+1=2\ell+2$ (since $\ExpQ=2s+2\ell+1$ and $\Expr=-s$).

\subsection{Small-frequency expansion of BC modes for $r\to r_+$}

Putting together the expansions in  Eq.~\eqref{eq:QBtra-generic} for $\qWtra{\indmode}$ and in Eq.~\eqref{eq:Rin-generic} for $\Rin{s}(r)$ into the expression Eq.~\eqref{Eq:GBC,rp}, we  have that, for $r'\in (r_+,\infty)$ and $r\to r_+$,
the small-$\sigma$ expansion of $\Disc \Glm$ with finite $\sigma r_*$ is, readily: 
\begin{align}\label{eq:BC-GFmodes-Hor}
&
\Disc \Glm(r,r';\fNIA)\sim
\\&
-2 i\qWtra{\text{pre}} 
\sigma^{\ExpQTra+\Expr+1}
\left(\gtra{0}(r')+\gtra{1}(r')\sigma+\left(\gtra{2}(r')+\gtralog{2}(r')\log(\sigma)\right)\sigma^2+o(\sigma^2)\right)\left(\Delta(r)\right)^{-s}e^{im\Omega_H r_*}e^{-{\sigma}r_*},\quad r\to r_+,
\nn
\end{align}

where
\begin{align}
  \gtra{0}(r)= & R_0(r),\\
 \gtra{1}(r)= & R_1(r)+\qWtra{1} R_0(r),\nonumber\\
 \gtra{2}(r)= & R_2(r) +\qWtra{1} R_1(r)
   +
  \left( \qWtra{2}+\qWtralog{2}\log 2\right) R_0(r)\nonumber\\
 \gtralog{2}(r) =  & \qWtralog{2}  R_0(r).\nonumber
\end{align}
The leading power in Eq.~\eqref{eq:BC-GFmodes-Hor} is $\ExpQTra+\Expr+1=2\ell+2+b$
(since $\Expr=-s$ and $\ExpQTra=2\ell+s +1+b$).

\subsection{Small-frequency expansion of BC modes for $r\to \infty$}

Putting together the expansions in Eq.~\eqref{eq:QBref-generic} for $\qWref{\indmode}$ and in Eq.~\eqref{eq:Rin-generic} for $\Rin{s}$ into the expression Eq.~\eqref{eq:dGlm,r->infty}  for $\Disc \Glm$, we  have that, for $r'\in (r_+,\infty)$ and $r\to \infty$:
\begin{align}\label{eq:BC-GFmodes-Inf}
\Disc \Glm(r,r';\fNIA)
\sim
& 2i  \qWref{\text{pre}} \sigma^{\ExpQRef+\Expr+1}\left(\gref{0}(r')+\left(\gref{1}(r')+\greflog{1}(r')\log(\sigma)\right)\sigma+
 \right. \nonumber \\ &\left.
\left(\gref{2}(r')+\greflog{2}(r')\log(\sigma)+\grefloglog{2}(r')\log^2(\sigma)\right)\sigma^2+o(\sigma^2)\right) 
 %\nonumber \\ \times & 
 r^{-2s-1}e^{\sigma r_*}, \quad r\to \infty
\end{align}

where
\begin{align}
  \gref{0}(r) = & R_0(r),\\
 \gref{1}(r)= & R_1(r)+\left(\qWref{1}+\qWreflog{1}\log 4 \right)R_0(r),\nonumber\\
 \greflog{1}(r)= & \qWreflog{1}R_0(r) ,\nonumber\\
 \gref{2}(r)= & R_2(r) +\left(\qWref{1}+\qWreflog{1}\log 4 \right) R_1(r)
   +
   \left(\qWref{2}+\qWrefloglog{2}\log^2 4 \right)R_0(r),\nonumber\\
 \greflog{2}(r)=  & \left( \qWreflog{2} +4\qWrefloglog{2} \log 2 \right) R_0(r)+ \qWreflog{1}  R_1(r),\nonumber\\
  \grefloglog{2}(r) =  & \qWrefloglog{2}  R_0(r).\nonumber
\end{align}
The leading power in Eq.~\eqref{eq:BC-GFmodes-Inf} is $\ExpQRef+\Expr+1=\ell-s+1$ (since $\ExpQRef=\ell$ and $\Expr=-s$).

%---------------------------------------------------------------------------------------------------------
%---------------------------------------------------------------------------------------------------------

\section{Late-time tails}\label{sec:tails}

In this section, we use the small-frequency expansions of the BC modes $\Disc \Glm$ in Sec.~\ref{sec:Small-w BC modes} and of the SWSHs in Sec.~\ref{sec:SWSH} in order to obtain,  via Eq.~\eqref{eq:Disc-m}, the late-time expansions of the $\ell$ \& $m$ modes $\Disc \mathcal{G}_{\ell m}$ of the BC contribution to the GF.
As explained in Sec.~\ref{sec:Radial} generically to all orders and for all spins $s$, the small frequency expansion of the BC integrand in Eq.~\eqref{eq:Disc-m} will contain terms of the type $\sigma^\lambda\log^\mu\sigma$ with $\lambda,\mu\in\mathbb{Z}$.
This is also seen explicitly up to the three leading orders in Sec.~\ref{sec:Small-w BC modes} for $s=-2$, where, further, $\lambda\ge 1$ and 
$\mu\geq 0$.
In order to obtain the late-time asymptotics via Eq.~\eqref{eq:Disc-m}, we thus use a generalized Watson’s lemma (e.g.,~\cite{doi:10.1137/1.9780898719260}):
\begin{equation}\label{eq:Watson}
\int_0^\infty \sigma^\lambda (-\log\sigma)^{\mu}e^{-\sigma t}=t^{-\lambda-1}\log^\mu t \left(\sum_{j=0}^{N}(-1)^j\binom{\mu}{j}\Gamma^{(j)}(\lambda+1)\left(\log t\right)^{-j}+O\left(\log^{-N-1} t\right)\right).
\end{equation}

As argued in Sec.~\ref{sec:deform}, the late-time asymptotics of $\Disc \mathcal{G}_{\ell m}$  should  provide the late-time asymptotics of the $\ell$ and $m$ modes $\mathcal{G}_{\ell m}$ of the GF.
Furthermore, we shall see that the  powers in the leading late-time decays of $\Disc \mathcal{G}_{\ell m}$ (or, equivalently, $\mathcal{G}_{\ell m}$)
  will be independent of $m$ (except for $s>0$ along $\mathcal{H}^+$, in which case the $m=0$ mode decays one power faster) and  will diminish with $\ell$.
  This implies that the leading order of the late-time asymptotics for $G_{m}$, $G_{\ell}$ and $\Gret$  in, respectively, Eqs.~\eqref{eq:Gm}, \eqref{eq:Gl} and \eqref{eq:Gret=lmode} (or \eqref{eq:Gret=mmode}), is given by those of, respectively, $\mathcal{G}_{\ell_0 m}$, $\sum_{m=-\ell}^{\ell}e^{im\phi}\mathcal{G}_{\ell m}$ and $\sum_{m=-|s|}^{|s|}e^{im\phi}\mathcal{G}_{|s|, m}$ (where all three expressions should be multiplied by $2i \Delta^s(r')$, and the $m$-sums should exclude $m=0$ for $s>0$ along $\mathcal{H}^+$).
 On the other hand, for the asymptotics beyond leading order   of $G_{m}$, $G_{\ell}$ and $\Gret$, one would need to combine, in a completely straight-forward manner (which we do not explicitly do), different orders in  $\mathcal{G}_{\ell m}$ for different values of $\ell$.

\begin{figure}[htbp]
\centering
\begin{tikzpicture}[scale=4.5]

    % Define consistent custom colors
    \definecolor{myblue}{RGB}{28, 97, 196}
    \definecolor{myred}{RGB}{200, 30, 30}
    \definecolor{mygreen}{RGB}{30, 140, 30}

    % Define text label styles
    \tikzset{
        blue_label/.style={text=myblue, font=\small, anchor=east},
        green_label/.style={text=mygreen, font=\small, anchor=west},
        red_label/.style={text=myred, font=\small, anchor=west}
    }

    % --- Core Coordinates ---
    \coordinate (i_minus) at (0, 0);
    \coordinate (i_plus)  at (0, 2);
    \coordinate (i_zero)  at (1, 1);
    \coordinate (H_mid)   at (-1, 1);

    % --- Boundaries ---
    % Future Event Horizon H+ and Past Event Horizon H-
    \draw[thick] (i_plus) -- (H_mid) node[midway, left=4pt] {$\mathcal{H}^+$};
    \draw[thick] (H_mid) -- (i_minus) node[midway, left=4pt] {$\mathcal{H}^-$};
    
    % Future Null Infinity J+ and Past Null Infinity J-
    \draw[thick, dotted] (i_plus) -- (i_zero) node[midway, right=4pt] {$\mathcal{J}^+$};
    \draw[thick, dotted] (i_zero) -- (i_minus) node[midway, right=4pt] {$\mathcal{J}^-$};

    % --- Conformal Infinity Points ---
    \fill (i_minus) circle (1pt) node[below=3pt] {$i^-$};
    \fill (i_plus)  circle (1pt) node[above=3pt] {$i^+$};
    \fill (i_zero)  circle (1pt) node[right=3pt] {$i^0$};
    \fill (H_mid)   circle (1pt);

    % --- Colored Trajectories/Arrows ---
    % Blue arrow spanning the upper left side (parallel to H+)
    \draw[-latex, myblue, thick] ([xshift=10.5,yshift=10]H_mid)--([xshift=-.8,yshift=-1.3]i_plus);

    % Red arrow spanning the upper right side (parallel to J+)
    \draw[-latex, myred, thick] ([xshift=-10.5,yshift=10]i_zero)--([xshift=.8,yshift=-1.3]i_plus);
    %\draw[-latex, myred, thick] ([xshift=0.85cm, yshift=-0.85cm]i_plus) -- ([xshift=0.035cm, yshift=-0.035cm]i_plus);

    % Green curved arrow in the interior
    \draw[-latex, mygreen, semithick] 
        (-0.1, 1) to[out=80, in=260] ([xshift=-0.01cm, yshift=-0.06cm]i_plus);

    % --- Text Label Columns ---

    % 1. Blue Labels (Left column, stacked vertically)
    \begin{scope}[xshift=-0.3cm, yshift=2.1cm]
        \node[blue_label] at (0, 0)      {$e^{im \Omega_H v} v^{-2 \ell -3 -b}$};
        \node[blue_label] at (0, -0.09)  {$e^{im \Omega_H v} v^{-2\ell -4 -b}$};
        \node[blue_label] at (0, -0.18)  {$e^{im \Omega_H v} v^{-2 \ell -5 -b}$};
        \node[blue_label] at (0, -0.27)  {$e^{im \Omega_H v} v^{-2 \ell -5 -b} \ln v$};
    \end{scope}

    % 2. Green Labels (Center column)
    \begin{scope}[xshift=0.0cm, yshift=1.42cm]
        \node[green_label] at (0, 0)      {$t^{-2\ell-3}$};
        \node[green_label] at (0, -0.11)  {$t^{-2\ell-4}$};
        \node[green_label] at (0, -0.22)  {$t^{-2\ell-5}$};
        \node[green_label] at (0, -0.34)  {$t^{-2\ell-5} \ln t$};
    \end{scope}

    % 3. Red Labels (Right column)
    \begin{scope}[xshift=0.44cm, yshift=2.0cm]
        \node[red_label] at (0, 0)      {$u^{-\ell+s-2}$};
        \node[red_label] at (0, -0.09)  {$u^{-\ell+s-3}$};
        \node[red_label] at (0, -0.18)  {$u^{-\ell+s-3} \ln u$};
        \node[red_label] at (0, -0.28)  {$u^{-\ell+s-4}$};
        \node[red_label] at (0, -0.37)  {$u^{-\ell+s-4} \ln^2 u$};
    \end{scope}

\end{tikzpicture}
\caption{\textbf{Time functionalities:} Penrose diagram of the exterior of  Kerr spacetime, together with a summary of the  time functionalities  appearing in the three leading orders of the late-time tails of the GF modes $\mathcal{G}_{\ell m}$ (in the subextremal black hole case). The tails are for the approach to future timelike infinity $i^+$ : (i) along the future horizon $\mathcal{H}^+$ (in blue; see Eq.~\eqref{eq:tail H^+}), (ii) at a finite radius away from the horizon (in green; see Eq.~\eqref{eq:tail,finite r}); and (iii) along future null infinity $\scri^+$ (in red; see Eq.~\eqref{eq:tail,inf}). It is $b=0$ except for $s>0$ and $m=0$, in which case it is $b=1$. }
\label{fig:Penrose}
\end{figure}
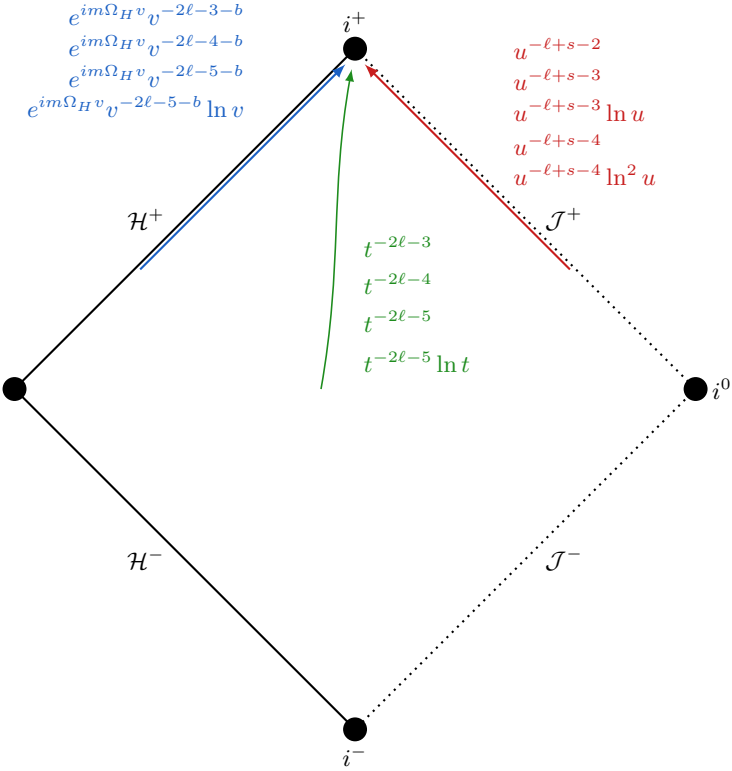

 We obtain the late-time asymptotics up to the first three orders separately in the cases of (A) finite radius away from the horizon; (B) along the future horizon $\mathcal{H}^+$; and (C) along future null infinity $\scri^+$.
 In Fig.~\ref{fig:Penrose} we show a Penrose diagram of the exterior of Kerr spacetime, with a summary  of the main time functionalities of these three late-time tails, which we next derive.

\subsection{Late-time tail at finite radius (away from horizon)}

We put together the small-$|a\omega|$ expansion in Eq.~\eqref{eq:SWSH,small-c} of the SWSH and the small-$\sigma$  in Eq.~\eqref{eq:BC-GFmodes-generic} of the BC discontinuity $\Disc \Glm$ into the expression in Eq.~\eqref{eq:Disc-m} for $\Disc \mathcal{G}_{\ell m}$, integrate and then take the large-$t$ asymptotics. The result is, for generic integer spin $s$ and finite $r,r'>r_+$:
\begin{equation}\label{eq:tail,finite r}
\Disc \mathcal{G}_{\ell m}(r,r',t)=
2\qW{\text{pre}}\,
t^{-2\ell-3}
%t^{-\ExpQ-2\Expr-2}
\left(\G{0}+\frac{\G{1}}{t}+\frac{\G{2}+\GLog{2}\log(t)}{t^2}+o\left(t^{-2}\right)\right),
\end{equation}
%if $\Re(\alpha)>-2$, 
with 
\begin{align}\label{eq:coeffs,late-t}
\G{0} =&  \Gamma (\ExpQ +2) \St{0}(\theta ) \St{0}(\theta') \g{0},\\
\G{1} =&  \Gamma (\ExpQ +3) \left(i\left( \St{1}(\theta') \St{0}(\theta )-\St{1}(\theta ) \St{0}(\theta')\right) \g{0}+\St{0}(\theta ) \St{0}(\theta') \g{1}\right),\nonumber\\
\G{2} =&  \Gamma (\ExpQ +4) \Big( \left(\St{1}(\theta ) \St{1}(\theta')-\St{2}(\theta') \St{0}(\theta ) -\St{2}(\theta )
   \St{0}(\theta')\right) \g{0}+i \left(\St{1}(\theta') \St{0}(\theta ) -
   \St{1}(\theta ) \St{0}(\theta')
   \right)\g{1}+
        \nonumber \\&
   \St{0}(\theta ) \St{0}(\theta')\left( \g{2}+\psi ^{(0)}(\ExpQ +4) \gLog{2}\right)\Big) ,\nonumber\\
\GLog{2} =&  -\Gamma (\ExpQ +4) \St{0}(\theta ) \St{0}(\theta') \gLog{2},\nonumber
\end{align}
where all $\g{0\to 2}, \gLog{2}$ are functions of both $r$ and $r'$,
and all $\G{0\to 2},\GLog{2}$ are functions  of $r$, $r'$, $\theta$ and $\theta'$.

The power law decay $t^{-2\ell-3}$ (resulting from $t^{-\ExpQ-2\Expr-2}$ with $\ExpQ=2s+2\ell+1$ and $\Expr=-s$)
of the leading order in Eq.~\eqref{eq:tail,finite r} was already obtained in Refs.~\cite{1999PhRvD..61b4033H,hod2000mode,PhysRevLett.84.10} (where it was obtained only for $r,r'\gg M$, whereas our expression is valid for generic $r,r'\in (r_+,\infty)$)\footnote{The power in the late-time tail in the final expressions in Refs.~\cite{1999PhRvD..61b4033H,hod2000mode,PhysRevLett.84.10} were in fact for a given {\it spherical} (not spheroidal) mode $\ell$ of the field point of the GF and for a fixed {\it spherical} (not spheroidal) mode of the source. However, if one implemented ${}_s\Slm(\theta)\to {}_sY_{\indmode}(\theta)$ for $a\omega\to 0$ (as done here in Eq.~\eqref{eq:SWSH,small-c}) in the 
intermediate expressions therein,
%in terms of ${}_s\Slm$, 
%it would readily follow  that then 
the same power as in Eq.~\eqref{eq:tail,finite r} here would readily  follow.
See also Refs.~\cite{PhysRevD.61.024026,1999PhRvD..61b4001O} for the late-time-$t$ tail when fixing spherical  modes.
}.
The leading power 
%$-\ExpQ-2\Expr-2=$
``$-2\ell-3$" is independent of $m$ and it decreases with $\ell$. 
This means that, at finite $r>r_+$, and generally, the leading tail of $G_m$ will be, via \eqref{eq:Gm}, $t^{-2\ell_0-3}$, that of $G_{\ell}$ will be, via \eqref{eq:Gl}, $t^{-2\ell-3}$, and that of 
$\Gret$ will be, via either \eqref{eq:Gret=lmode} or \eqref{eq:Gret=mmode}, $t^{-2|s|-3}$.
The first logarithm in $\mathcal{G}_{\ell m}$ in Eq.~\eqref{eq:tail,finite r} generically appears  at third leading order: $t^{-2\ell-5}\log t$ (the same order for the  first logarithm along finite $r'>r_+$ was derived in Schwarzschild in Ref.~\cite{Casals:Ottewill:2015}).

In the most physically relevant case of the dominant mode $\ell=2$ for  $s=-2$, we have obtained the explicit coefficients in the late-time asymptotics in Eq.~\eqref{eq:tail,finite r} for $\Disc \mathcal{G}_{\ell m}$ as resulting from an 8PN expansion. We compare these late-time asymptotics for $\Disc \mathcal{G}_{\ell m}$ against numerics in Sec.~\ref{sec:numerics}. As explained in Sec.~\ref{sec:Small_freq}, an expansion to 8PN order, means that it is up to $\eta^{16}$ beyond leading order. Since the late-time expansion in Eq.~\eqref{eq:tail,finite r} starts at $t^{-2\ell-3}=t^{-7}$ for $\ell=2$, this means that an 8PN expansion of $\Disc \mathcal{G}_{\ell m}$ for $\ell=2$ is up to $t^{-23}$, not just up to $t^{-9}$ as per Eq.~\eqref{eq:tail,finite r}. Here we give the coefficients for $\ell=2$ up to  2.5PN@2PS for readability. (Note in the context of our PN expansion we allow $t$ to carry scaling of $\eta^{-3}$ which we absorb into the $G_i$ functions)
\begin{subequations}
\label{[eq:Tail_G20_PN]}
\begin{align}
    Q_\mathrm{pre} =& -\frac{384 \pi  }{5}\eta ^5 r^4 r'^4 {}_{-2}Y_{2m}(\theta ,0) {}_{-2}Y_{2m}(\theta' ,0)
    \,,\\
    G_0 =&1+ \eta^2 \frac{4+i a m}{r} + \eta^4 \left[\frac{-\frac{1}{2} a^2 \left(m^2-4\right)+3 i a m+4}{r^2} -\frac{(a m-4 i)^2}{2 r r'}\right]
    +\mathcal{O}(\eta^6) + (r \leftrightarrow r')
    \,, \\
    G_1=& \eta \frac{14 r}{3}-\eta^3 \left[\frac{7 i r (a m-4 i)}{3 r'} -\frac{i \sqrt{7} a \sqrt{9-m^2}}{9} 
    \mathcal{Y}_{m}
    %\left(\frac{{}_{-2}Y_{3m}(\theta ,0)}{{}_{-2}Y_{2m}(\theta ,0)} + \frac{{}_{-2}Y_{3m}(\theta' ,0)}{{}_{-2}Y_{2m}(\theta' ,0)}  \right)
    -\frac{56}{9} i a m-\frac{311}{60} \right]
    \notag\\
    &+ \eta^5 \left[\frac{7 r \left(-\left(a^2 \left(m^2-4\right)\right)+6 i a +8\right)}{3 r'^2}
    + \frac{a^2 \left(840-910 m^2\right)+3943 i a m-48}{r} \right.
    \notag\\
    & -\left. \frac{20 \sqrt{7} a \sqrt{9-m^2} (a m-4 i)}{r} 
    \mathcal{Y}_{m}
    %\left(\frac{{}_{-2}Y_{3m}(\theta ,0)}{{}_{-2}Y_{2m}(\theta ,0)} + \frac{{}_{-2}Y_{3m}(\theta' ,0)}{{}_{-2}Y_{2m}(\theta' ,0)}  \right) 
    \right]
    +\mathcal{O}(\eta^6) + (r \leftrightarrow r')
    \,,\\
    G_2 =& \eta^2 \left[\frac{44 r^2}{3} + \frac{112 r r'}{9} \right] 
    + \eta^4 \left[  -\frac{44 i r^2 (a m-4 i)}{3 r' } + \frac{832 r}{45}-\frac{1556}{27} i a m r \right.
    \notag\\
    &+\left.40 \sqrt{7} a \sqrt{9-m^2}  
    \mathcal{Y}_{m}
    %\left(\frac{{}_{-2}Y_{3m}(\theta ,0)}{{}_{-2}Y_{2m}(\theta ,0)} + \frac{{}_{-2}Y_{3m}(\theta' ,0)}{{}_{-2}Y_{2m}(\theta' ,0)}  \right)
    \right]
    +\mathcal{O}(\eta^6) + (r \leftrightarrow r')
    \,,\\
    \GLog{2} =& \mathcal{O}(\eta^6)
    \,,
\end{align}
\end{subequations}
where
\begin{equation}
\mathcal{Y}_{m}(\theta,\theta')\equiv\left(\frac{{}_{-2}Y_{3m}(\theta ,0)}{{}_{-2}Y_{2m}(\theta ,0)} + \frac{{}_{-2}Y_{3m}(\theta' ,0)}{{}_{-2}Y_{2m}(\theta' ,0)}  \right).
\end{equation}
In \eqref{[eq:Tail_G20_PN]}, ``$(r \leftrightarrow r')$" means that we add all the terms preceding it in the expression, with $r$ and $r'$ swapped.

We attach two Mathematica notebooks. One stores  generic-$\ell$ expressions for various quantities used in this paper: the BC strength factor $Q$, the scattering amplitudes $B^{\text{inc/ref/tra}}$, the BC modes $\delta \mathcal{G}_{\ell m}$, the radial solution $\Rin{-2}$ to low PN, or PS order. The other notebook enables the user to calculate these quantities, and in particular the BC modes $\delta \mathcal{G}_{\ell m}$ for specific $s$ and $\ell$ to higher orders. The second notebook requires the \texttt{PN} branch of the BHPT's \cite{BHPToolkit} \texttt{Teukolsky} package which can be obtained from the respective github and will soon be merged into the main one. We have found 8PN to give accurate results down to $r=r'=3$ (cf. Fig.~\ref{fig:PN_tail_convergence_r3}). If one requires very high PN orders one could use a cluster to run our notebook, where we expect its limit of feasibility to be somewhere around 15PN based on previous experiences.

%---------------------------------------------------------------------------------------------------------

\subsection{Late-time tail along $\hor^+$}

We put together the small $|a\omega|$ expansion in Eq.~\eqref{eq:SWSH,small-c} of the SWSH and the small-$\sigma$  in Eq.~\eqref{eq:BC-GFmodes-Hor} of the BC discontinuity $\Disc \Glm$ for $r\to r_+$ into the expression Eq.~\eqref{eq:Disc-m} for $\Disc \mathcal{G}_{\ell m}$.
The $\sigma$-integrand contains a factor $e^{im\Omega_H r_*}e^{-\sigma r_*}$ from Eq.~\eqref{eq:BC-GFmodes-Hor} and a factor $e^{-\sigma t}$ from Eq.~\eqref{eq:Disc-m}.
On the horizon, $\phi$ is not a regular  coordinate whereas $\phi_+$ is regular (e.g., Eq.~(250) in Ch.6~\cite{Chandrasekhar}). Thus, in the late-time asymptotics here we also include the factor $e^{im\phi}=e^{im\phi_+}e^{im\Omega_H t}$ from \eqref{eq:Gl} or \eqref{eq:Gret=mmode}. Combining all these exponential factors together yields $e^{im\phi}e^{im\Omega_H r_*}e^{-\sigma r_*}e^{-\sigma t}=e^{im\phi_+}e^{im\Omega_H v}e^{-\sigma v}$. 
We then integrate, carry out large-$v$ asymptotics of the integral and obtain, for generic integer spin $s$ and finite $r'>r_+$:
\begin{equation}\label{eq:tail H^+}
\left.
\left(\Delta(r)\right)^{s}
e^{im\phi}\Disc \mathcal{G}_{\ell m}(r,r',t)
\right|_{x\in\mathcal{H}^+}=
2\qWtra{\text{pre}} 
v^{-2\ell-3-b}
\left(\Grp{0}+\frac{\Grp{1}}{v}+\frac{\Grp{2}+\GrpLog{2}\log(v)}{v^2}+o\left(v^{-2}\right)\right)
e^{im\phi_+}e^{im\Omega_H v},
%\quad r\to r_+,
\end{equation}
where the coefficients
$\Grp{0\to 2}$ and $\GrpLog{2}$ are given by, respectively, $G_{0\to 2}$ and $\GLog{2}$ in Eq.~\eqref{eq:coeffs,late-t} with $\ExpQ$, $g_{0\to 2}(r,r')$ and $\gLog{2}(r,r')$  replaced by, respectively, $\ExpQTra$, $\gtra{0\to 2}(r')$ and $\gtralog{2}(r')$. Explicitly, 
\begin{align}
\Grp{0} =&  \Gamma (\ExpQTra +2) \St{0}(\theta ) \St{0}(\theta') \gtra{0},\\
\Grp{1} =&  \Gamma (\ExpQTra +3) (i \left(\St{1}(\theta') \St{0}(\theta ) -\St{1}(\theta ) \St{0}(\theta')\right) \gtra{0}+\St{0}(\theta ) \St{0}(\theta') \gtra{1}),\nonumber\\
\Grp{2} =&  \Gamma (\ExpQTra +4) \Big( \left(\St{1}(\theta ) \St{1}(\theta') -\St{2}(\theta') \St{0}(\theta )-\St{2}(\theta )
   \St{0}(\theta')\right) \gtra{0}+i \left(\St{1}(\theta') \St{0}(\theta )-
    \St{1}(\theta ) \St{0}(\theta')
 \right)  \gtra{1}+
        \nonumber \\&
   \St{0}(\theta ) \St{0}(\theta')\left( \gtra{2}+\psi ^{(0)}(\ExpQTra +4) \gtralog{2}\right)\Big) ,\nonumber\\
\GrpLog{2} =&  -\Gamma (\ExpQTra +4) \St{0}(\theta ) \St{0}(\theta') \gtralog{2},\nonumber
\end{align}
where all $\gtra{0\to 2}, \gtralog{2}$ are understood to be functions of $r'$ and all $\Grp{0\to 2},\GrpLog{2}$ are functions  of $r'$, $\theta$ and $\theta'$.

The power law decay is $v^{-2\ell-3-b}$ (resulting from $v^{-\ExpQTra-\Expr-2}$ with $\Expr=-s$ and $\ExpQTra=2\ell+s +1+b$), with $b=0$ except for $s>0$ and $m=0$, in which case it is $b=1$)
of the leading order in Eq.~\eqref{eq:tail H^+} was already obtained in Eq.~(21) in Ref.~\cite{hod2000mode}
(see Refs.~\cite{PhysRevD.61.024026,1999PhRvD..61b4001O} for the corresponding tail when fixing spherical --instead of spheroidal-- modes).
The leading power 
``$-2\ell-3-b$" is independent of $m$ (except for $s>0$, in which case the $m=0$ mode decays one power faster) and it decreases with $\ell$. 
This means that, along $\mathcal{H}^+$, the leading tail of $G_m$ will be, via \eqref{eq:Gm}, $v^{-2\ell_0-3}$, that of $G_{\ell}$ will be, via \eqref{eq:Gl}, $v^{-2\ell-3}$, and that of 
$\Gret$ will be, via either \eqref{eq:Gret=lmode} or \eqref{eq:Gret=mmode}, $v^{-2|s|-3}$.
The first logarithm in $\mathcal{G}_{\ell m}$  in Eq.~\eqref{eq:tail H^+} generically appears  at third leading order: $v^{-2\ell-5}\log v$
(the same order for the  first logarithm along $\mathcal{H}^+$ was derived in Schwarzschild in Ref.~\cite{Casals:Ottewill:2015} -- see footnote \ref{ftn:Schw,H}).

Note that, apart from the power law and logarithmic behaviour in $v$, Eq.~\eqref{eq:tail H^+} also presents the factor $e^{im\Omega_H v}$, which oscillates with $v$ for nonaxisymmetric perturbations (i.e., $m\neq 0$), yielding a `wagging' of the tail.
This oscillatory power law behaviour along $\hor^+$ translates (when combining this result for $s=-2$ with the analogous one for $s=+2$) into an oscillatory behaviour in the way the quadratic curvature scalar blows up when approaching the Cauchy horizon (see Refs.~\cite{1999PhRvL..83.5423O,1999PhRvD..61b4001O}).

%---------------------------------------------------------------------------------------------------------

\subsection{Late-time tail along $\scri^+$}

We  put together the small-$|a\omega|$ expansion in Eq.~\eqref{eq:SWSH,small-c} of the SWSH and the small-$\sigma$ expansion in Eq.~\eqref{eq:BC-GFmodes-Inf} of the BC discontinuity $\Disc \Glm$ for $r\to \infty$ into the expression Eq.~\eqref{eq:Disc-m} for $\Disc \mathcal{G}_{\ell m}$.
The $\sigma$-integrand contains a factor $e^{\sigma r_*}$ from Eq.~\eqref{eq:BC-GFmodes-Inf} and a factor $e^{-\sigma t}$ from Eq.~\eqref{eq:Disc-m}, which we combine into $e^{-\sigma u}$.
We then integrate, carry out large-$u$ asymptotics of the integral and obtain, for generic integer spin $s$ and finite $r'>r_+$:
\begin{align}\label{eq:tail,inf}
&
\left.r^{2s+1}\Disc \mathcal{G}_{\ell m}(r,r',t)\right|_{x\in \scri^+}=
\\ &
-2\qWref{\text{pre}} 
u^{-\ell+s-2}
\left(\Ginf{0}+\frac{\Ginf{1}+\GinfLog{1}\log(u)}{u}+\frac{\Ginf{2}+\GinfLog{2}\log(u)+\GinfLogLog{2}\log^2(u)}{u^2}+o\left(u^{-2}\right)\right),
%\quad r\to \infty,
\nonumber
\end{align}
where
\begin{align}
\Ginf{0} & =  \Gamma (\ExpQRef +2) \St{0}(\theta ) \St{0}(\theta')   \gref{0},\\
\Ginf{1} & =  \Gamma (\ExpQRef +3) \left(i \left(\St{1}(\theta') \St{0}(\theta ) -\St{1}(\theta ) \St{0}(\theta') \right)  \gref{0}+\St{0}(\theta ) \St{0}(\theta') \left(  \gref{1}+\psi^{(0)}\left(\beta+3\right)\right)\right),\nonumber\\
\GinfLog{1} & =  -\Gamma (\ExpQRef +3) \St{0}(\theta ) \St{0}(\theta') \greflog{1},\nonumber\\
\Ginf{2} & =  
\Gamma (\beta +3) \left(\St{1}(\theta ) ((\beta +3) \gref{0} \St{1}(\theta')-i \St{0}(\theta') ((\beta +3) \gref{1}+(\beta +3) \greflog{1} \psi ^{(0)}(\beta
   +3)+\greflog{1}))
   \right.\nonumber\\&\left.
  - \beta  \gref{0} \St{2}(\theta') \St{0}(\theta )-\beta  \gref{0} \St{2}(\theta ) \St{0}(\theta')-3 \gref{0} \St{2}(\theta') \St{0}(\theta )-3
   \gref{0} \St{2}(\theta ) \St{0}(\theta')+
      \right.\nonumber\\&\left.
   i \St{1}(\theta') \St{0}(\theta ) ((\beta +3) \gref{1}+(\beta +3) \greflog{1} \psi ^{(0)}(\beta +3)+\greflog{1})+\beta  \gref{2}
   \St{0}(\theta ) \St{0}(\theta')+
      \right.\nonumber\\&\left.
   3 \gref{2} \St{0}(\theta ) \St{0}(\theta')+
   3 \greflog{2} \psi ^{(0)}(\beta +3) \St{0}(\theta ) \St{0}(\theta')+\beta  \greflog{2} \psi ^{(0)}(\beta +3)
   \St{0}(\theta ) \St{0}(\theta')+
      \right.\nonumber\\&\left.
   \greflog{2} \St{0}(\theta ) \St{0}(\theta')+3 \grefloglog{2} \psi ^{(0)}(\beta +4)^2 \St{0}(\theta ) \St{0}(\theta')+\beta  \grefloglog{2} \psi ^{(0)}(\beta
   +4)^2 \St{0}(\theta ) \St{0}(\theta')+
      \right.\nonumber\\&\left.
   3 \grefloglog{2} \psi ^{(1)}(\beta +4) \St{0}(\theta ) \St{0}(\theta')+\beta  \grefloglog{2} \psi ^{(1)}(\beta +4) \St{0}(\theta ) \St{0}(\theta')\right)
,\nonumber\\
\GinfLog{2} & = 
-\frac{1}{\beta +3}\left(i \Gamma (\beta +4) ((\beta +3) \greflog{1} \St{1}(\theta')\St{0}(\theta )-\St{0}(\theta') ((\beta +3) \greflog{1} \St{1}(\theta )+
\right.\nonumber\\&\left.
i \St{0}(\theta ) ((\beta +3) \greflog{2}+
2
   (\beta +3) \grefloglog{2} \psi ^{(0)}(\beta +3)+2 \grefloglog{2})))\right)
   ,\nonumber\\
   \GinfLogLog{2} & =  \Gamma(\ExpQRef+4)\grefloglog{2}\St{0}(\theta )\St{0}(\theta'),
   \nonumber
\end{align}
where all $\gref{0\to 2}, \greflog{1\to 2}, \grefloglog{2}$ are understood to be functions of $r'$  and all $\Ginf{0\to 2},\GinfLog{2}, \GinfLogLog{2}$ are functions  of $r'$, $\theta$ and $\theta'$.

The power law decay $u^{-\ell+s-2}$ (resulting from $u^{-\ExpQRef-\Expr-2}$ with $\ExpQRef=\ell$ and $\Expr=-s$)
of the leading order in Eq.~\eqref{eq:tail,inf} agrees with Refs.~\cite{1999PhRvD..61b4033H,hod2000mode,PhysRevLett.84.10} (where it was obtained only for $r'\gg M$, whereas we have obtained it for generic $r'\in (r_+,\infty)$) (see Ref.~\cite{PhysRevD.61.024026} for the corresponding tail when fixing spherical --instead of spheroidal-- modes).
The leading power 
%$-\ExpQRef-\Expr-2=$
``$-\ell+s-2$" is independent of $m$ and it decreases with $\ell$. 
This means that, at $r\to \infty$, the leading tail of $G_m$ will be, via \eqref{eq:Gm}, $u^{-\ell_0+s-2}$, that of $G_{\ell}$ will be, via \eqref{eq:Gl}, $u^{-\ell+s-2}$, and that of 
$\Gret$ will be, via either \eqref{eq:Gret=lmode} or \eqref{eq:Gret=mmode}, $u^{s-|s|-2}$.

As opposed to the tails in Eqs.~\eqref{eq:tail,finite r} and \eqref{eq:tail H^+} at finite radius, where the first logarithm appeared at third leading order, the first logarithm in the tail for  $\mathcal{G}_{\ell m}$ in Eq.~\eqref{eq:tail,inf} at infinity appears already at {\it second} leading order: $u^{-\ell+s-3}\log u$
(the same order for the  first logarithm along $\scri^+$ was derived in Schwarzschild in Ref.~\cite{Casals:Ottewill:2015} -- see footnote \ref{ftn:Schw,scri}).
In Eq.~\eqref{eq:tail,inf},  a quadratic logarithmic term (as well as another linear one) also appears at third leading order: $u^{-\ell+s-4}\log^2 u$.

%---------------------------------------------------------------------------------------------------------

\subsection{Numerical Validation}\label{sec:numerics}

We now wish to validate our analytic expressions for the tail at finite radius by numerically computing the GF modes for representative sample values. Our claim thus far is that the BC contribution 
$\Disc \mathcal{G}_{\ell m}$ (given in  \eqref{eq:Disc-m}) to the GF modes $\mathcal{G}_{\ell m}$ (given in \eqref{eq:Glm_real_axis})  should represent the late time behaviour of $\mathcal{G}_{\ell m}$ themselves,
specifically 
\begin{align}
    \mathcal{G}_{\ell m}(r,r';\theta,\theta';t)= \, \Disc \mathcal{G}_{\ell m}(r,r';\theta,\theta';t)+O\left(e^{-\Lambda t}\right),\quad t\rightarrow\infty,
\end{align}
for some $\Lambda>0$.
We will do the check
for the choice of $\{s,\ell,m,a,r,\theta\}=\{-2,2,0\, \text{and}\, 2,\tfrac{9}{10},10,\tfrac{\pi}{2}\}$, and source point $r'=r$, $\theta'=\theta$.
On the analytical side, we will use the late-time tail in Eq.~\eqref{eq:tail,finite r} with the coefficients from the
explicit 8PN expansion given in Eq.~\eqref{[eq:Tail_G20_PN]}. On the numerical side, we will
evaluate the real frequency inverse Fourier transform (IFT) in \eqref{eq:Glm_real_axis}, which we repeat for clarity 
\begin{equation*} 
\mathcal{G}_{\ell m}\equiv  \int_{\mathbb{R}}d\omega\, e^{-i\omega t}\Glm(r,r';\omega){}_s\Slm(\theta){}_s\Slm
%^{*}
(\theta') .
%\label{eq:Glm_real_axis}
\end{equation*}
We use the \texttt{Teukolsky} package of the BHPT to numerically generate both the radial Green function $\Glm(r,r';\omega)$ and the SWSHs for real $\omega$ and then numerically integrate as per \eqref{eq:Glm_real_axis}. The results of our comparison 
between the analytical BC modes modes $\Disc \mathcal{G}_{\ell m}$
and the numerical (real-frequency integration) GF modes $\mathcal{G}_{\ell m}$ 
are found in Figs.~\ref{fig:G20}--\ref{Fig:residuals m=2,with/without cc}.
%and \ref{fig:residuals}. 
While the data generation using the BHPT is straightforward, the specifics of ensuring the convergence of the Fourier integral are worth discussing. 

Previous works \cite{CKO} by some of us numerically computed the IFT for the $\ell=0,1$ modes of the scalar ($s=0$) field, where the falloff with $\omega$ of the radial Green function was sufficiently fast to ensure convergence. However, as is discussed at length in the context of the Schwarzschild gravitational GF \cite{aruquipa2026greenfunctionsreggewheelerteukolsky}, the falloff with $\omega$ of the Green function of the (nonzero-spin) Teukolsky equation can be very slow, roughly $\sim1/\omega$. This is problematic, since the computation of the radial Green function with increasing $\omega$ becomes prohibitively slow and numerically unstable. Moreover, the numerical integrations of such highly oscillatory integrals (note the factor $e^{-i\omega t}$) also becomes delicate. In the specific case of $r=r'$ in Schwarzschild, Ref.~\cite{aruquipa2026greenfunctionsreggewheelerteukolsky} remedied this issue by applying a trick through the  analytical understanding of the large-$\omega$ behavior of the radial Green function (for the Teukolsky and Regge-Wheeler  equations, this trick improved the large-$\omega$ asymptotics of the non-oscillatory factor in the integrand from $1/\omega$ to, respectively, $1/\omega^2$ and exponential decay).

\begin{figure}[htbp]
    \centering
    \includegraphics[width=.7\linewidth]{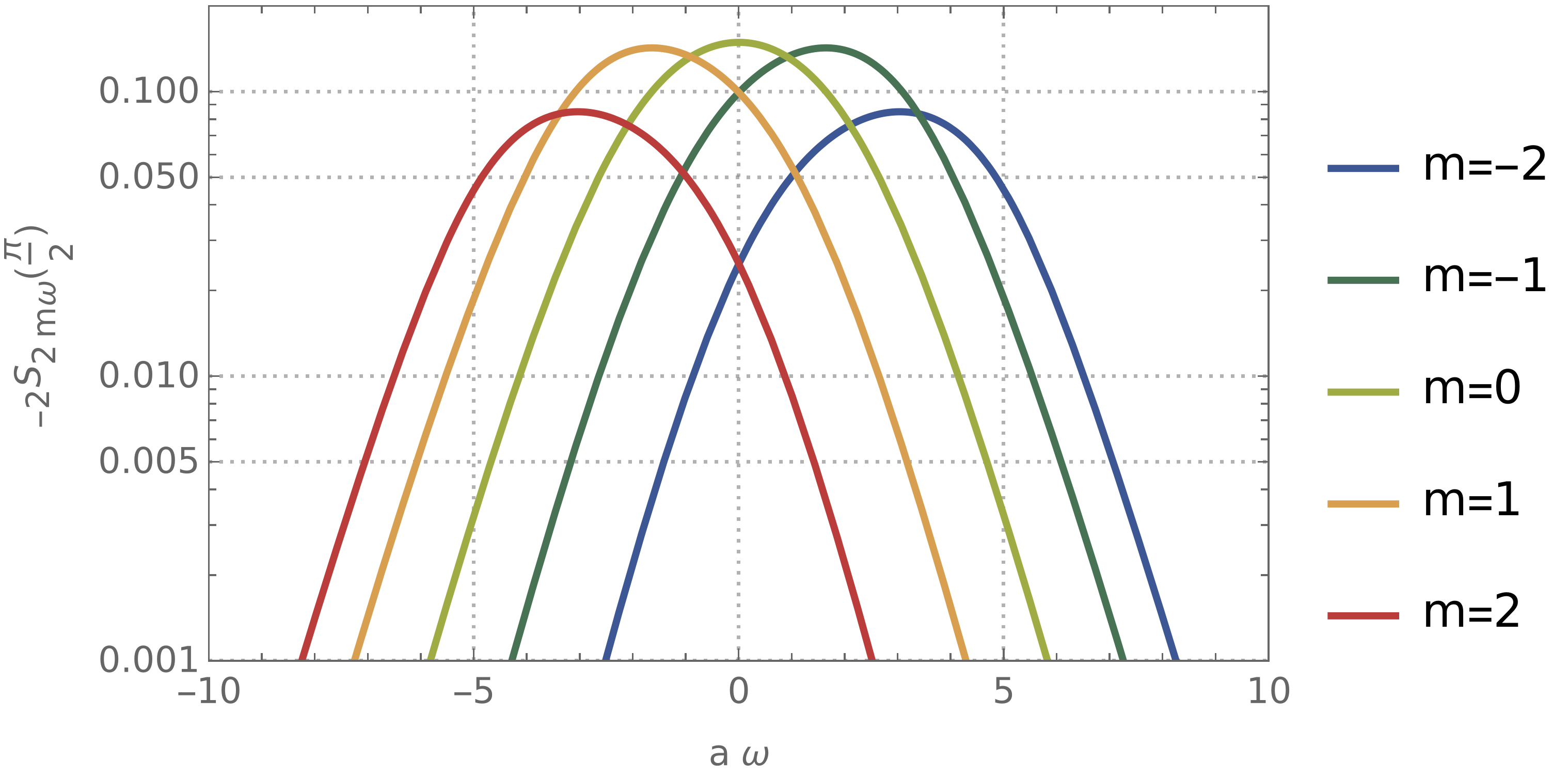}
    \caption{\textbf{SWSH supression:} SWSH ${}_{s}S_{\ell m\omega}(\theta)$ for $s=-2$, $\ell=2$ at $\theta=\pi/2$ over spheroidicity $a \omega$ for varying values of $m$. The plot shows the suppression in the frequency domain for higher $|a\omega|$.}
    \label{fig:SWSH falloff}
\end{figure}

\begin{figure}[h]
    \centering
        \begin{subfigure}{0.48\textwidth}
        \centering
        \includegraphics[width=\linewidth]{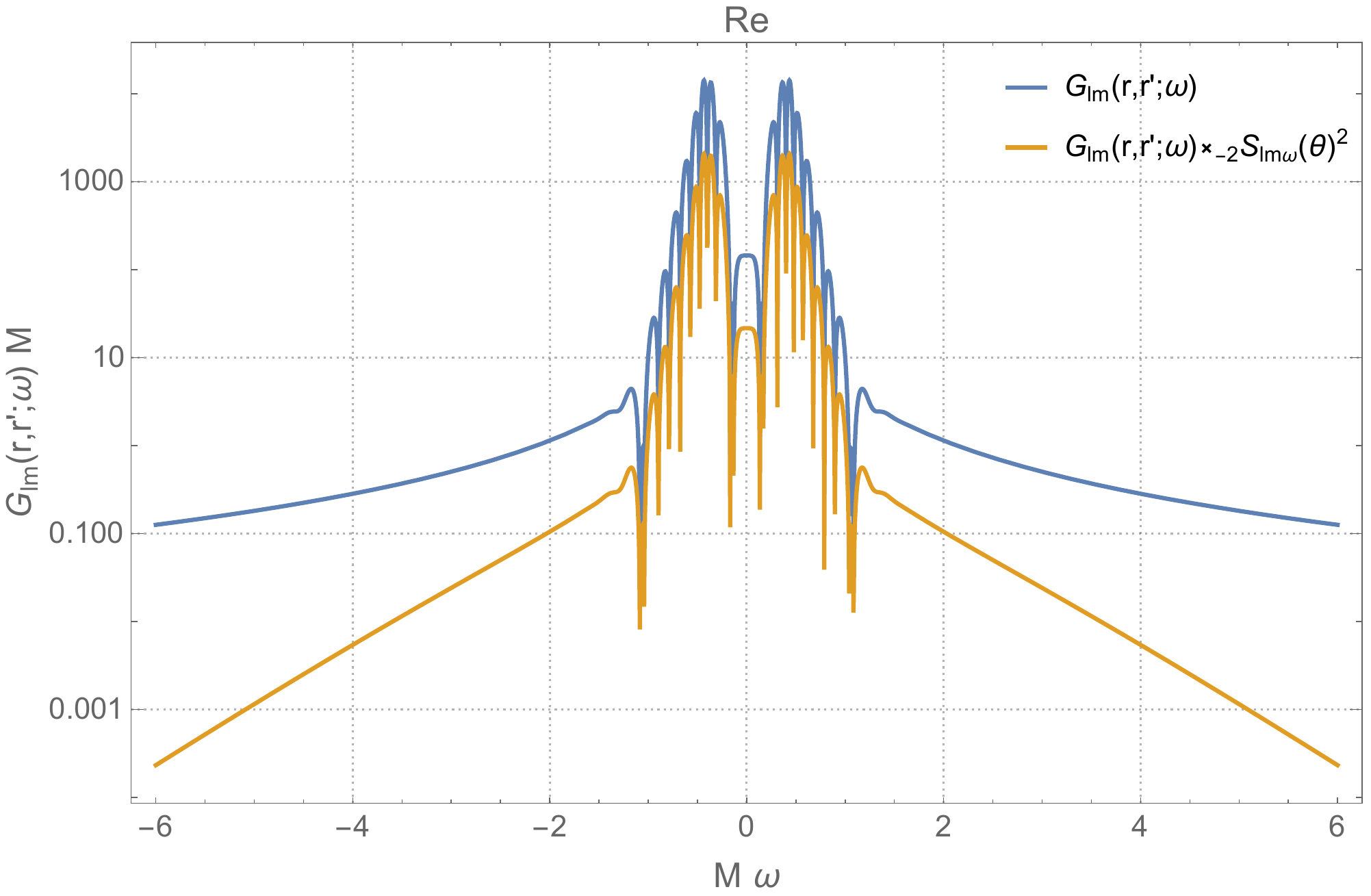}
        \label{fig:left}
    \end{subfigure}
    \hfill
    \begin{subfigure}{0.48\textwidth}
        \centering
        \includegraphics[width=\linewidth]{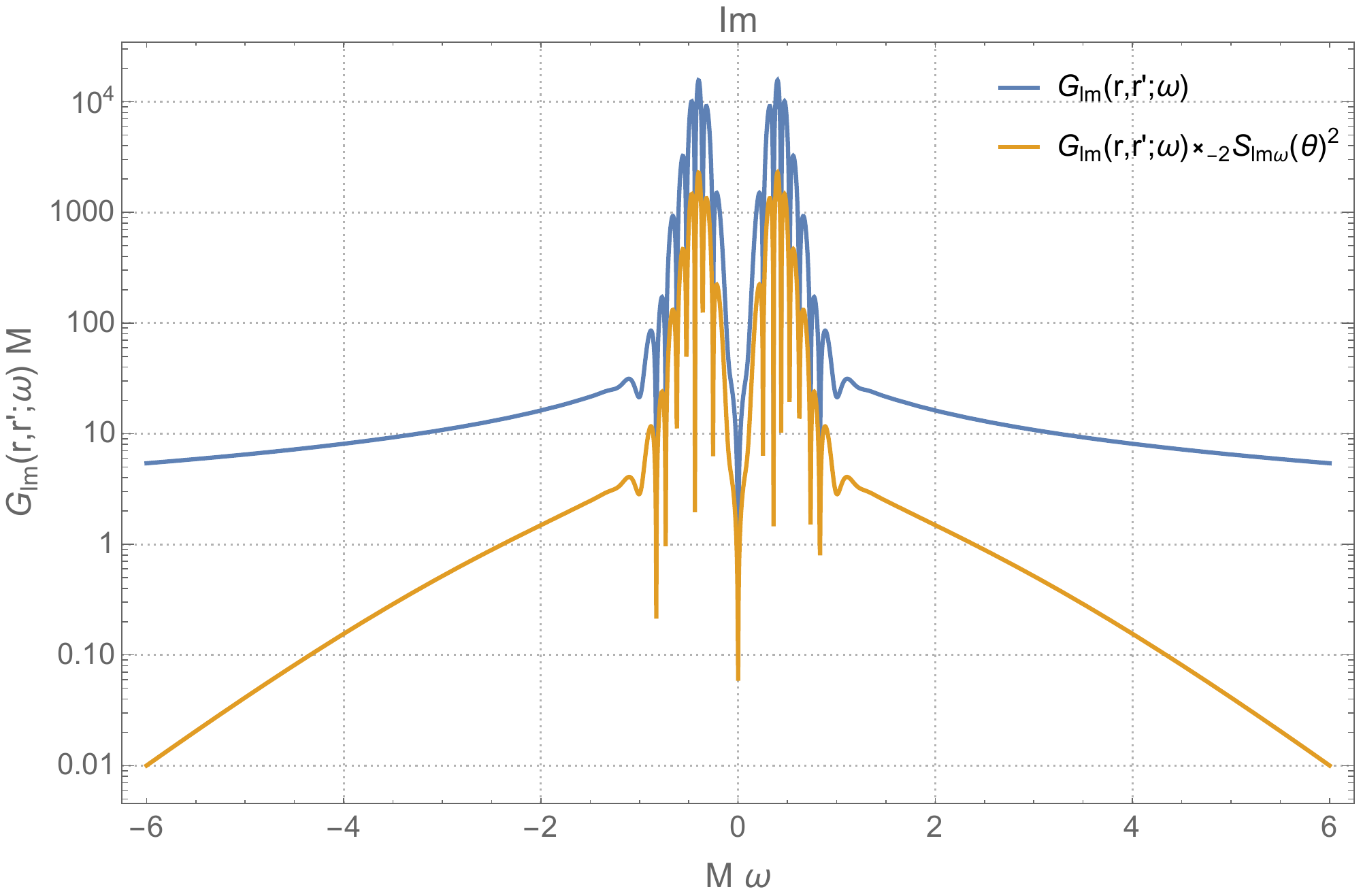}
        \label{fig:right}
    \end{subfigure}
    \caption{\textbf{Frequency domain integrand:} Log-plot of the real (left) and imaginary (right) part of the frequency domain Green function mode $G_{\ell m}(r,r';\omega)$ along the real frequency axis for $s=-2$, $\ell=2$, $m=0$ at a radius $r'=r=10$ and black hole spin $a=0.9$. In each figure, the blue curve is the radial Green function, and the orange curve is the radial Green function times the SWSHs (as per \eqref{eq:Glm_real_axis}) evaluated at $\theta'=\theta=\pi/2$, illustrating the suppression caused by the SWSHs at larger frequency. }
    \label{fig:FDGFwHarmonic}
\end{figure}

\begin{figure}[!h]
    \centering
    \includegraphics[width=.8\linewidth]{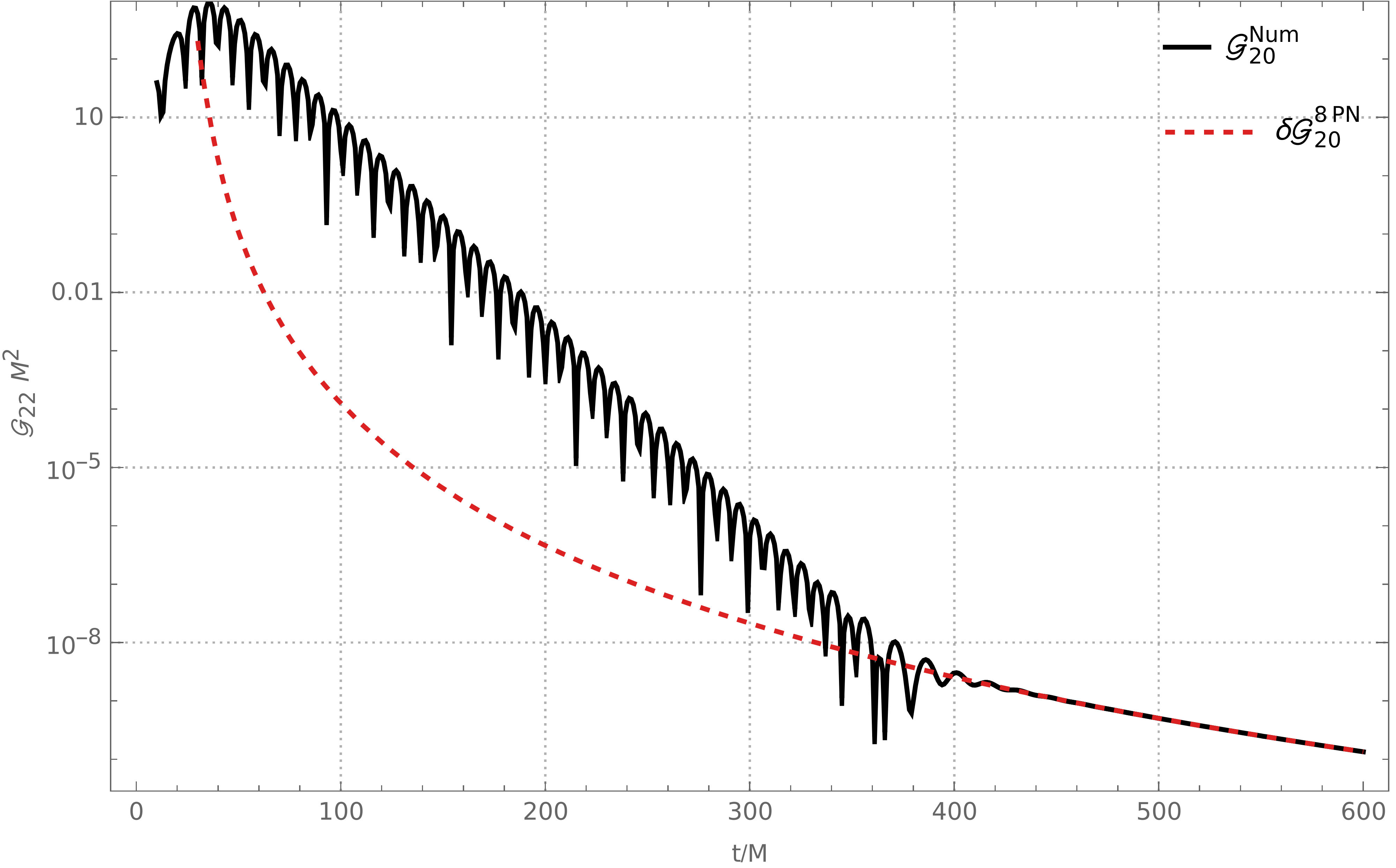}
    \caption{
    \textbf{Numerics vs. PN tail for $m=0$ and $r=10$:} Log-plot of the real part of the GF mode $\mathcal{G}_{\ell m}$ over time $t$ for $s=-2$, $\ell=2$, $m=0$ at a radius $r=r'=10$, at angles $\theta=\theta'=\pi/2$ and black hole spin $a=0.9$. In black is the numerical computation integrated along the real axis according to \eqref{eq:Glm_real_axis}. In red dashed is the low frequency computation of the tail using \eqref{eq:Disc-m} (i.e. \eqref{[eq:Tail_G20_PN]}) expanded to 8PN. The PN tail is not PS truncated and thus contains orders up to $t^{-23}$. However, as can be seen from Fig.~\ref{fig:residuals}, the higher powers of $1/t$ play virtually no role in the late time behavior.}
    \label{fig:G20}
\end{figure}

In our calculation, the inclusion of the frequency-dependent SWSHs in the integrand (in contrast to the Schwarzschild calculation where the harmonics are independent of the frequency) has a surprising benefit. 
For $\theta=\pi/2$ we find the harmonics fall off for large $|\omega|$ (c.f. Fig.~\ref{fig:SWSH falloff}). One may hope (optimistically) that this asymptotic behavior may serve as a regulator of the IFT integral, and in our test case we find that it does, see Figure~\ref{fig:FDGFwHarmonic}.  Perhaps paradoxically, the integral is far more convergent for large Kerr spin parameter $a$, since in the harmonics the frequency always appears in combination $a\omega$, so that the large $\omega$ asymptotic regime is reached sooner for higher $a$.

Practically speaking, in our numerical IFT we sample the radial Green function and SWSHs over a range $-6<\omega<6$. We also include a further high-$\omega$ cutoff function to avoid Gibbs ringing at the boundaries of our numerical integration. Explicitly we use 
\begin{equation}
   \mathrm{cut}(x)=\frac{1}{16} (1-\mathrm{erf}(2 (-x-\omega )))^2 (1-\mathrm{erf}(2 (\omega -x)))^2 
   \,,
\end{equation}
where $x$ is the variable width of the cutoff and $\mathrm{erf} (x)$ is the error function. Though the results are not very sensitive to its explicit choice, for most cases we choose $x=3.5$.

In Figs.~\ref{fig:G20} and \ref{fig:G22}
we compare the results of our numerical integration for the GF modes $\mathcal{G}_{\ell m}$ with a computation of the BC modes  $\Disc \mathcal{G}_{\ell m}$ using the mentioned PN zone approximation of the radial functions. We do it for $m=0$  in Fig.~\ref{fig:G20} using an 8PN approximation and for $m=2$  in Fig.~\ref{fig:G22} using a 5PN approximation, (see Appendix~\ref{sec:more_plots} for comparisons at lower PN orders, at different radii and at different black hole spins $a$). 
While $\mathcal{G}_{\ell m}$ is real-valued for $m=0$, it is complex-valued for $m=2$; in Fig~\ref{fig:G22} for $m=2$ we only plot the real part of $\mathcal{G}_{\ell m}$, since it is already sufficiently illustrative of the various features.

\begin{figure}[h]
    \centering
    \includegraphics[width=\linewidth]{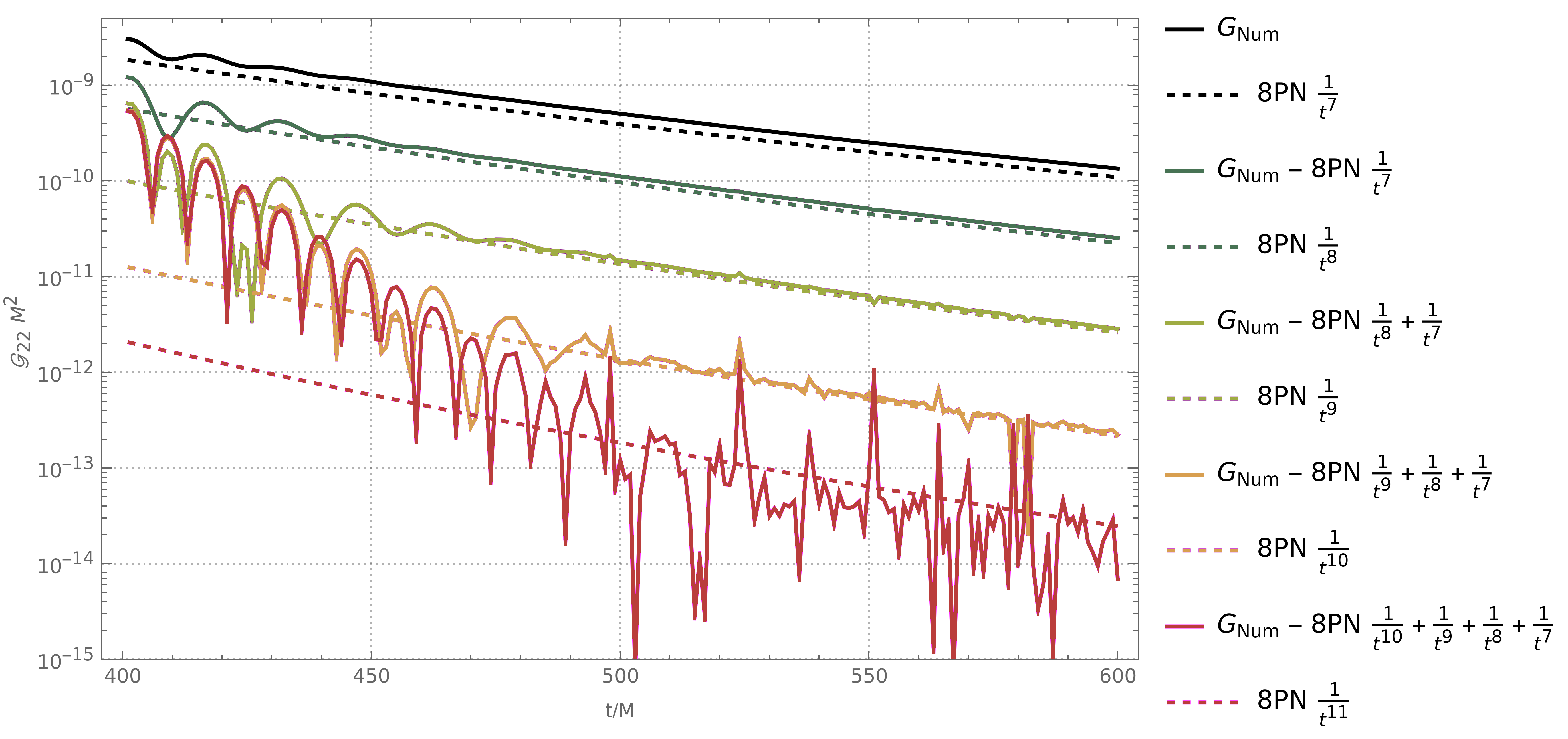}
    \caption{\textbf{Residuals for $m=0$ and $r=10$:} Log-plot of the residuals of the real part of the GF mode $\mathcal{G}_{\ell m}$ over time $t$ for $s=-2$, $\ell=2$, $m=0$ at radii $r=r'=10$, at angles $\theta=\theta'=\pi/2$ and black hole spin $a=0.9$. In black is the numerical computation integrated along the real axis according to \eqref{eq:Glm_real_axis}. In colors of the rainbow spectrum are the residuals from successively subtracting $t^n$ terms from the 8PN (17 terms in $\eta$) tail (c.f. \eqref{[eq:Tail_G20_PN]} and the text above). In dashed is the respective $t^n$ component expanded up to 8PN. The components $t^{-9}$, $t^{-10}$ and $t^{-11}$ contain $\log t$ terms, that are not treated as extra orders here. After subtracting $t^{-10}$ we are hitting the noise floor. However, between $t=400$ and $t=500$ we can see how stripping away the tail lays bare the QNMs.}
    \label{fig:residuals}
\end{figure}

\begin{figure}
    \centering
    \includegraphics[width=.8\linewidth]{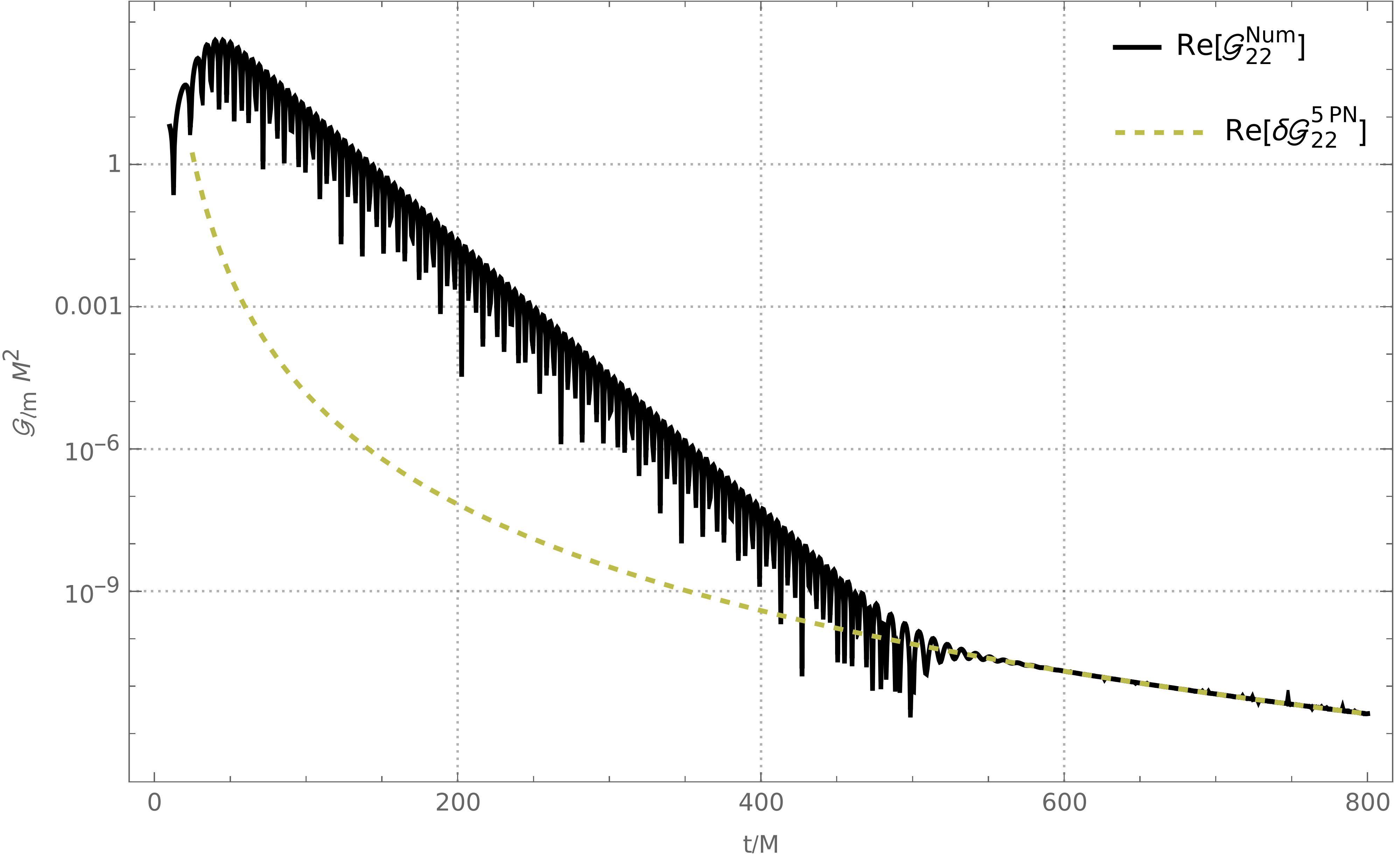}
    \caption{
    \textbf{Numerics vs. PN tail for $m=2$ and $r=10$:} Log-plot of the real part of the GF mode $\mathcal{G}_{\ell m}$ over time $t$ for $s=-2$, $\ell=2$, $m=2$ at a radius $r=r'=10$, at angles $\theta=\theta'=\pi/2$ and black hole spin $a=0.9$. In black is the numerical computation integrated along the real axis according to \eqref{eq:Glm_real_axis}. In red is the low frequency computation of the tail using \eqref{eq:Disc-m} where we expanded the radial functions up to 5PN.} 
    \label{fig:G22}
\end{figure}

We find that the tail dominates over the QNM ringing at dimensionless time $t>t_{\rm tail}$, where we begin to see agreement between the numerics for $\mathcal{G}_{\ell m}$ and the analytic BC modes   $\Disc \mathcal{G}_{\ell m}$, with $t_{\rm tail}\approx 400$ for $m=0$ and $t_{\rm tail}\approx 460$ for $m=2$.  
Figs.~\ref{fig:G20} and \ref{fig:G22} for the azimuthal mode $m=0$ and $m=2$, respectively, show very good visual agreement between the real-frequency numerical integration via \eqref{eq:Glm_real_axis} and our analytical PN expansion.
We remind the reader that the expansion to 8PN order of $\Disc \mathcal{G}_{\ell m}$ in Eq.~\eqref{[eq:Tail_G20_PN]} for $s=-\ell=-2$ is up to $t^{-23}$, so 16 orders beyond the leading $t^{-7}$, not just three orders as per Eq.~\eqref{[eq:Tail_G20_PN]}.
However, the fine agreement between the 8PN and the numerical result is mostly due to the fact that high order of 8PN means going to high order for large radius (not only to high order for small frequency). 
Correspondingly holds for the 5PN expansion for $m=2$.
It is the higher order expansions for large radius that are achieving the fine results, as is shown in Figs.~\ref{fig:residuals} and \ref{Fig:residuals m=2,with/without cc} by subtracting the PN expansion for each power of time separately. Specifically, 
in Figs.~\ref{fig:residuals} and \ref{Fig:residuals m=2,with/without cc} we investigate the sub-leading late-time asymptotics of $\mathcal{G}_{\ell m}$ by successively subtracting off powers of $1/t$ from the numerical $\mathcal{G}_{\ell m}$ in the tail-dominated regime $t>t_{\rm tail}$.  We see that each successive subtraction leaves a residual with the expected subleading power law behaviour consistent with our analytic results being correct. Interestingly, one also nicely sees the emergence of the `buried' QNM ringing at later and later times as we subtract tail contributions. Upon subtracting the next-to-next-to-next-to-leading order ($t^{-10}$) we appear to reach the numerical noise floor of the real frequency IFT. 

Finally, we note that Fig.~\ref{Fig:residuals m=2,with/without cc} for $m=2$ contains a comparison of the residues with the PN calculated in two ways: (a) by including a complex conjugation in the factor ${}_s\Slm(\theta')$ in Eq.~\eqref{eq:Disc-m} (top plot); and (b) by using Eq.~\eqref{eq:Disc-m} as is (so no complex conjugation of any SWSH) (bottom plot).
This comparison is in relation to the discussion below Eq.~\eqref{eq:Disc-m}.
The results for the first few order residuals are very similar with and without the complex conjugation on the SWSH, but 
comparison of the two plots for the highest-order residuals clearly supports Eq.~\eqref{eq:Disc-m} as is.

\begin{figure}
\begin{subfigure}{\linewidth}
    \centering
    \includegraphics[width=\linewidth]{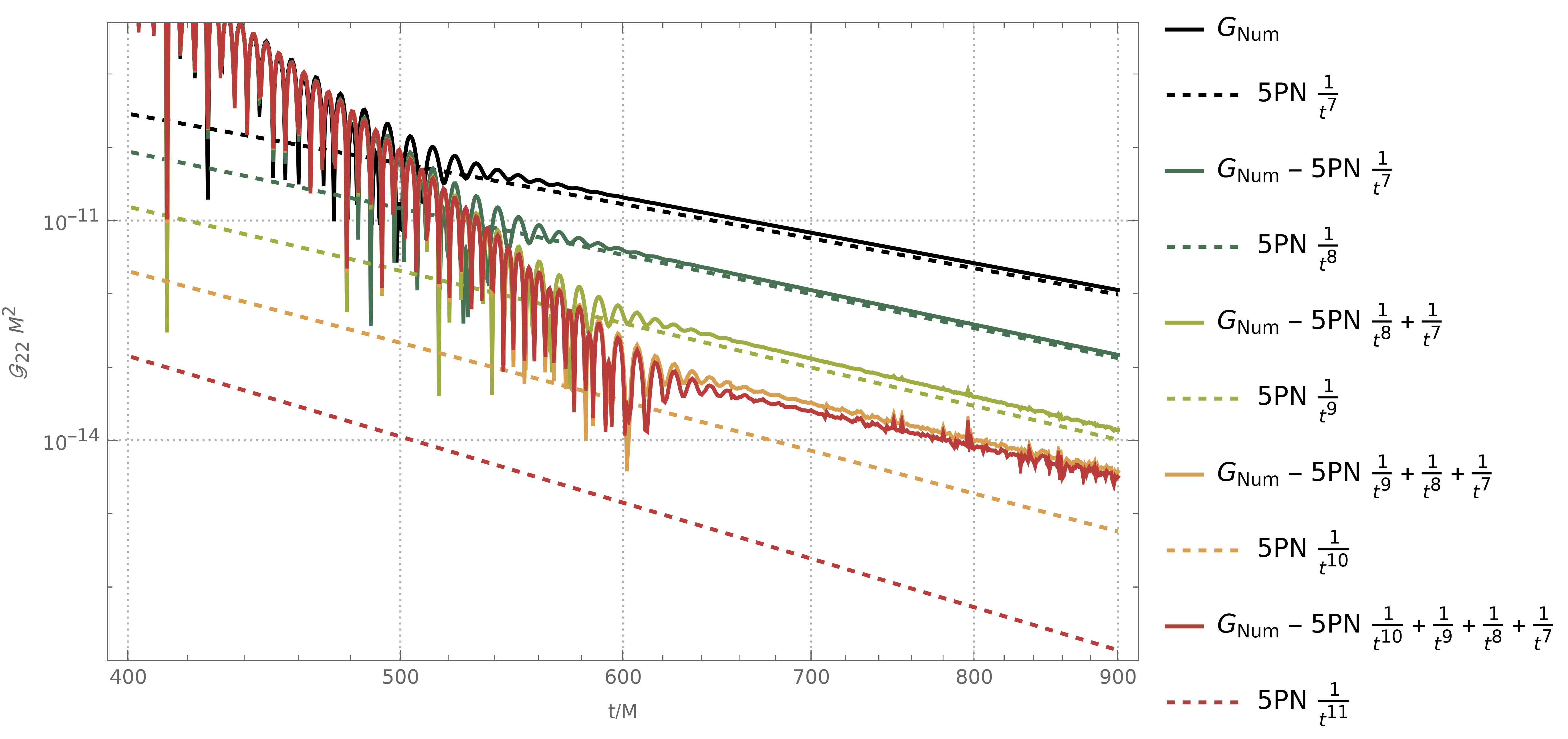}
    \caption{Using Eq.~\eqref{eq:Disc-m} with ${}_s\Slm(\theta')$
    complex conjugated in the PN tails. The yellow and red curve shows that after subtracting $t^{-10}$ the residual follows a $t^{-9}$ power law.}
    \label{fig:placeholder}
\end{subfigure}
\begin{subfigure}{\linewidth}
    \centering
    \includegraphics[width=\linewidth]{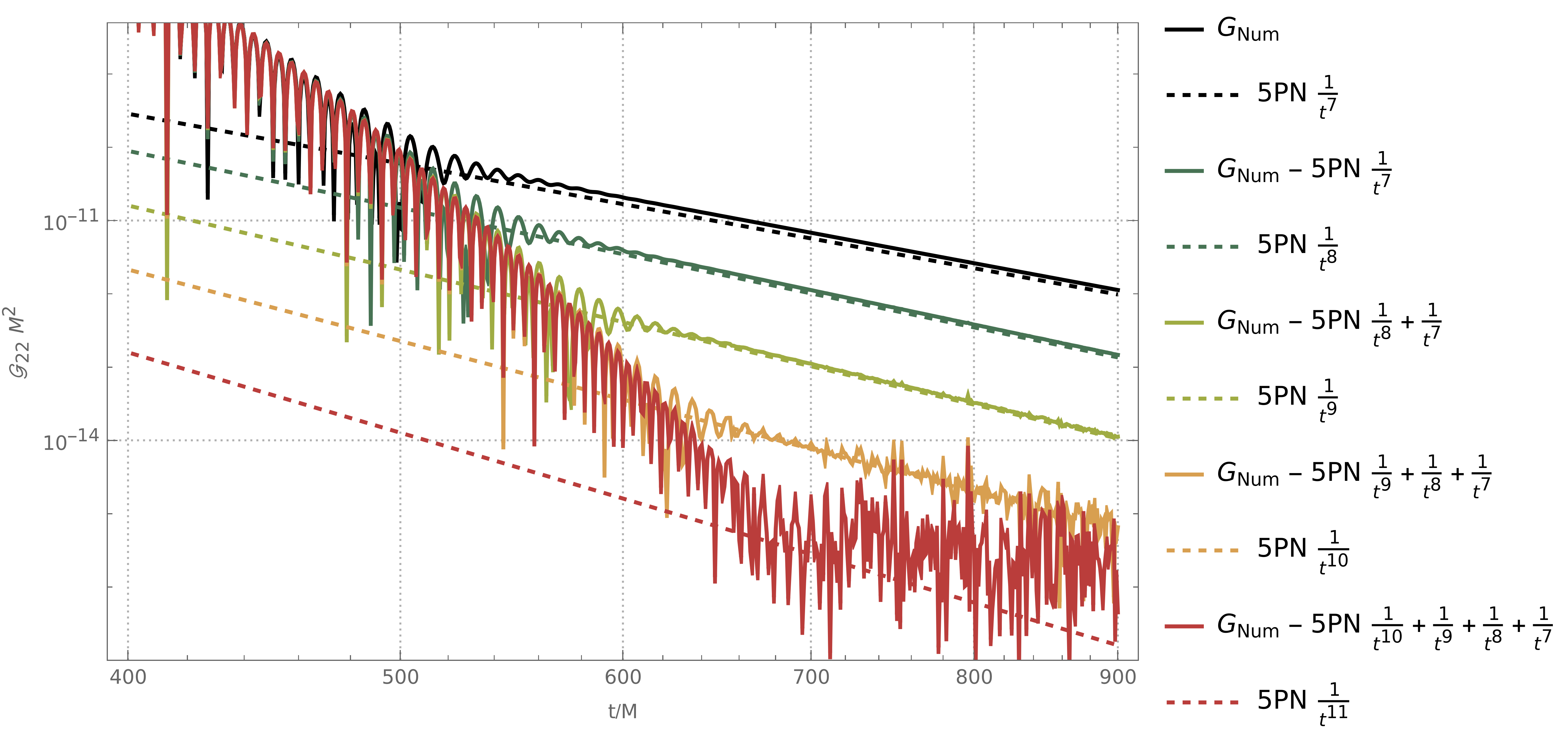}
    \caption{Using Eq.~\eqref{eq:Disc-m} as is for the PN tails. We can see that the residuals continuously become smaller. The red curve does become dominated by noise around $t=700$.}
    \label{fig:placeholder}
\end{subfigure}
\caption{\textbf{Numerics vs. PN tail for $m=2$ and $r=10$:} Log-plot of the residuals of the real part of the GF mode $\mathcal{G}_{\ell m}$ over time $t$ for $s=-2$, $\ell=2$, $m=2$ at radii $r=r'=10$, angles $\theta=\theta'=\pi/2$ and black hole spin $a=0.9$. In black is the numerical computation integrated along the real axis according to \eqref{eq:Glm_real_axis}. In colors of the rainbow spectrum are the residuals from successively subtracting $t^n$ terms from the 5PN (11 terms in $\eta$) tail (c.f. \eqref{[eq:Tail_G20_PN]}). In dashed is the respective $t^n$ component expanded up to 5PN. The components $t^{-9}$, $t^{-10}$, and $t^{-11}$ contain $\log t$ terms, that are not treated as extra orders here.  (a) Residuals where we complex conjugated the factor ${}_s\Slm(\theta')$ in Eq.~\eqref{eq:Disc-m} to compute the PN tails. (b) Eq.~\eqref{eq:Disc-m} as is (so no complex conjugation of any SWSH) to compute the PN tails. Comparison of these plots indicate that taking the complex conjugate leads to a leftover $\frac{1}{t^9}$ residual where not taking it does not.}\label{Fig:residuals m=2,with/without cc}
\end{figure}

%---------------------------------------------------------------------------------------------------------

\section{Conclusions}\label{sec:conclusions}

%---------------------------------------------------------------------------------------------------------
%---------------------------------------------------------------------------------------------------------

In this paper we have obtained late-time tails up to the first three leading orders for the GF of the Teukolsky equation in subextremal Kerr. We have obtained the tails: (a) at finite radius away from the horizon (Eq.~\eqref{eq:tail,finite r}); (b) along $\hor^+$ (Eq.~\eqref{eq:tail H^+}); and (c) along $\scri^+$ (Eq.~\eqref{eq:tail,inf}).
The powers in the corresponding leading-order terms, for fixed multipolar $\ell$ and azimuthal $m$ modes, are, generically\footnote{In special cases of measure zero, the coefficient of the leading order term  might so happen to be zero.}, $t^{-2\ell-3}$, $v^{-2\ell-3-b}$ (with $b=1$ when $s>0,m=0$ and $b=0$ otherwise), and $u^{-\ell+s-2}$, respectively.
The tail along $\hor^+$ also includes oscillations in $v$ (via a factor $e^{im\Omega_H v}$) for the non-axisymmetric perturbations, as was previously well-known.
These tails per $(\ell,m)$-mode translate into the following tails for the full GF: $t^{-2|s|-3}$, $v^{-2|s|-3}$ and $u^{s-|s|-2}$ in, respectively, cases (a), (b) and (c).
We have obtained the powers for generic integer spin $s$ and the expansion coefficients explicitly for $s=-2$.

The main novel result in this paper is showing that logarithmic-in-time terms first appear at next-to-leading order in case (c) and at next-to-next-to-leading order in cases (a) and (b). In case (c), at next-to-next-to-leading order a quadratic logarithmic term also appears. 
Although we did not explicitly go beyond three leading orders, we have seen that the generic-order terms in the late-time expansion of the GF are of the type $t^{-\lambda-1}\log^\mu t$  (and similarly for $u$ or $v$ instead of $t$) with $\lambda\in \mathbb{Z}_{\geq 1}$ and $\mu\in \mathbb{Z}_{\geq 0}$.

We have obtained the late-time expansions of the GF in the time-domain via small-frequency (MST) expansions  of  quantities in the frequency domain. 
These explicit small-frequency expansions which we have provided can be readily used for scattering problems not only in the exterior of a Kerr black hole (as done here) but in the interior as well (see, e.g., App.~D in~\cite{AZCO:2026} for the generic-$\omega$ MST expressions for the In radial solution and its scattering coefficients in the interior of Kerr).

The case of extremal Kerr ($a=M$) deserves a  separate comment.
As mentioned, in the case of subextremal Kerr dealt with in this paper, the branch point $\omega=0$ is the highest-lying singular point of the  Fourier modes of the GF.
%This is so in the case of subextremal Kerr dealt with in this paper, whereas 
However, in extremal Kerr, the superradiant bound frequency $\omega=m\Omega_H$ on the real axis also becomes a branch point~\cite{detweiler1980black,PhysRevD.64.104021,Casals:2019vdb}.
The  competition between the two branch points ($\omega=0$ and $\omega= m\Omega_H$) gives rise to a rich spectrum of leading late-time tails in the extremal case (see~\cite{Casals:2018eev,casals2016horizon,Gralla:2017lto}), depending on the direction of approach to timelike infinity $i^+$ and on the field mode values of $\ell$ and $m$ and spin $s$. Notably, this includes an instability of the extremal event horizon (the leading instability, which comes from non-axisymmetric perturbations, was found in~\cite{casals2016horizon} and was later rigorously proven in~\cite{gajic2023azimuthalinstabilitiesextremalkerr}; \cite{Aretakis:2012ei} previously found a subleading instability from axisymmetric perturbations).
The tails in the extremal case also include a new specific type of oscillation in time, which appears not only on the horizon but away as well~\cite{casals2016horizon,Casals:2018eev,gajic2023azimuthalinstabilitiesextremalkerr}.
This oscillatory factor in the late-time asymptotics is thus a third type of `wagging of the tail', on the top of the already mentioned one along $\mathcal{H}^+$ (via $e^{im\Omega_H v}$), which is present even in the subextremal case, as well as a trivial one due to  a possible time-dependence of $\phi-\phi'$ (or $\theta$ or $\theta'$), and so which may appear in Schwarzschild already (see, e.g.,~\cite{CDOW13}) as well as Kerr~\cite{CKO}. 
The MST method which we have used in this paper for calculating small-frequency expansions is specific for subextremal Kerr. The MST method for extremal Kerr was developed in~\cite{Casals:2018eev} and we leave its application to obtaining higher-order late-time asymptotics in extremal Kerr for future work. 

%---------------------------------------------------------------------------------------------------------
%---------------------------------------------------------------------------------------------------------
\section*{Acknowledgments}
We thank Barry Wardell for technical support. CK and JN acknowledge support from Research Ireland under Grant number 21/PATH-S/9610. MC is thankful to Junquan Su, Dejan Gajic, Stefan Hollands and Lionor Kehrberger for useful discussions. This work makes use of the Black Hole Perturbation Toolkit.
%---------------------------------------------------------------------------------------------------------
%---------------------------------------------------------------------------------------------------------
\newpage
\appendix

%---------------------------------------------------------------------------------------------------------
%---------------------------------------------------------------------------------------------------------

\section{Further numerical comparisons}\label{sec:more_plots}

\begin{figure}[h!]
\begin{subfigure}{.49\textwidth}
    \centering
    \includegraphics[width=\linewidth]{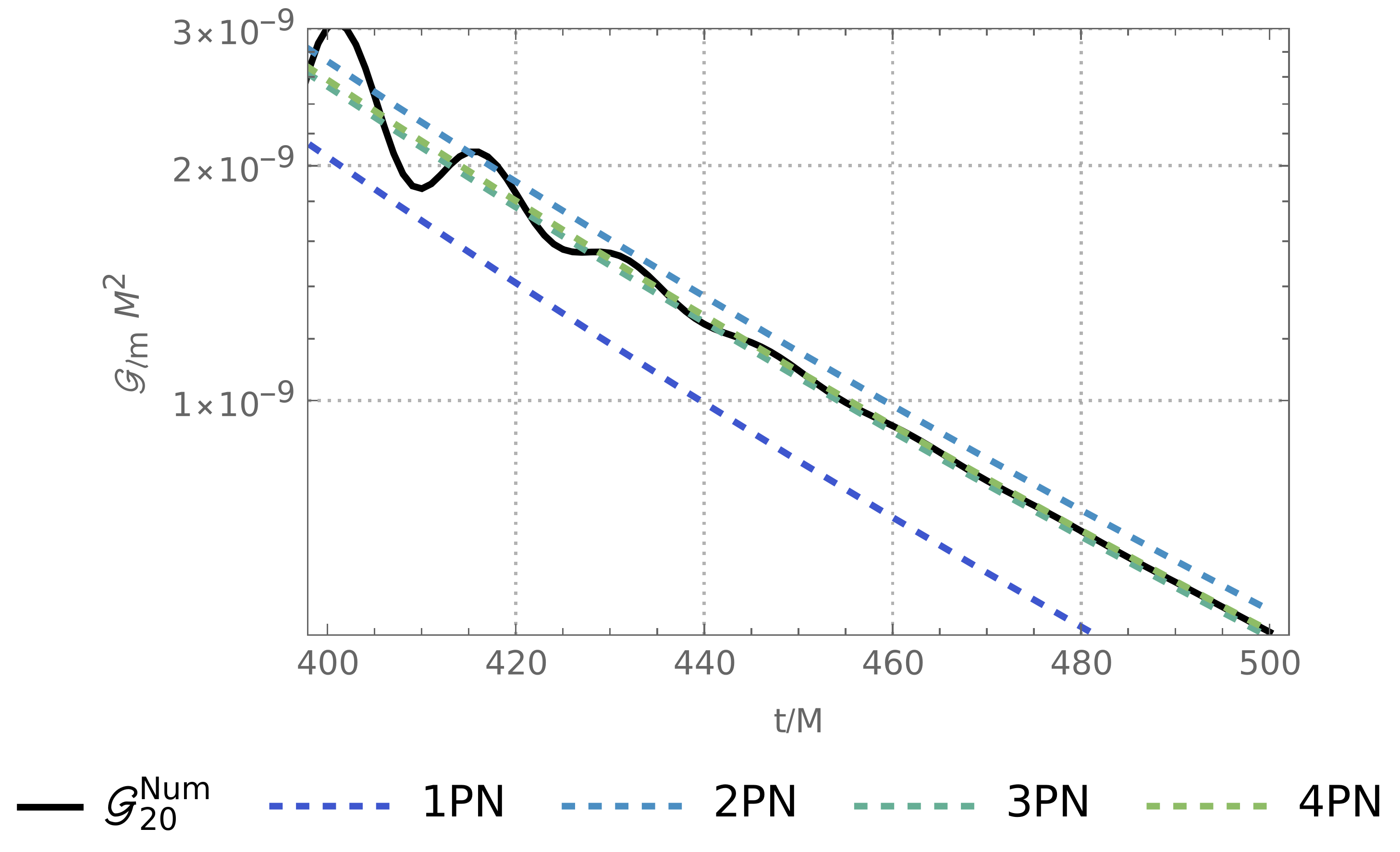}
    \caption{$r=r'=10$. 4PN yields visual agreement in the tail.}
    \label{fig:placeholder}
\end{subfigure}
\begin{subfigure}{.49\textwidth}
    \centering
    \includegraphics[width=\linewidth]{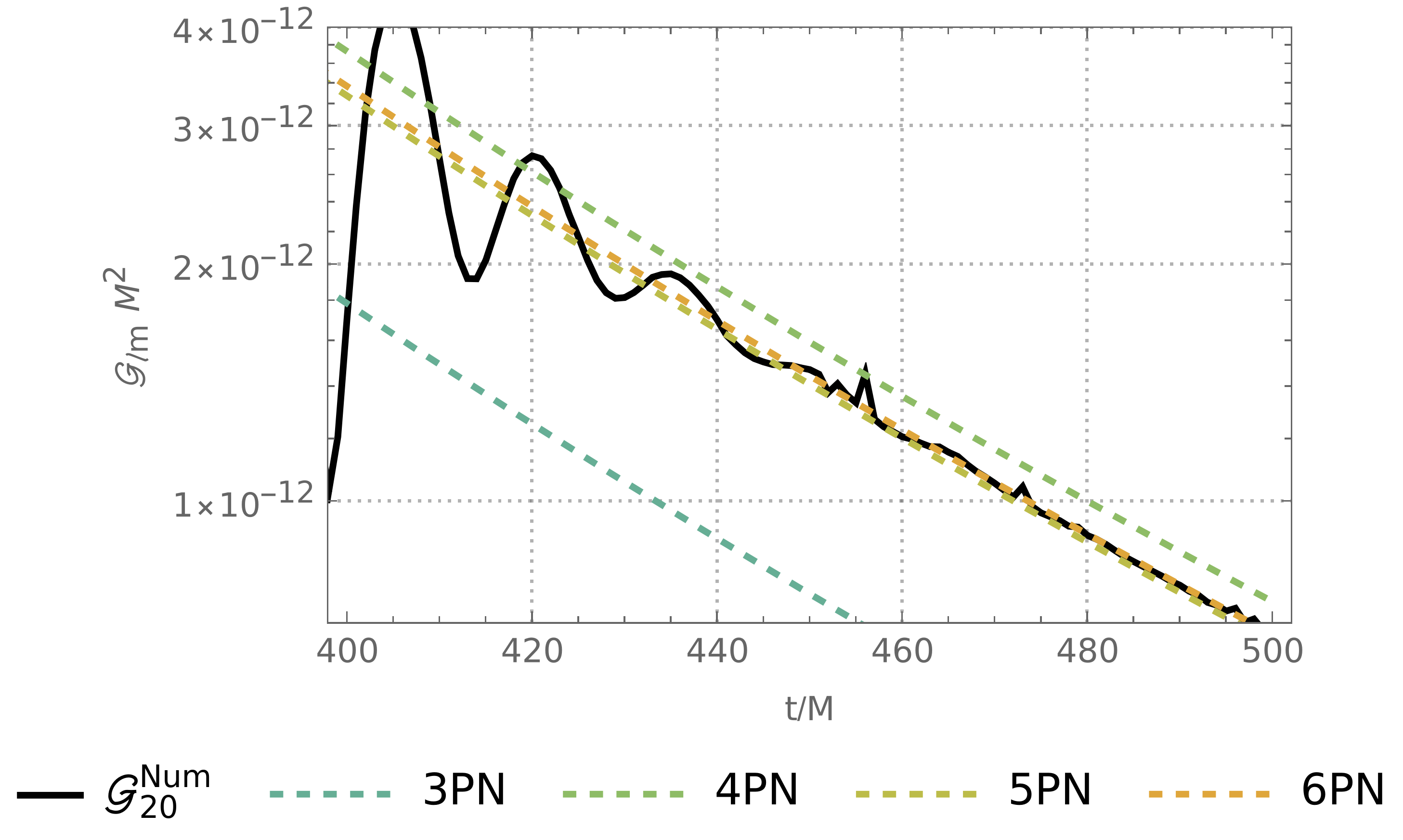}
    \caption{$r=r'=5$. 6PN yields visual agreement  in the tail.}
    \label{fig:placeholder}
\end{subfigure}
\begin{subfigure}{.49\textwidth}
    \centering
    \includegraphics[width=\linewidth]{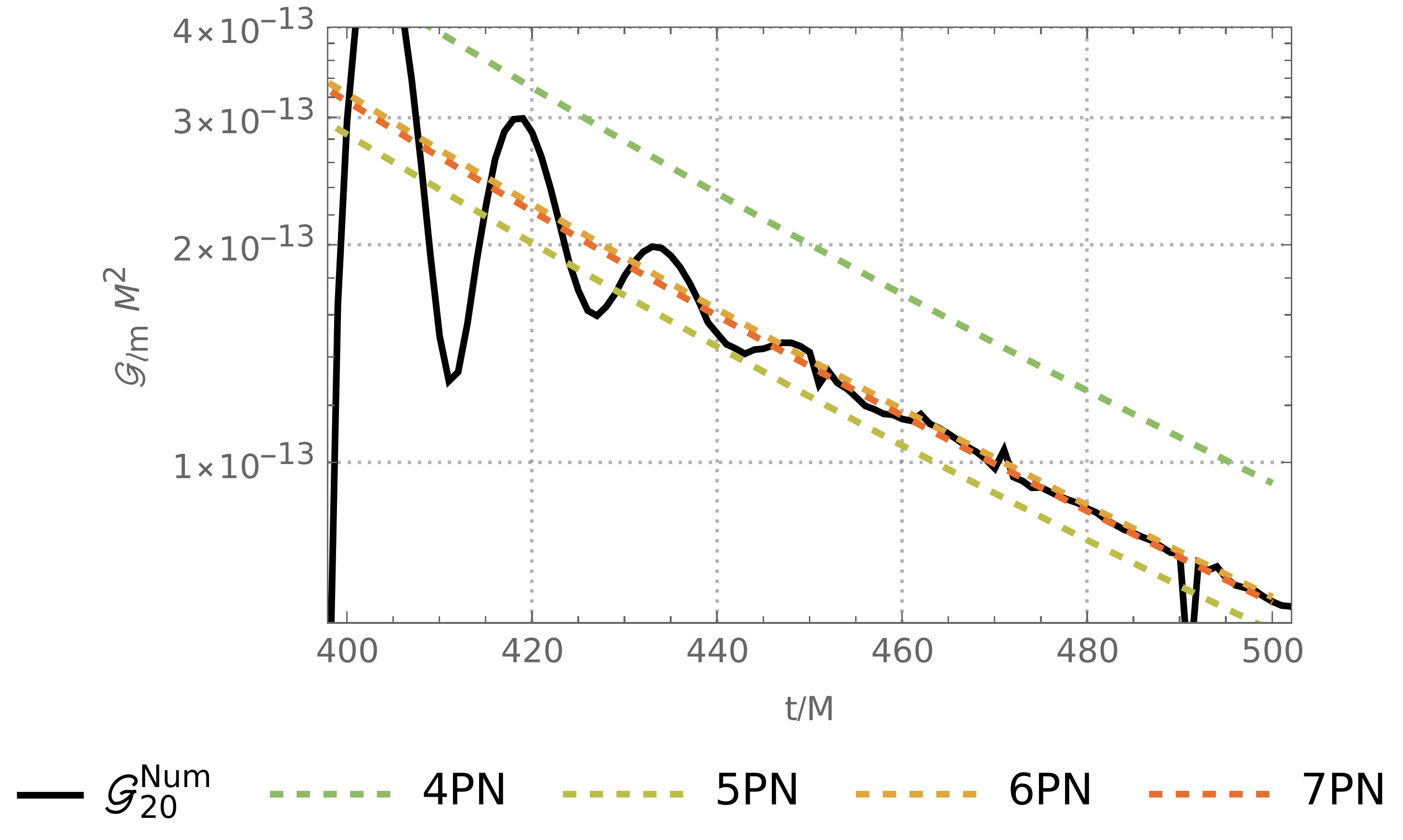}
    \caption{$r=r'=4$. 7PN yields visual agreement  in the tail.}
    \label{fig:placeholder}
\end{subfigure}
\begin{subfigure}{.49\textwidth}
    \centering
    \includegraphics[width=\linewidth]{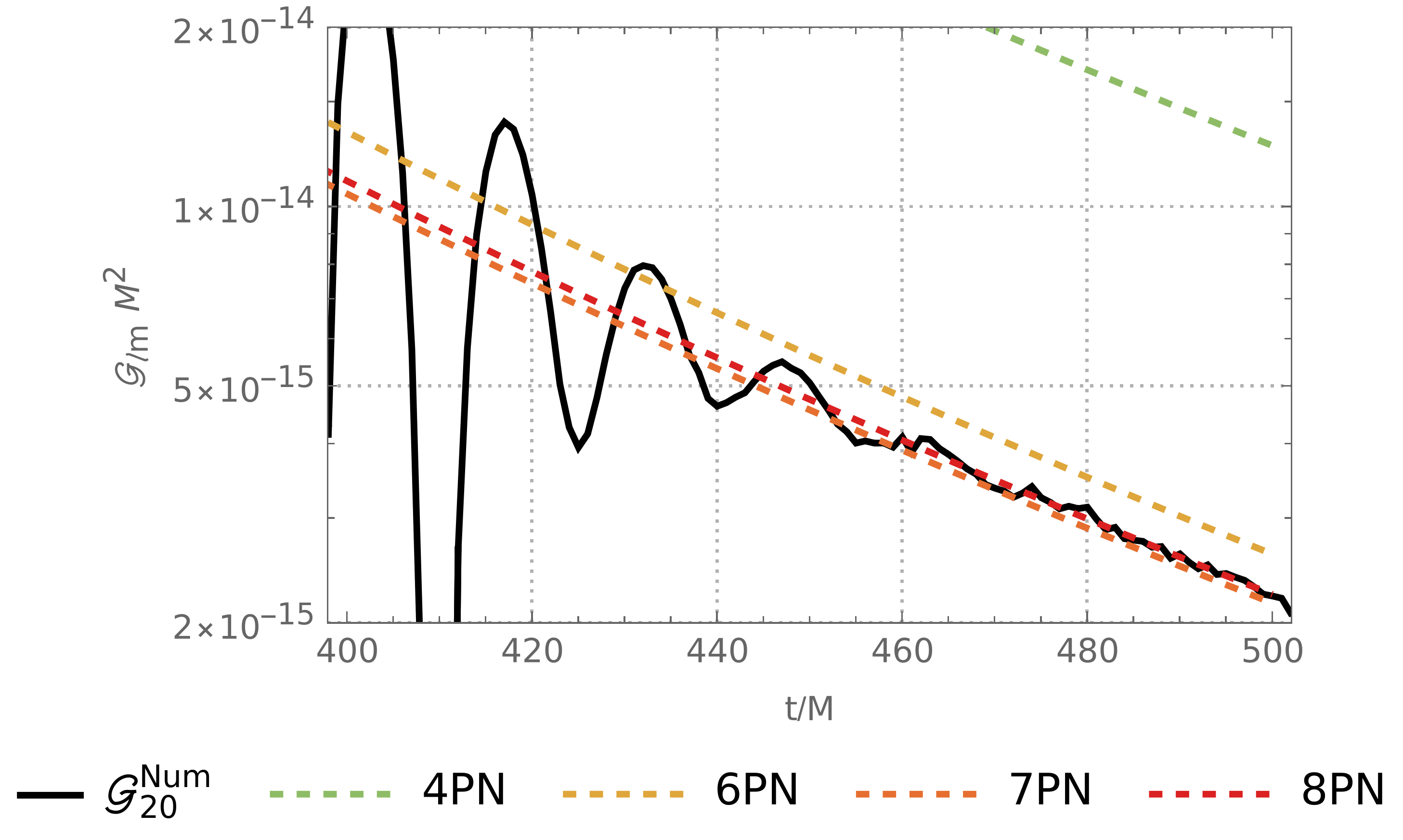}
    \caption{$r=r'=3$. 8PN yields visual agreement  in the tail. 5PN omitted due to zero-crossing.}
    \label{fig:PN_tail_convergence_r3}
\end{subfigure}
\caption{\textbf{PN convergence}: Log-plots of the GF modes $\mathcal{G}_{\ell m}$ and $\delta\mathcal{G}_{\ell m}$ over time $t$ for $s=-2$, $\ell=2$, $m=0$, $r=r'$ for a range of values of $r$, at angles $\theta=\theta'=\pi/2$ and black hole spin $a=0.9$. In black is the numerical computation of $\mathcal{G}_{\ell m}$ by integrating along the real axis according to \eqref{eq:Glm_real_axis}. In dashed are our analytical tail results for $\delta\mathcal{G}_{\ell m}$ using \eqref{eq:Disc-m} together with a PN approximation. The colour coding of the various PN orders is consistent across plots, e.g. 4PN is the same colour in all four plots. }
\label{Fig:varying r}
\end{figure}

In this appendix we give further Figs.~\ref{Fig:varying r}--\ref{fig:residuals_r3} which compare the GF mode $\mathcal{G}_{\ell m}(r,r';\theta,\theta';t)$ calculated using the numerical integration along the real-$\omega$ axis in \eqref{eq:Glm_real_axis} with our analytical  tail results for $\delta\mathcal{G}_{\ell m}(r,r';\theta,\theta';t)$ using \eqref{eq:Disc-m} together with a PN approximation.
All plots in these figures are as functions of time $t$ for $s=-2$, $\ell=2$, $m=0$,
angles $\theta=\theta'=\pi/2$ and $r=r'$.
These plots highlight different physical and technical aspects.

Fig.~\ref{Fig:varying r} shows various plots of $\mathcal{G}_{\ell m}$ and $\delta\mathcal{G}_{\ell m}$ for black hole spin $a=0.9$ and varying values of $r=r'$ and varying PN orders for $\delta\mathcal{G}_{\ell m}$.
The plots show that increasing PN orders are converging towards the numerical results. They also show that higher PN orders are required to maintain a certain precision as we get closer to the event horizon.

Fig.~\ref{Fig:varying a} shows various plots of $\mathcal{G}_{\ell m}$ and $\delta\mathcal{G}_{\ell m}$ for varying values of $r=r'$, with  $a=0.9$ for $\mathcal{G}_{\ell m}$ and using 8PN for  $\delta\mathcal{G}_{\ell m}$ in the two cases of $a=0.9$ and $a=0$ (Schwarzschild).
\label{Fig:varying a}
The plots show that the effect of Kerr spin $a$ on the tail becomes more relevant closer to the event horizon.
%\newpage

%
\begin{figure}[h!]
\begin{subfigure}{.49\textwidth}
    \centering
    \includegraphics[width=\linewidth]{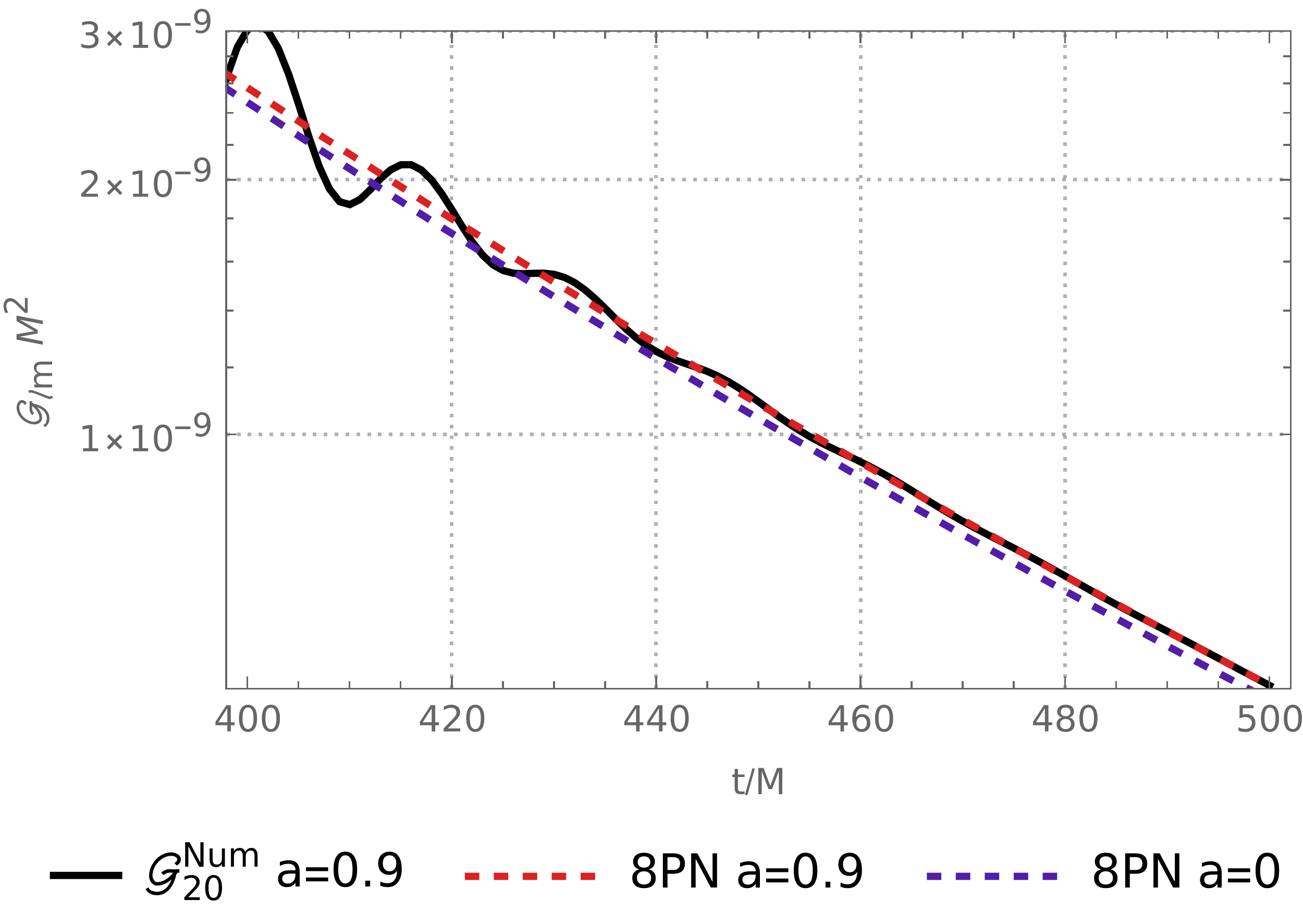}
    \caption{$r=r'=10$.}
    \label{fig:placeholder}
\end{subfigure}
\begin{subfigure}{.49\textwidth}
    \centering
    \includegraphics[width=\linewidth]{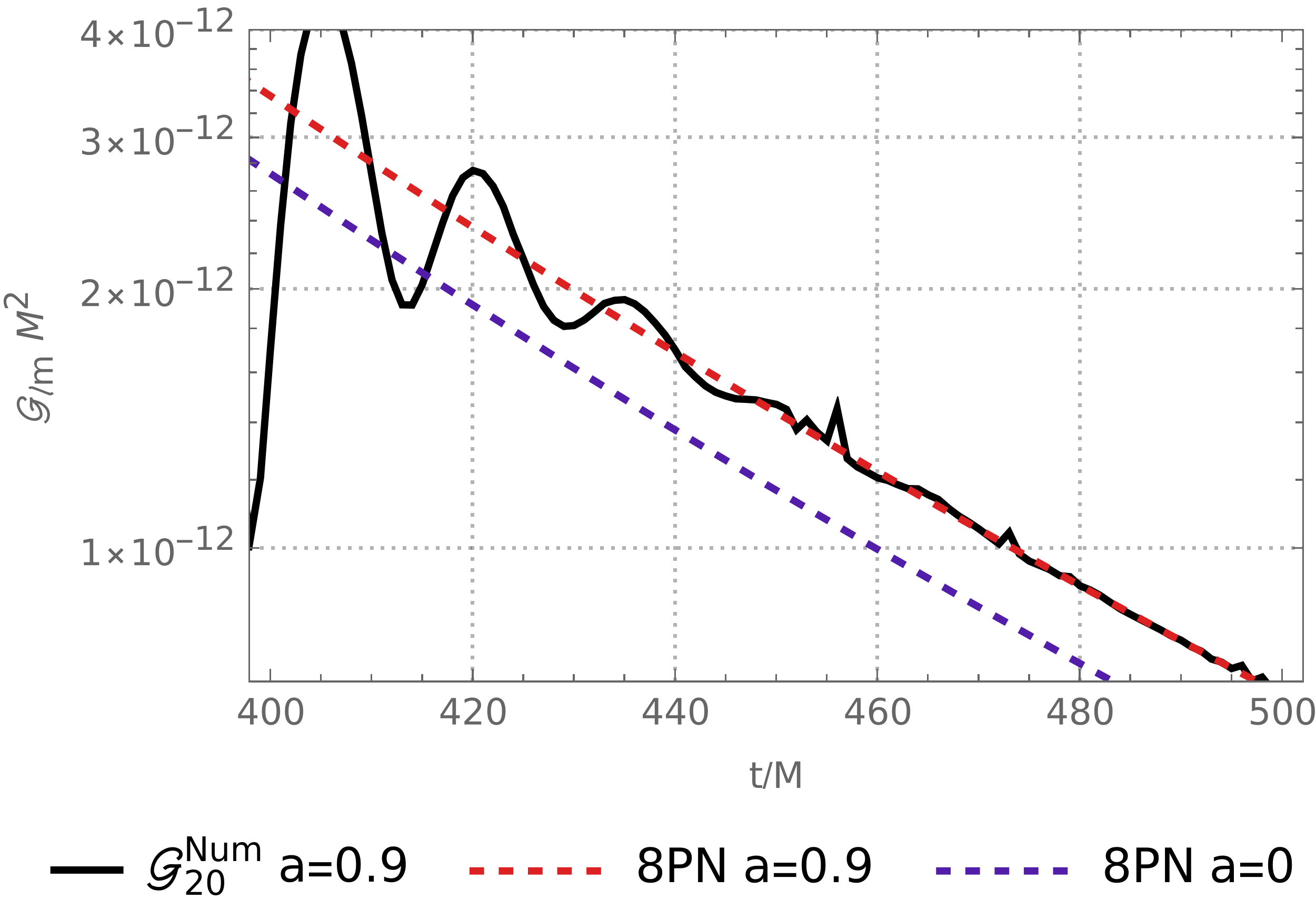}
    \caption{$r=r'=5$.}
    \label{fig:placeholder}
\end{subfigure}
\begin{subfigure}{.49\textwidth}
    \centering
    \includegraphics[width=\linewidth]{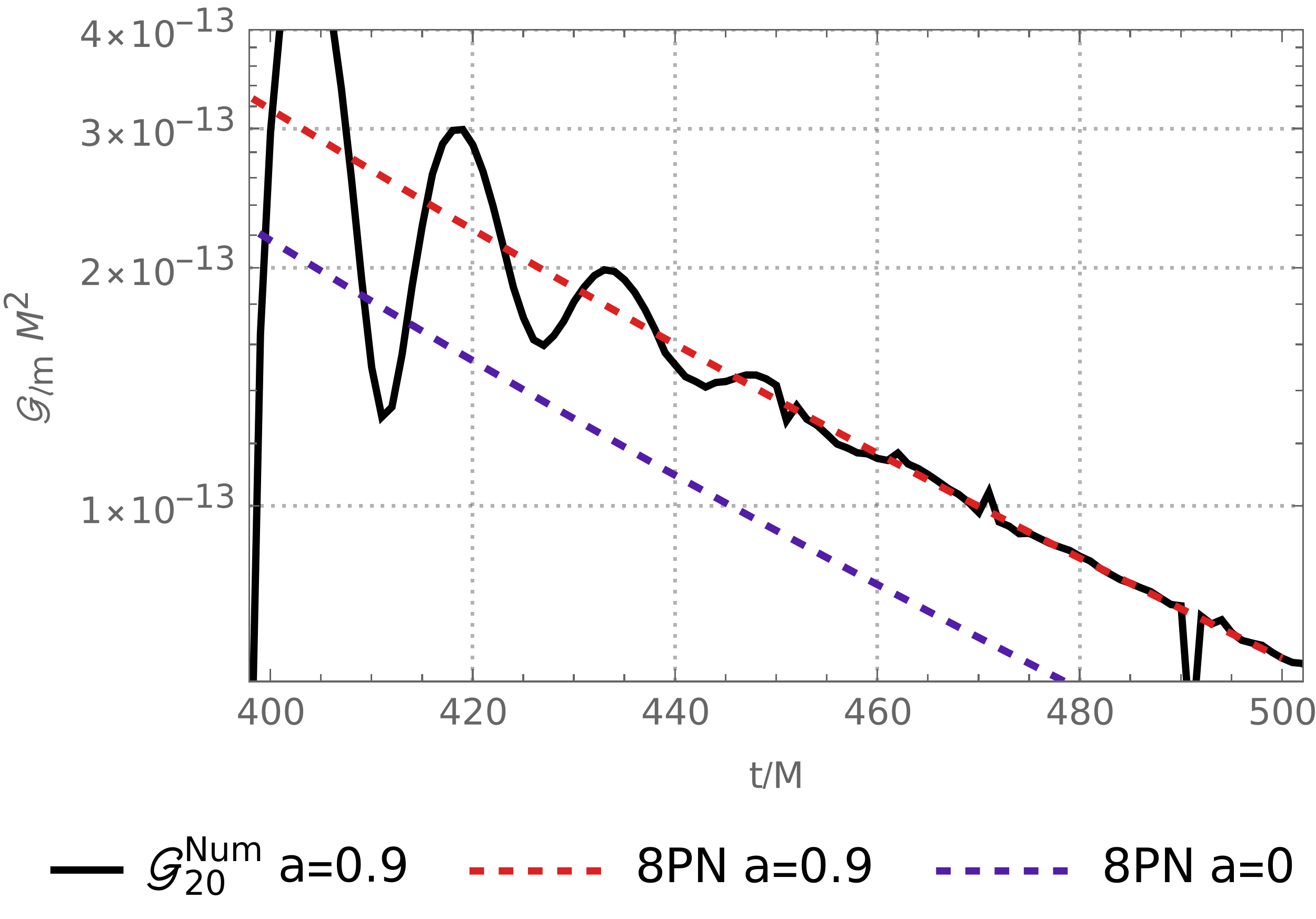}
    \caption{$r=r'=4$. }
    \label{fig:placeholder}
\end{subfigure}
\begin{subfigure}{.49\textwidth}
    \centering
    \includegraphics[width=\linewidth]{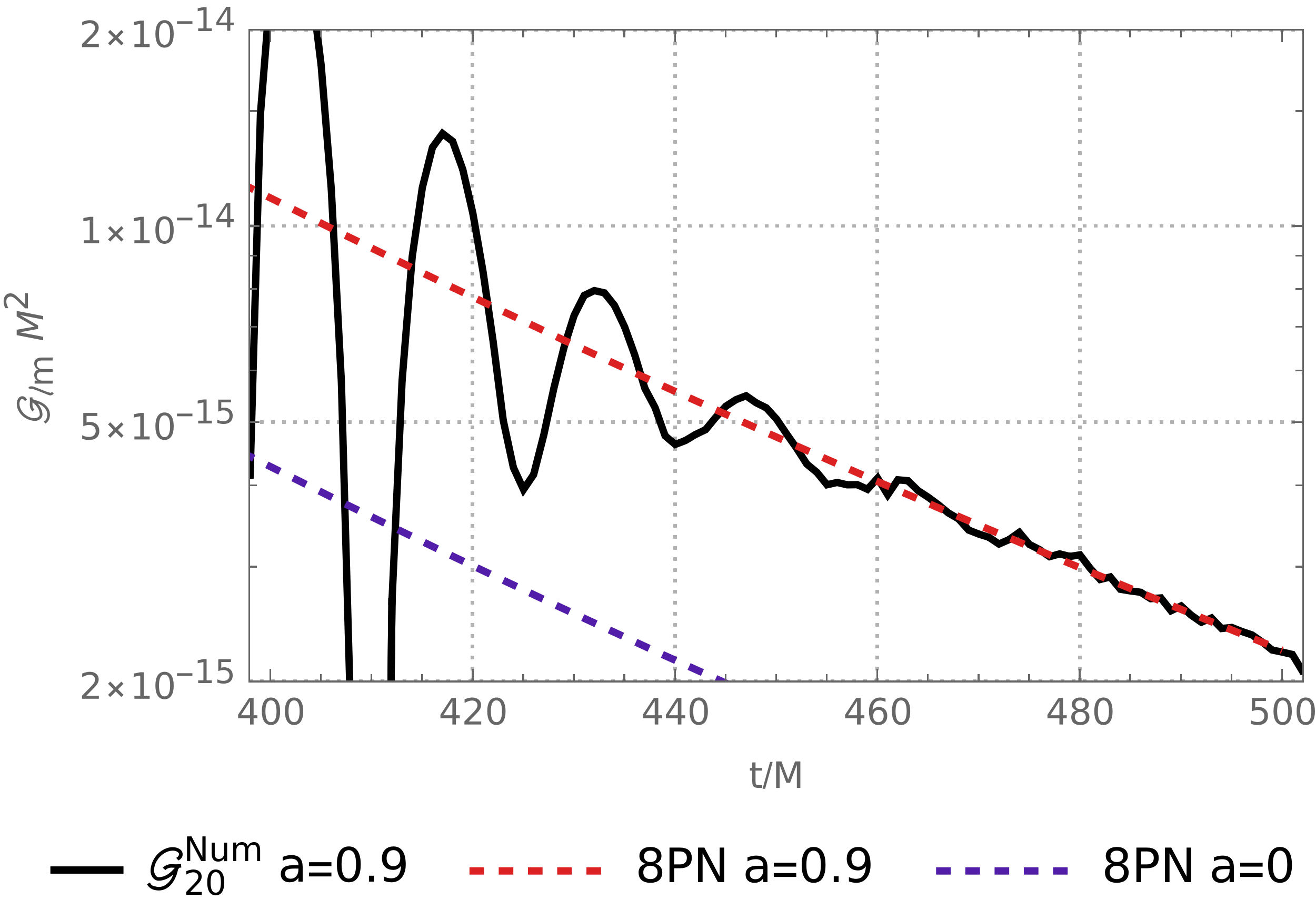}
    \caption{$r=r'=3$.}
    \label{fig:placeholder}
\end{subfigure}
\caption{\textbf{Effect of Kerr}: Log-plots of the GF modes $\mathcal{G}_{\ell m}$ and BC modes $\delta\mathcal{G}_{\ell m}$ over time $t$ for $s=-2$, $\ell=2$, $m=0$, $r=r'$ for a range of values of $r$, at angles $\theta=\theta'=\pi/2$.
In black is the numerical computation of $\mathcal{G}_{\ell m}$ by integrating along the real axis according to \eqref{eq:Glm_real_axis} for black hole spin $a=0.9$. In dashed are our analytical tail results for $\delta\mathcal{G}_{\ell m}$ using \eqref{eq:Disc-m} together with a PN approximation:
In dashed red is the 8PN tail for $a=0.9$; In dashed blue is the same tail but for Schwarzschild $a=0$. }
\label{Fig:varying a}
\end{figure}

Fig.~\ref{fig:residuals_r3} shows the residuals of the GF mode $\mathcal{G}_{\ell m}$  at radii $r=r'=3$ (as opposed to $r=r'=10$ in Fig.~\ref{fig:residuals}),   and $a=0.9$. 
These plots show that stripping away the tail lays bare the QNMs. They also show that even though the PN  approximation involves a large-$r$ expansion, it converges reasonably well near the event horizon.

%
%\newpage
\begin{figure}[h!]
    \centering
    \includegraphics[width=\linewidth]{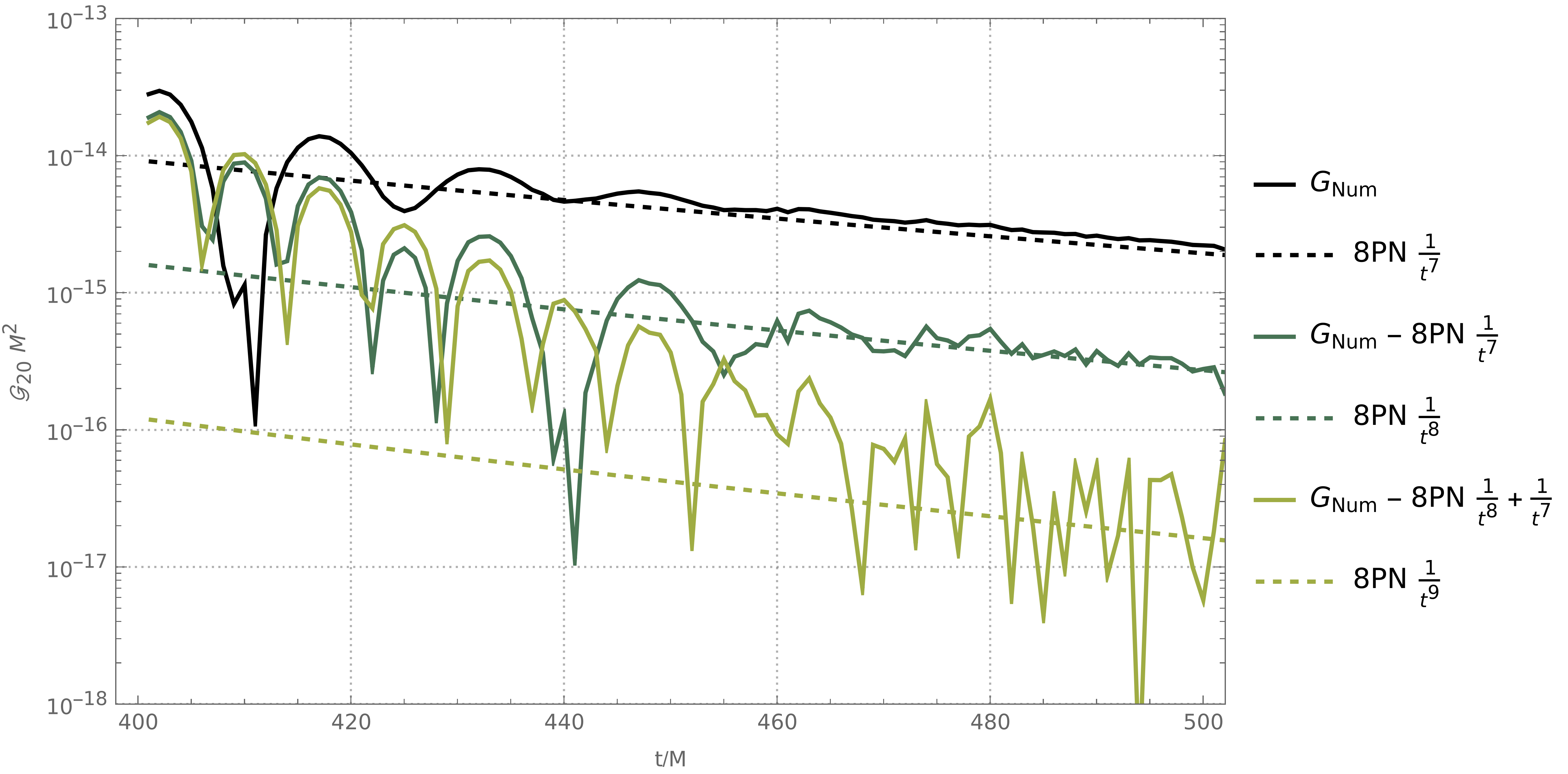}
    \caption{\textbf{Residuals for $m=0$ and $r=3$}: Log-plots of the residuals of the GF mode $\mathcal{G}_{\ell m}$ over time $t$ for $s=-2$, $\ell=2$, $m=0$ at radii $r=r'=3$,  angles $\theta=\theta'=\pi/2$ and black hole spin $a=0.9$. In black is the numerical computation of $\mathcal{G}_{\ell m}$ by integrating along the real axis according to \eqref{eq:Glm_real_axis}. In green are the residuals from successively subtracting $t^n$ terms from the 8PN (17 terms in $\eta$) tail (c.f. \eqref{[eq:Tail_G20_PN]}). In dashed is the respective $t^n$ component expanded up to 8PN. The component $t^{-9}$ contains $\log t$ terms, that are not treated as extra orders here. After subtracting $t^{-8}$ we are hitting the noise floor. This is an earlier order than in Fig.~\ref{fig:residuals} for $r=r'=10$ because the numerical computation was run to significantly lower precision.}
    \label{fig:residuals_r3}
\end{figure}

%---------------------------------------------------------------------------------------------------------

\newpage
\section{Small-frequency expansions of the In scattering coefficients}\label{sec:small-w InCoeffs}

In this appendix we provide the small-frequency expansions of the scattering coefficients $\Binc$, $\Bref$ and $\Btra$ of the In radial solution $\Rin{s}$, as defined via Eq.~\eqref{eq:bc Rin} with the normalization determined by Eq.~\eqref{eq:RIn}.
We provide the coefficients in the expansions below explicitly for spin $s=-2$. However, we have also determined the leading orders for all $s=0,\pm 1,\pm 2$, $\ell$, and $m$ in the  small-$\sigma$ expansions for the following coefficients: (i) $\Binc$ has leading order $\sigma^{-\ell-1}$; (ii) $\Bref$ has leading  order $\sigma^{-2s -\ell -1}$; (iii) $\Btra$ has leading  order
 $\sigma^{-s+b}$, where $b=0$ except for $s>0$ and $m=0$, in which case it is $b=1$.

The small-$\sigma$ expansions are calculated using the MST series representations in Eq.~\eqref{eq:Binc/ref/tra} and using the expansions for the MST series coefficients $\an{n}$ and the renormalized angular momentum $\nu$ provided in Sec.~\ref{sec:small-w an-nu}.
The expansions are in principle valid for $\text{Re}(\omega)\geq 0$ and one may use the symmetries in \eqref{eq:symm-In} to obtain the expansions for $\text{Re}(\omega)< 0$.

First, the In incidence scattering coefficient for $s=-2$:
 \begin{equation}\label{eq:Binc}
%\BiW{\indmode}
\Bincp
=
\sigma ^{-\ell -1}
\BiW{\text{pre}} \left(1 +\left(\BiW{1}+\BiW{1,l}\,\log(2 \sigma)\right)\sigma+\left(\BiW{2}+\BiW{2,l}\,\log(2 \sigma)+\BiW{2,ll}\,\log(2 \sigma)^2\right)\sigma^2+o(\sigma^2)\right), 
\end{equation}
 where
\begin{subequations}\label{:eq:Binc-exp}
\begin{align}
\BiW{\text{pre}} &= (-1)^{ \ell }2^{-\ell -3} (\ell -1) \ell  (\ell +1) (\ell +2), \\
\BiW{1}&=\frac{4 i a m \left(\ell ^2+\ell -1\right)}{\ell ^2 (\ell +1)^2}-2
   \log 2+\frac{\ell ^4+6 \ell ^3-\ell ^2-14 \ell -4}{(\ell -1) \ell  (\ell +1)
   (\ell +2)}+\frac{4 \kappa }{\ell  (\ell +1)}+4 \psi ^{(0)}(\ell )+2 i \pi, \\
   \BiW{1,l}&=2,\\
   \BiW{2}&=-a^2 m^2 \left(\frac{15 \left(124 \ell ^2+124 \ell -153\right)}{(2 \ell -1)^2 (2 \ell
   +3)^2}-\frac{4 \left(29 \ell ^6+87 \ell ^5+94 \ell ^4+43 \ell ^3-\ell ^2-8 \ell
   -8\right)}{\ell ^4 (\ell +1)^4}\right)
   \nonumber\\
   &
   +\frac{15 a^2 \left(124 \ell ^2+124 \ell
   -153\right)}{4 (2 \ell -1)^2 (2 \ell +3)^2}-\frac{32 a^2}{\ell  (\ell
   +1)}+\frac{1}{4} \left(a^2+1\right)
   \nonumber\\
   &
   +\log{2} \left(-\frac{8 i a m \left(\ell
   ^2+\ell -1\right)}{\ell ^2 (\ell +1)^2}-\frac{2 \left(\ell ^4+6 \ell ^3-\ell ^2-14
   \ell -4\right)}{(\ell -1) \ell  (\ell +1) (\ell +2)}-\frac{8 \kappa }{\ell  (\ell
   +1)}-8 \psi ^{(0)}(\ell )-4 i \pi \right)
   \nonumber\\
   &
   +\kappa  \left(i a m \left(\frac{225}{(2
   \ell -1) (2 \ell +3)}-\frac{8 \left(7 \ell ^4+14 \ell ^3+11 \ell ^2+4 \ell
   +4\right)}{\ell ^3 (\ell +1)^3}\right)+\frac{4 \left(\ell ^4+6 \ell ^3-\ell ^2-14
   \ell -4\right)}{(\ell -1) \ell ^2 (\ell +1)^2 (\ell +2)}+\frac{8 i \pi }{\ell  (\ell
   +1)}\right)
   \nonumber\\
   &
   +\psi ^{(0)}(\ell ) \left(\frac{16 i a m \left(\ell ^2+\ell
   -1\right)}{\ell ^2 (\ell +1)^2}+\frac{4 \left(\ell ^4+6 \ell ^3-\ell ^2-14 \ell
   -4\right)}{(\ell -1) \ell  (\ell +1) (\ell +2)}+\frac{16 \kappa }{\ell  (\ell +1)}+8
   i \pi \right)
   \nonumber\\
   &
   +m \left(-\frac{8 \pi  a \left(\ell ^2+\ell -1\right)}{\ell ^2 (\ell
   +1)^2}+\frac{4 i a \left(\ell ^2+\ell -1\right) \left(\ell ^4+6 \ell ^3-\ell ^2-14
   \ell -4\right)}{(\ell -1) \ell ^3 (\ell +1)^3 (\ell +2)}\right)+2
   (\log{2})^2-\frac{225 \left(4 \ell ^2+4 \ell -7\right)}{4 (2 \ell -1)^2 (2 \ell
   +3)^2}
   \nonumber\\
   &
   +2 i \pi  \left(\frac{143 \ell ^2+143 \ell -51}{(2 \ell -1) (2 \ell +1) (2
   \ell +3)}-\frac{12 \left(\ell ^3+2 \ell ^2-\ell -1\right)}{(\ell -1) \ell  (\ell +1)
   (\ell +2)}+1\right)+\frac{4 \left(\ell ^5+7 \ell ^4+3 \ell ^3-12 \ell ^2-9 \ell
   +4\right)}{(\ell -1) \ell ^2 (\ell +1)^2 (\ell +2)}
   \nonumber\\
   &
   +8 \psi ^{(0)}(\ell )^2-2 \pi ^2,\\
   \BiW{2,l}&=-\frac{2
   \left(143 \ell ^2+143 \ell -51\right)}{(2 \ell -1) (2 \ell +1) (2 \ell +3)}+\frac{24
   \left(\ell ^3+2 \ell ^2-\ell -1\right)}{(\ell -1) \ell  (\ell +1) (\ell +2)}-8 \psi ^{(0)}(\ell )+4 \log{2}-4 i \pi -2
   \nonumber\\
   &
  -\frac{8
   \kappa }{\ell  (\ell +1)} -\frac{8 i a m \left(\ell ^2+\ell -1\right)}{\ell ^2 (\ell +1)^2},
   \\
    \BiW{2,ll}&=2.
 \end{align}
 \end{subequations}

As for the In reflection scattering coefficient for $s=-2$:
 \begin{equation}\label{eq:Bref}
%\BrW{\indmode}
\Bref
=
\sigma ^{3-\ell }
\BrW{\text{pre}} \left(1 +\left(\BrW{1}+\BrW{1,l}\,\log(2 \sigma)\right)\sigma+\left(\BrW{2}+\BrW{2,l}\,\log(2 \sigma)+\BrW{2,ll}\,\log(2 \sigma)^2\right)\sigma^2+o(\sigma^2)\right), 
\end{equation}
 where
\begin{subequations}\label{:eq:Bref-exp}
\begin{align}
\BrW{\text{pre}} &=-2^{1-\ell },\\
\BrW{1}&=-\frac{4 i a m}{\ell ^2 (\ell +1)^2}+\frac{4 \kappa }{\ell  (\ell +1)}-1+2 \log 2,\\
\BrW{1,l}&=2,\\
\BrW{2}&=-a^2 m^2 \left(\frac{225 \left(4 \ell ^2+4 \ell -7\right)}{(2 \ell -1)^2 (2 \ell
   +3)^2}-\frac{4 \left(14 \ell ^6+42 \ell ^5+39 \ell ^4+8 \ell ^3-19 \ell ^2-16 \ell
   -8\right)}{\ell ^4 (\ell +1)^4}\right)
   \nonumber\\
   &
   +a^2 \left(\frac{225 \left(4 \ell ^2+4 \ell
   -7\right)}{4 (2 \ell -1)^2 (2 \ell +3)^2}-\frac{16}{\ell  (\ell
   +1)}+\frac{1}{4}\right)+\log{2} \left(-\frac{8 i a m}{\ell ^2 (\ell +1)^2}+\frac{8
   \kappa }{\ell  (\ell +1)}-2\right)+\frac{4 i a m (\ell  (\ell +1)+1)}{\ell ^2 (\ell
   +1)^2}
   \nonumber\\
   &
   +i a \kappa  m \left(\frac{225}{(2 \ell -1) (2 \ell +3)}-\frac{8 \left(7 \ell
   ^4+14 \ell ^3+13 \ell ^2+6 \ell +4\right)}{\ell ^3 (\ell +1)^3}\right)+2
   (\log{2})^2+\frac{8 \left(2 \ell ^2+2 \ell +1\right)}{\ell ^2 (\ell +1)^2}
   \nonumber\\
   &
   -\frac{225
   \left(4 \ell ^2+4 \ell -7\right)}{4 (2 \ell -1)^2 (2 \ell +3)^2}-\frac{4 \kappa
   }{\ell  (\ell +1)}+\frac{1}{4},\\
\BrW{2,l}&=\frac{8 i a m}{\ell ^2 (\ell +1)^2}+4 \log{2}-2 \left(\frac{143 \ell ^2+143 \ell
   -51}{(2 \ell -1) (2 \ell +3) (2 \ell +1)}-\frac{8 (2 \ell +1)}{\ell  (\ell
   +1)}+1\right)+\frac{8 \kappa }{\ell  (\ell +1)},\\
\BrW{2,ll}&=2.
 \end{align}
 \end{subequations}
 
 Finally, the In transmission scattering coefficient for $s=-2$:
\begin{align}\label{eq:Btra-exp,s=-2}
%\BtW{\indmode}(\sigma)
\Btra
&= 
\sigma^{2}\BtW{\text{pre}}
\left(1+\BtW{1}\sigma+\BtW{2}\sigma^2+O(\sigma^3)\right), 
\end{align}
where
\begin{subequations}\label{eq:Qref coeffs}
\begin{align}
    %\beta=&\ell
   % \\
    \BtW{\text{pre}} &= -\frac{2^{\ell -4} (\ell +1)^2 (\ell
   +2)^2 \kappa ^{\ell -2} \Gamma (\ell )^2
   e^{-\frac{1}{2} i a m \left(\frac{2 \log
   (\kappa )}{\kappa +1}+1\right)} \Gamma
   \left({i a m}/{\kappa }+\ell
   +1\right)}{(2 \ell +1)
   \Gamma (2 \ell )^2 \Gamma \left({i a
   m}/{\kappa }+3\right)},\\
   \BtW{1} &=\frac{2}{\kappa } \left((\kappa +1) \psi
   ^{(0)}\left({i a m}/{\kappa
   }+1\right)-\psi ^{(0)}\left({i a
   m}/{\kappa }+\ell \right)+\kappa  \log
   (\kappa )\right)+
   \notag\\
   &
   \frac{1}{\ell ^2 (\ell +1) (\kappa
   +i a m) (2 \kappa +i a m) (\kappa  \ell +i a
   m)}
  \left(4 a^4 m^4 (\ell -2)+a^4 m^2 \left(\ell
   ^4+4 \ell ^3+23 \ell ^2+8 \ell -16\right)
 \right.
   \notag\\
   &
   +2
   a^4 \ell ^2 \left(\ell ^2+\ell +8\right)+a^3
   m^3 \left(-i (\ell +1) \ell ^2-i \kappa 
   \left(\ell ^3+5 \ell ^2+12 \ell
   -24\right)\right)+
  \notag\\
   &
   a^3 m \left(-i \kappa 
   \left(3 \ell ^2+5 \ell +34\right) \ell ^2-i
   (\ell +1) (7 \ell +8) \ell ^2\right)+a^2 m^2
   \left(-\ell ^4-6 \ell ^3-\kappa  (\ell +1)
   (\ell +7) \ell ^2-25 \ell ^2-8 \ell
   +16\right)+
   \notag\\
   &
  %  \left. 
   a^2 \left(-8 \kappa  (\ell +1)
   \ell ^3-2 \left(5 \ell ^2+3 \ell +14\right)
   \ell ^2\right)+
   \notag\\
   &
    \left. 
   a m \left(i \kappa  \left(7
   \ell ^2+9 \ell +34\right) \ell ^2+i (\ell
   +1) (7 \ell +8) \ell ^2\right)+8 \kappa 
   \ell ^3 (\ell +1)+4 \ell ^2 \left(2 \ell
   ^2+\ell +3\right)\right),\label{eq:B1tra-exp}
\end{align}
%\end{subequations}
where the combination in brakets in the first line is regular in the limit $|a|\to 1$ (implying $\kappa\to 0$) for $m\neq 0$,
\begin{align}
 \BtW{2} &=2\left(\frac{2}{\kappa } \left((\kappa +1) \psi
   ^{(0)}\left({i a m}/{\kappa
   }+1\right)-\psi ^{(0)}\left({i a
   m}/{\kappa }+\ell \right)+\kappa  \log
   (\kappa )\right)\right)^2+
   \notag\\
   &
 \frac{2 }{\kappa ^2}\psi ^{(1)}\left(\ell+{i a m}/{\kappa
   } \right) -\frac{2 (\kappa
   +1)^2 }{\kappa ^2}\psi ^{(1)}\left({i a m}/{\kappa
   }+1\right)+ 4 \psi ^{(1)}(\ell )+\frac{4 \ell  (\ell +1) (15 \ell^2 
   +5 \ell+13)+96}{\ell  (\ell +1) (2 \ell -1) (2
   \ell +1) (2 \ell +3)}\left(\psi ^{(0)}(\ell )-2 \psi ^{(0)}(2\ell )\right)+
    \notag\\
   &  
   \frac{2 \left(15 \ell ^4+30 \ell ^3+28 \ell
   ^2+13 \ell +24\right)}{\ell  (\ell +1) (2
   \ell -1) (2 \ell +1) (2 \ell +3) }\left(
   (\kappa +1) \psi ^{(0)}\left({i a
   m}/{\kappa }+1\right)+(1+\kappa) \log \kappa + \log 2\right) +
     \notag\\
   & 
 \frac{c_1}{ (a m-i  \kappa ) (a m-2 i \kappa ) (a m - i \ell \kappa)}
  \left(\frac{2}{\kappa } \left((\kappa +1) \psi
   ^{(0)}\left({i a m}/{\kappa
   }+1\right)-\psi ^{(0)}\left({i a
   m}/{\kappa }+\ell \right)+\kappa  \log
   (\kappa )\right)\right)+   
       \notag\\
   & 
 \frac{c_2}{ (a m-i  \kappa )^2 (a m-2 i \kappa )^2 (a m - i  \ell \kappa )},
 \label{eq:B2tra-exp}
\end{align}
\end{subequations}

\begin{align}
c_1&= \frac{1}{8} \left( a^3 m^3+i a^2 m^2 \left(a^2 (8 \ell +9)-8
   \ell -25\right)+8 a \left(a^2-1\right) m
   (7 \ell +8)+2 i \left(a^2-1\right)
   \left(a^2 (8 \ell -15)-32 \ell
   +31\right)+
 \right.
      \notag\\
   & \qquad \qquad
 \left.
  \left(8 a^3 m^3-8 i
   a^2 m^2 (\ell +7)+a m \left(a^2 (24 \ell
   -29)-56 \ell +29\right)-64 i
   \left(a^2-1\right) \ell \right)   \kappa 
   \right) +
        \notag\\
   &
 \frac{4}{\ell ^2 (\ell +1)} 
 \left(
 i a^4 m^4 (\ell -2)+i m^2 \left(4 a^2
   \left(1-a^2\right) (1-2 \ell )-15 a^2
   \left(1-a^2\right) \ell ^2\right)+
 \right.
      \notag\\
   & \qquad \qquad
 \left.
\left(a^3 m^3 \left(5 \ell ^2+5 \ell
  -6\right)-2 a \left(1-a^2\right) m \ell 
  (5 \ell +2)\right) \kappa \right)
       \notag\\
   &
  \frac{1}{8 (2 \ell -1) (2 \ell +1) (2 \ell
   +3)} \left(
   5 i a^2 \left(1-a^2\right) m^2 \left(572
   \ell ^2+662 \ell -159\right)+2 i
   \left(1-a^2\right)^2 \left(572 \ell
   ^2+122 \ell -429\right)+
   \right.
       \notag\\
   & \qquad \qquad
   \left.
    \left(a
   \left(1-a^2\right) m \left(572 \ell
   ^2+1922 \ell +471\right)-8 a^3 m^3
   \left(143 \ell ^2+143 \ell
   -51\right)\right)  \kappa
   \right),
\end{align}
and
\begin{align}
c_2&=
\frac{15}{4 (2 \ell -1)^2 (2 \ell +3)}
\left(-3 a^7 m^5 \left(576 \ell ^2+1006 \ell
   -1277\right)+3 \left(1-a^2\right) a^5 m^3
   \left(1908 \ell ^2-352 \ell -541\right)
   \right.
       \notag\\
   &  \qquad
   -4
   \left(1-a^2\right)^2 a^3 m \left(54 \ell
   ^2+\ell +61\right)+ \left(3 i a^6
   m^4 \left(1534 \ell ^2+631 \ell
   -1374\right)-
   \right.
       \notag\\
   & \qquad \qquad
\left.
   \left.
   4 i \left(1-a^2\right)^2 a^2
   \ell  (62 \ell -61)-21 i
   \left(1-a^2\right) a^4 m^2 \left(146 \ell
   ^2-67 \ell +12\right)
   \right) 
   \kappa\right)+
    \notag\\
   &
   \frac{1}{12 (2 \ell -1)^2 (2 \ell +1)^2 (2
   \ell +3)}
   \left(
   a^5 m^5 \left(361216 \ell ^4+896488 \ell
   ^3-116980 \ell ^2-555994 \ell
   -167595\right)+
     \right.
       \notag\\
   & \qquad
    5 a^3
   \left(a^2-1\right) m^3 \left(313328 \ell
   ^4+266288 \ell ^3-138272 \ell ^2-75836
   \ell -6303\right)+
      \notag\\
   & \qquad
   4 a \left(a^2-1\right)^2 m
   \left(1912 \ell ^4+23356 \ell ^3+46430
   \ell ^2+3497 \ell -2895\right)+
       \notag\\
   & \qquad
   \left(-i a^4 m^4 \left(1110088 \ell
   ^4+1412764 \ell ^3-457450 \ell ^2-671887
   \ell -160170\right)+
  \right.
        \notag\\
   & \qquad      
           i a^2
   \left(1-a^2\right) m^2
   \left(902728 \ell^4+680524 \ell ^3-298690 \ell ^2-67627   \ell +23700\right)+
           \notag\\
   & \qquad
 \left.
 \left.
   4 i
   \left(1-a^2\right)^2 
   \left(30488 \ell^4+9764 \ell ^3-35270 \ell ^2-977 \ell   +3480\right)
%   +
%              \notag\\
%   & \qquad
%\left.  
%\left.    
  \right)\kappa 
   \right)+
   \notag\\
   & 
   \frac{1}{4 \ell ^4 (\ell +1)^3}
   \left(
   -a^7 m^5 \left(\ell ^7-3213 \ell ^6-13581
   \ell ^5-16463 \ell ^4-6480 \ell ^3+2048
   \ell ^2+4608 \ell +4160\right)+
 \right.
                 \notag\\
   & \qquad
\left.    
   a^7 m^3
   \left(-6 \ell ^8-31 \ell ^7+10167 \ell
   ^6+23475 \ell ^5+21923 \ell ^4+11520 \ell
   ^3+5056 \ell ^2-768 \ell -1280\right)+
  \right.
                 \notag\\
   & \qquad
\left.     
   4
   a^7 m \ell ^2 \left(-3 \ell ^6-10 \ell
   ^5-60 \ell ^4-38 \ell ^3+15 \ell ^2+192
   \ell +320\right)+32 i a^6 m^6 (\ell -1)
   \ell ^2 (\ell +1)^2
   \right)
    \notag\\
   &
   \frac{1}{12 \ell ^4 (\ell +1)^3 (2 \ell
   -1)^2 (2 \ell +3)}
   \left(
   -12 a^7 m^7 (\ell -2) \left(2 \ell ^7+7 \ell
   ^6+35 \ell ^5-61 \ell ^4+59 \ell ^3+322
   \ell ^2+196 \ell -120\right)+
    \right.
                 \notag\\
   & \qquad
\left.   
   12 i a^6
   \kappa  m^6 \left(2 \ell ^9+51 \ell ^8+65
   \ell ^7+25 \ell ^6-979 \ell ^5+736 \ell
   ^4+1188 \ell ^3-2648 \ell ^2-3120 \ell
   +1440\right)
   \right)
 \notag\\
   &  
   \frac{1}{\ell ^2 (\ell +1)}
   \left(
   -i a^2 \left(1-a^2\right) m^2 \left(a^2
   \left(31 \ell ^4+69 \ell ^3+486 \ell
   ^2+80 \ell -128\right)-73 \ell ^4-147
   \ell ^3-522 \ell ^2-80 \ell +128\right)+
       \right.
                 \notag\\
   & \qquad  
   4
   i \left(1-a^2\right)^2 \ell ^2 \left(4
   a^2 \left(\ell ^2+\ell +8\right)-21 \ell
   ^2-13 \ell -24\right)+
%        \right.
                 \notag\\
   & \qquad  
   i a^4 m^4 \left(a^2
   \left(\ell ^4+11 \ell ^3+82 \ell ^2+176
   \ell -248\right)
   -\ell ^4-13 \ell ^3-84
   \ell ^2-176 \ell +248\right)+
                    \notag\\
   & \qquad
%\left.    
   \left(
   	a^5 m^5 \left(\ell ^3+9 \ell ^2+64   \ell -80\right)
   	-2 a \left(1-a^2\right) m
   \ell ^2 \left(
   a^2 \left(19 \ell ^2+27 \ell +200\right)
   -4 \left(17 \ell ^2+18 \ell +49\right)
   \right)+
   \right.
                     \notag\\
   & \qquad
\left.
\left.       
   a^3 m^3 \left(a^2
   \left(10 \ell ^4+41 \ell ^3+287 \ell
   ^2+200 \ell -304\right)-14 \ell ^4-67
   \ell ^3-309 \ell ^2-200 \ell
   +304\right)\right)   \kappa 
   \right)
   \notag\\
   &  
   \frac{1}{12 \ell ^4 (\ell +1)^3 (\ell +2)^2}
   \left(
   a^5 m^5 \left(9 \ell ^9-11153 \ell ^8-89125
   \ell ^7-274239 \ell ^6-404816 \ell
   ^5-280348 \ell ^4-40896 \ell ^3+
 \right.
  \right.
     \notag\\
   &  \qquad
\left.
   92160
   \ell ^2+105216 \ell +49920\right)+4 a m
   \ell ^2 \left(a^4 \left(309 \ell ^8+2262
   \ell ^7+7828 \ell ^6+15830 \ell ^5+16491
   \ell ^4+2644 \ell ^3 
    \right.
      \right.
     \notag\\
   &  \qquad
\left.  
   -13876 \ell ^2-19008   \ell -10176\right)
   -a^2 \left(699 \ell
   ^8+5010 \ell ^7+16196 \ell ^6+30130 \ell
   ^5+29517 \ell ^4+5468 \ell ^3-17708 \ell
   ^2
   \right.
       \notag\\
   &  \qquad 
   \left.  
   \left.  
   -19584 \ell -8832\right)+399 \ell
   ^8+2814 \ell ^7+8704 \ell ^6+15254 \ell
   ^5+14157 \ell ^4+2524 \ell ^3-7276 \ell
   ^2-6720 \ell -2496\right)+
       \notag\\
   &  \qquad
   a^3 m^3 \left(4
   a^2 \left(42 \ell ^{10}+441 \ell ^9-17477
   \ell ^8-118693 \ell ^7-294201 \ell
   ^6-356684 \ell ^5-237748 \ell ^4-98496
   \ell ^3-21936 \ell ^2 +
   \right.
   \right.
         \notag\\
   &  \qquad
  \left.
   12288 \ell
   +7680\right)-150 \ell ^{10}-1647 \ell
   ^9+39515 \ell ^8+281803 \ell ^7+706131
   \ell ^6+846632 \ell ^5+534316 \ell
   ^4+
        \notag\\
   &  \qquad 
 %  \left.
   \left.
   197376 \ell ^3+40128 \ell ^2-24576
   \ell -15360\right)
+\left(
i a^4 m^4 \left(3 a^2 (\ell +2)^2 \left(\ell
   ^8+9 \ell ^7-8443 \ell ^6-24477 \ell
   ^5-26026 \ell ^4
   \right.
     \right.
       \right.
           \notag\\
   &  \qquad 
   \left.
   -11568 \ell ^3-928 \ell
   ^2+3776 \ell +3840\right)-9 \ell
   ^{10}-189 \ell ^9+33011 \ell ^8+228037
   \ell ^7+602946 \ell ^6+781004 \ell
   ^5+515368 \ell ^4+
            \notag\\
   &  \qquad 
   \left.
   143040 \ell ^3-46848
   \ell ^2-91392 \ell -46080\right)+i a^2
   m^2 \ell  \left(3 a^4 (\ell +2)^2
   \left(13 \ell ^7+51 \ell ^6-4885 \ell
   ^5-9999 \ell ^4-8404 \ell ^3
 \right.  
 \right.
      \notag\\
   &  \qquad 
   \left.
   -5760 \ell
   ^2-4480 \ell -256\right)+4 a^2 \left(-183
   \ell ^9-1503 \ell ^8+4601 \ell ^7+51307
   \ell ^6+134358 \ell ^5+163172 \ell
   ^4+113800 \ell ^3+
  \right. 
       \notag\\
   &  \qquad 
   \left.
   60912 \ell ^2+25536
   \ell +1536\right)+837 \ell ^9+6711 \ell
   ^8-1781 \ell ^7-113623 \ell ^6-331308
   \ell ^5-413996 \ell ^4-271792 \ell
   ^3
        \notag\\
   &  \qquad 
\left.
   -120000 \ell ^2-45312 \ell
   -3072\right)+4 i \left(a^2-1\right) \ell
   ^3 (\ell +1) \left(3 a^4 (\ell +2)^2
   \left(\ell ^4+2 \ell ^3+161 \ell ^2+96
   \ell +64\right)
   \right.
       \notag\\
   &  \qquad 
     -8 a^2 \left(21 \ell
   ^6+123 \ell ^5+479 \ell ^4+1283 \ell
   ^3+1672 \ell ^2+796 \ell +72\right)+
     \notag\\
   &  \qquad 
   \left.  
    \left. 
     \left. 
   249
   \ell ^6+1398 \ell ^5+3973 \ell ^4+8092
   \ell ^3+9572 \ell ^2+4160 \ell
   -192\right)
\right)\kappa
   \right).
\end{align}

%---------------------------------------------------------------------------------------------------------
%---------------------------------------------------------------------------------------------------------

\bibliography{references}{}

\bibliographystyle{apsrev}

\end{document}